\documentclass[12pt]{report}
\usepackage[nodraft]{uncthesis_sp07}			
\usepackage{setspace}						
\usepackage{amsmath}
\usepackage{amssymb}
\usepackage{multirow}
\usepackage[noadjust]{cite}
\usepackage{tocloft}              		
\usepackage{ifthen}
\usepackage{longtable}
\usepackage{palatino}			
\usepackage{bm}
\usepackage{indentfirst}
\usepackage{graphics,graphicx}				
\usepackage{appendix}
\usepackage[linktocpage=true]{hyperref}			
\usepackage[nonumberlist,hyperfirst=false,sort=def]{glossaries}

\newcommand{\gsim}{\, \raisebox{-0.8ex}{$\stackrel{\textstyle >}{\sim}$ }}
\newcommand{\lsim}{\, \, \raisebox{-0.8ex}{$\stackrel{\textstyle <}{\sim}$ }}

\newcommand{\beq}{\begin{equation}}
\newcommand{\eeq}{\end{equation}}
\newcommand{\ba}{\begin{array}}
\newcommand{\ea}{\end{array}}
\newcommand{\bea}{\begin{eqnarray}}
\newcommand{\eea}{\end{eqnarray}}
\newcommand{\bi}{\begin{itemize}}  
\newcommand{\ei}{\end{itemize}}
\newcommand{\ben}{\begin{enumerate}} 
\newcommand{\een}{\end{enumerate}}
\newcommand{\bc}{\begin{center}}
\newcommand{\ec}{\end{center}}

\mathchardef\mhyphen="2D

\let\oldbibliography\thebibliography
\renewcommand{\thebibliography}[1]{%
 \oldbibliography{#1}%
 \singlespacing                     
 \setlength{\itemsep}{15pt}   
}

\newcommand*{\noaddvspace}{\renewcommand*{\addvspace}[1]{}}
\addtocontents{lof}{\protect\noaddvspace} 
\addtocontents{lot}{\protect\noaddvspace} 

\makeglossaries
\newacronym{A4}{${\bf A_4}$}{Tetrahedral Symmetry Group}
\newacronym{BSM}{BSM}{Beyond the Standard Model}
\newacronym{CKM}{CKM}{Cabbibo-Kobayashi-Maskawa matrix}
\newacronym{CP}{${\bf CP}$}{Charge-Parity Symmetry}
\newacronym{IH}{IH}{Inverted Neutrino Mass Hierarchy}
\newacronym{irrep}{irrep}{Irreducible Representation}
\newacronym{MRTM}{MR${\bf T^{'}}$M}{Minimal Renormalizable ${\bf T^{'}}$ Model}
\newacronym{MSM}{MSM}{Minimal Standard Model}
\newacronym{NH}{NH}{Normal Neutrino Mass Hierarchy}
\newacronym{NMRTM}{NMR${\bf T^{'}}$M}{Next-to-Minimal Renormalizable ${\bf T^{'}}$ Model}
\newacronym{PMNS}{PMNS}{Pontecorvo-Maki-Nakagawa-Sakata matrix}
\newacronym{QCD}{QCD}{Quantum Chromodynamics}
\newacronym{QED}{QED}{Quantum Electrodynamics}
\newacronym{SUSY}{SUSY}{Supersymmetry}
\newacronym{T'}{${\bf T^{'}}$}{Binary Tetrahedral Symmetry Group}
\newacronym{TBM}{TBM}{Tribimaximal Mixing}
\newacronym{VEV}{VEV}{Vacuum Expectation Value}
\newacronym{WIMP}{WIMP}{Weakly Interacting Massive Particle}
\newacronym{Z2}{${\bf Z_2}$}{Cyclic Symmetry Group, of Order 2}

\renewcommand{\footnoterule}{%
  \kern -3pt\
  \hrule width 2in       
  \kern 2.6pt}

\makeatletter     
\renewcommand*\l@chapter[2]{
\vskip 1.0em \@plus\p@  
\begingroup
   \bfseries   
   \@dottedtocline{1}{0em}{1.5em}{#1}{\textbf{#2}}\par  
\endgroup}
\renewcommand*\l@section{\addvspace{1em}\@dottedtocline{1}{1.5em}{2.3em}}
\renewcommand*\l@subsection{\addvspace{1em}\@dottedtocline{2}{3.8em}{3.2em}}
\renewcommand*\l@subsubsection{\addvspace{1em}\@dottedtocline{3}{7.0em}{4.1em}}
\renewcommand*\l@paragraph{\addvspace{1em}\@dottedtocline{4}{10em}{5em}}
\renewcommand*\l@subparagraph{\addvspace{1em}\@dottedtocline{5}{12em}{6em}}
\renewcommand{\@biblabel}[1]{[#1]\hfill}
\makeatother       

\allowdisplaybreaks[1]      
\begin{document}

\thesistitle{Binary Tetrahedral Flavor Symmetry}

\thesisauthor{David A. Eby}
\thesisadvisora{Y. Jack Ng (Chair)}
\thesisreadera{Charles R. Evans}
\thesisreaderb{Jonathan Engel}
\thesismembera{Thomas W. Kephart}
\thesismemberb{Amy L. Oldenberg}

\thesismaketitle 
\thesismakecopyright
\doublespace 
\begin{abstract}
A study of the $T^{'}$ Model and its variants utilizing Binary Tetrahedral Flavor Symmetry. We begin with a description of the historical context and motivations for this theory, together with some conceptual background for added clarity, and an account of our theory's inception in previous works. Our model endeavors to bridge two categories of particles, leptons and quarks, a unification made possible by the inclusion of additional Higgs particles, shared between the two fermion sectors and creating a single coherent system. This is achieved through the use of the Binary Tetrahedral symmetry group and an investigation of the Tribimaximal symmetry evidenced by neutrinos. Our work details perturbations and extensions of this $T^{'}$ Model as we apply our framework to neutrino mixing, quark mixing, unification, and dark matter. Where possible, we evaluate model predictions against experimental results and find excellent matching with the atmospheric and reactor neutrino mixing angles, an accurate prediction of the Cabibbo angle, and a dark matter candidate that remains outside the limits of current tests. Additionally, we include mention of a number of unanswered questions and remaining areas of interest for future study. Taken together, we believe these results speak to the promising potential of finite groups and flavor symmetries to act as an approximation of nature.
\end{abstract}


\begin{acknowledgments}
First and foremost, David Eby is grateful to Paul Frampton, his longtime advisor, co-author, and colleague for his guidance and his irreplaceable assistance in completing his research and this dissertation. David is also grateful to the rest of his committee, Y. Jack Ng in particular, for their backing despite trying circumstances. He would like to thank his friends for their support and the department for its counsel and instruction. He would also like to express his gratitude to Shinya Matsuzaki, the former group postdoc, for useful discussions and a productive collaboration. He wishes to thank his Family for their enduring love and, lastly, sends thanks to those unmentioned that have aided in his work and his life.
\end{acknowledgments}

%

\begin{spacing}{1.0}
\clearpage
\tableofcontents       
\clearpage
\listoftables                
{\addcontentsline{toc}{chapter}{\listtablename}} 
\clearpage
\listoffigures              
{\addcontentsline{toc}{chapter}{\listfigurename}} 
\clearpage
\renewcommand{\glossaryname}{List of Abbreviations}
\printglossaries
\end{spacing}

\cleardoublepage 

\pagenumbering {arabic}


\chapter{Introduction}
\label{intro}
\section{Overview}
Particle physics currently stands in transition. After a search lasting decades, the Higgs Boson, the final outstanding prediction of the Standard Model, has been discovered. And, despite lingering unexplained questions leading to dozens of innovative theories developed in the intervening years since the Standard Model's creation, the physics community has not settled on a likely successor, or even agreed on a single direction to pursue. Currently, there are numerous theories working on a multitude of problems in the hopes of uncovering the path to a more fundamental understanding of nature.

In this dissertation we hope to describe one of the relatively newer areas of study, and, more specifically, to explain the development, evolution, and means for evaluating our model of Binary Tetrahedral Flavor Symmetry. While this text includes background on several topics, we should note that it is written with the expectation that the reader has a working knowledge of both the particle content and the fundamental principles of the Standard Model. This is not written with the intention of being a text for instruction or a comprehensive resource, and merely describes the relevant sections of traditional particle theory. For the duration, except where noted, we will be using the natural units of $\hslash c=1$
 
In this first chapter, we give an extended background on several topics. We highlight the successes and ongoing shortcomings of the Standard Model, noting key features that tie in with our model. Group Theory, in particular, remains a key tool in the study of nature. We define mathematical groups and describe a series of groups, both familiar and practical examples, for later use. The uncharged leptons, known as neutrinos, have continued to prove surprising and notoriously difficult to explain in the 80 years since first proposed. Despite these challenges, neutrinos act as an entry point for most modern flavor models, including ours, and it behooves us to describe recent discoveries and suggested explanations in that area. This includes neutrino masses, mixings, and the proposed Majorana designation. Our background will conclude with a description of the \gls{MRTM}. This model serves as a starting point for all of our subsequent original work and we include both its approach to the dual sectors of quarks and leptons, as well as its pioneering Cabibbo Angle approximation.

In the second chapter, and first chapter of original work, we discuss a significant revision to the previously established model. By rearranging our assumptions of inputs and variables, we are able to determine new relations between the mixing parameters of quarks and leptons. We describe the manner by which we have determined a relation between the neutrino mixing angles as a result of the deviation between the experimental data and the prior Cabibbo Angle prediction. We assess estimates for both individual 3-neutrino mixing values, and correlated values, in light of the recent groundbreaking experimental results. While no experiments currently running have the express purpose of validating our model, we compare our predictions with the global fits of accumulated neutrino data.

The following chapter seeks to extend our model in order to encompass the full quark mixing matrix, rather than a simplified mixing of the first two quark families. The \gls{NMRTM} will introduce both many new parameters and a greatly expanded potential utility. Though our predictions of quark mixing, Cabibbo angle aside, remain limited in comparison with their neutrino counterparts, we hope this model will better enable testing of our ideas as mixing parameters become better defined in the coming years.

Next, we seek to integrate our prior work more closely with the physics of the Standard Model. Where, elsewhere in our research, we are primarily concerned with flavor physics, here we attempt to fuse our work with the famed physics of (${\bf SU(3)_{\rm C}}\times {\bf SU(2)_{\rm L}} \times {\bf U(1)_{\rm Y}}$). This is achieved by combining the binary tetrahedral group with multiple ${\bf SU(3)}$ groups, in an arrangement called quartification. While not as data driven as our other work, this discussion provides a successful test case of unification, and may lay the groundwork for connection with other theories, including GUTs and string theory. 

Our final chapter of original work modifies the symmetry groups and particle content of our model to create a potential explanation for dark matter. We first discuss the general mechanism that allows for a suitable dark matter particle candidate to arise out of the finite symmetry, as well as the specifics of how these methods are integrated into our theory. We describe several properties of this new particle and show how it currently remains outside current testing limits.

We end our discussion with some concluding thoughts. These include a description of which ongoing and proposed experiments will provide results suitable for testing our theories in the coming decade, as well as a summary of the limitations of our model and the remaining outstanding questions, wrapping up with a few reflections on the greater significance of this work.

Not since the years prior to the invention of general relativity and quantum mechanics, nearly a century ago, has physics been at such a turning point. With the exhaustion of old theories and a dawning generation of colliders and detectors poised to discover new physics, we have hope that the next few years will prove just as revelatory. 

\section{Historical Theory}
\label{general}
\subsection{Flavor}
The initial discovery of flavor physics, while highly remarked upon at the time, was unrecognized for its true significance. In 1936, physicists Carl Anderson and Seth Neddermeyer were using cloud chambers to examine the decay products of cosmic radiation that survived long enough (aided by relativistic effects) to reach ground level. Cloud Chambers are designed to indicate the curved (due to a magnetic field) trail of electrically charged particles in water droplets, which, via the equations of centripetal motion, can determine a particle's mass. They discovered a particle that had a mass between that of the electron/positron and the proton (the only other charged subatomic particles known at the time).\cite{Street:1937me} One year earlier, Hideki Yukawa had proposed a new particle, dubbed the meson, to mediate the strong nuclear force and have a mass approximately the same as the newly discovered particle. This led to the muon's name, a conjunction of $\mu$ (at the time, the symbol used for mesons) and meson. Later on, physicists would recognize that the muon more closely resembled an unstable, heavier version of the electron, and would repurpose the name meson to mean bound states of quark-antiquark pairs. While the $\pi$ meson, not the muon, would turn out to be Yukawa's predicted particle, both discoveries marked a fundamental step forward in physics. 

To say the muon was unexpected would be an understatement: Isidor Rabi is reported to have said, "Who ordered $that?$'' upon hearing of its discovery, but the muon acted as a harbinger of the future of particle physics in several ways. It was the first indication that the family of leptons, fundamental fermions that do not interact via the strong nuclear force, was larger than simply the electron. It was also the first indication of flavor, and of forthcoming particle discoveries exhibiting similar interactions and ever-increasing mass scales. Thus began an 80-year struggle to understand and explain this odd corner of physics.\cite{Muons}

Our modern understanding of flavor contains much more variety. We currently know of 6 flavors of quark, organized into 3 families, each of which consists of an up-type quark and a down-type quark. The three up-types, from lightest to heaviest, are named the up, charm, and top quarks. The three down-types, from lightest to heaviest, are named the down, strange, and bottom quarks. Each quark has a flavor charge (with corresponding antiquarks given opposing flavor charges), and, as fermions, have a spin of $\tfrac{1}{2}$. Leptons, the other fermions, also have spin of $\tfrac{1}{2}$ and several flavors, one for each of the three families, each of which contains a single charged lepton ($e^{-}$, $\mu^{-}$, and $\tau^{-}$) and neutrino ($\nu_{e}$, $\nu_{\mu}$, and $\nu_{\tau}$).
 
\subsection{The Standard Model}
\subsubsection{Unification}
In the 20th century physicists developed several theories to deal with a profusion of new phenomena. By the 1930s, the community was aware of 4 fundamental forces: gravity, electromagnetism, the strong nuclear force, and the weak nuclear force. Gravity had been explained by general relativity, although it continues to resist consolidation with the other three. Electromagnetism, which had been explained in the 1800s by Maxwell's laws, was developed over the 1940s and 50s into the theory of \gls{QED}, explaining the interaction of electric and magnetic fields in terms of charged leptons and photons. The strong nuclear force continued to advance our understanding of quantum fields during the development of \gls{QCD}, which explained the inner workings of hadrons by proposing a new set of mediating bosons, the gluons. \gls{QCD} also managed to bring some order to the ever-growing number of mesons and baryons by reducing them to various combinations of the six currently known flavors of quark, in what came to be called the Eightfold Way.

In 1967, it was realized that \gls{QED} and the weak nuclear force could be unified under the gauge groups (${\bf SU(2)_{\rm L}}\times {\bf U(1)_{\rm Y}}$).\cite{Glashow:1961tr} This new electroweak symmetry was later combined with the ${\bf SU(3)_{\rm C}}$ of \gls{QCD} to form the basis of the \gls{MSM}.\cite{Weinberg:1967tq,Salam:1968rm} This model proved to be one of the most successful in the history of theoretical physics crafting dozens of accurate predictions including the $W^{\pm}$ and $Z^{0}$ bosons,\cite{Arnison:1983rp,Banner:1983jy,Arnison:1983mk,Bagnaia:1983zx} the charm,\cite{Augustin:1974xw, Aubert:1974js} bottom,\cite{Herb:1977ek} and top quarks,\cite{Abe:1995hr,Abachi:1995iq} as well as the gluon.\cite{Berger:1978rr,Berger:1979cj,Barber:1979yr,Brandelik:1979bd} With the discovery of the Higgs boson in 2012, it has now reached completion.\cite{Aad:2012tfa,Chatrchyan:2012ufa}

\subsubsection{28 Parameters}
Key parts of the overall \gls{MSM} theory are the 28 parameters (sometimes seen numbering 18, given the original assumption that neutrinos were massless and exhibited no mixing). These constants have no predicted value, and yet, are part of the theory. In a sense, it is a marvel that an accurate description of the known universe (on small scales, excluding gravity) can be described using a single framework with under thirty measurable quantities. These constants consist of:

\begin{spacing}{1.0}
\begin{itemize} 
\item $ {\rm 6~masses~for~the~quarks}$

\item $ {\rm 3~masses~for~the~charged~leptons}$

\item $ {\rm 3~mass~eigenstates~for~the~neutrinos}$

\item $ {\rm 3~gauge~couplings~for~each~of~{\bf U(1)_{\rm Y}},~{\bf SU(2)_{\rm L}},~and~{\bf SU(3)_{\rm C}}}$

\item $ {\rm 3~angles~from~the~\glsentrytext{CKM}~matrix~and~a~}\glsentrytext{CP}{\rm -violating~phase}$

\item $ {\rm 3~angles~from~the~\glsentrytext{PMNS}~matrix~and~a~}\glsentrytext{CP}{\rm -violating~phase}$\\
${\rm (assuming~Majorana~neutrinos)}$

\item $ {\rm Mass~of~the~Z^0~boson}$

\item $ {\rm Mass~of~the~Higgs~boson}$

\item $ {\rm The~\glsentrytext{QCD}~vacuum~angle}$
\end{itemize} 
\end{spacing}
These constants and their known values from Ref.~\citen{Beringer:1900zz} have been summarized in the following table.

\begin{table}[ht]
\renewcommand{\arraystretch}{1.75}
\centering
\resizebox{\columnwidth}{!}{%
\begin{tabular}{||c|c|c||}
\hline\hline
${\rm Description}$ & ${\rm Parameter}$ & ${\rm Value}$ \\
\hline\hline
${\rm Up{\it -}type~quark~masses}$ & $m_u$,~$m_c$,~$m_t$ & $2.3 {\rm MeV}$,~$1.28 {\rm GeV}$,~$173.5 {\rm GeV}$ \\
\hline
${\rm Down{\it -}type~quark~masses}$ & $m_d$,~$m_s$,~$m_b$ & $4.8 {\rm MeV}$,~$95 {\rm MeV}$,~$4.18 {\rm GeV}$ \\
\hline
${\rm Charged~lepton~masses}$ & $m_e$,~$m_{\mu}$,~$m_{\tau}$ & $511 {\rm keV}$,~$105.7 {\rm MeV}$,~$1.78 {\rm GeV}$ \\
\hline
${\rm Neutrino~mass~states}$ & $m_{\nu_{1,2,3}}$ & $\lsim 1 {\rm eV}$ \\
\hline
${\rm Gauge~couplings}$ & $g_1$,~$g_2$,~$g_3$ & $0.345$,~$0.630$,~$1.184$ \\
\hline
${\rm CKM~angles~\&~{\bf CP}{\it -}phase}$ & $\Theta_{12}$,~$\Theta_{13}$,~$\Theta_{23}$,~$\delta_{{\rm CP}}$ & $13.0^{\circ}$,~$0.2^{\circ}$,~$2.4^{\circ}$,~$0.995~{\rm rad}$ \\
\hline 
${\rm PMNS~angles~\&~{\bf CP}{\it -}phases}$ & $\theta_{12}$,~$\theta_{13}$,~$\theta_{23}$,~$\delta_{i}~i{\it =}1,2,3$ & ${\it \backsim}33.9^{\circ}$,~${\it \backsim}9.1^{\circ}$,~${\it \gsim}38.5^{\circ}$,~$\delta_{i}=?$ \\
\hline
${\rm Electroweak~scales}$ & $M_{Z^0}$,~$M_H$ & $91.2 {\rm GeV}$,~${\it \backsim}125 {\rm GeV}$ \\
\hline
${\rm QCD~vacuum~angle}$ & $\theta_{{\rm QCD}}$ & ${\it \backsim}0$ \\
\hline\hline
\end{tabular}
\label{28para}
}
\caption[Standard Model Constant Summary]{Determined under $\overline{\rm MS}$ scheme. See Ref.~\citen{Beringer:1900zz} for the individual renormalization scales used.}
\end{table}

Despite this compact form, it remains an ongoing effort among physicists to simplify and combine these items. These efforts take several forms and fall under several different searches.

\subsection{\texorpdfstring{Lingering Mysteries and \glsentrytext{BSM} Physics}{BSM}}
\label{sec:BSM}
Amazingly successful, prescient for its time, a roadmap for 40 years of particle physics, and indisputably incomplete${\it -}$these all describe the \glsreset{MSM}\gls{MSM} of particle physics. Experimentalists have spent much of the past half-century searching for the last pieces of this theory, while also hunting for indications of where it falls short. After such a long search, physicists are not entirely empty handed in their quest for so-called \gls{BSM} physics, and these problems generally fall into two categories. The first details places where experiments have deviated from \gls{MSM} expectations; the second might be generally classified as theoretical inconsistencies of \gls{MSM} theory, where idiosyncrasies (either problems or coincidences) indicate that we do not yet have a complete understanding.

\subsubsection{Experimental Contradictions}
For decades, neutrinos were thought to be massless. This was, in part, a result of how little information has been gathered in the 80 years since they were initially described. It should be noted that this is in no way an indication of a lack of interest on the part of the physics community, on the contrary, it is due to the weak nature of neutrino interactions, and the extreme means that experimentalists must go to in order to obtain statistically viable data. Consequently, when the \gls{MSM} was being formulated in the 1970s, it was believed that neutrinos were simply massless. Though it had been suggested for years that neutrinos might have a small but non-zero mass, or that additional massive neutrinos might be hidden from experiments via some hypothetical mechanism, it was not until 1998 that the Super-Kamiokande Neutrino Detector was able to measure neutrino flavor oscillation (i.e.~that a neutrino could change its own flavor, from electron neutrino to muon neutrino, for example).\cite{Fukuda:1998fd} If that were the end of it, the \gls{MSM} could be briefly amended to include neutrino masses with little other consequence. However, there were a number of additional mysteries borne out of this discovery. There is the question of why neutrino masses are so very light, of why they do not exhibit the nearly diagonal mixing exhibited by most other fermions, and why, unlike all other fermions, they do not appear to be Dirac particles. These mysteries provide much of the motivation for the models detailed in later chapters.

During the 1970s astronomers began to examine galactic rotation curves, a measure of the relative rotational velocity of galactic plane segments.\cite{Rubin:1970zza} These curves were expected to peak at a small radius and have long diminishing tails at higher radii as one traveled further from the galactic core. This would roughly correspond to a quickly spinning area fairly close to the center of the galaxy with the remainder dragging behind. Instead, they observed that the galactic rotation remained fairly constant out to large distance.\cite{Begeman:1991iy} This indicated that the density in most galaxies was steady to a much greater degree than was visibly indicated. Combined with an observation of galaxies in the Coma Cluster dating from the 1930s,\cite{Zwicky:1933gu} this missing mass was dubbed Dark Matter, as both indicated evidence for significant mass in excess of what could be visibly observed. Now there are many elements to a galaxy that are not observable from one part of the spectrum or another, but we have continued to catalog a multitude of galaxies that otherwise conform to our established models, yet still manage to have this unseen and unexplained excess of mass. Astrophysicists have also made use of gravitational lensing to indicate the presence of dark matter.\cite{Clowe:2006eq} Gravitational lensing is the practice of measuring the mass of a large stellar body by calculating the curvature of light emitted by an objects on the far side and bent around by gravity. These studies have resulted, on several occasions, in evidence for dark matter. Though there have been many suggested explanations for dark matter over the years, the most prominent one in the modern physics community is the \gls{WIMP}.\cite{Goldberg:1983nd} These are hypothetical rare heavy particles that interact only via the weak force (thus shielding them from detection by conventional means) and by gravity (corresponding with astronomical observations).

When Einstein originally formulated general relativity he inserted a cosmological constant, $\Lambda$, in order to avoid a dynamic universe size (largely due to philosophical objections). Following the discovery of Hubble expansion,\cite{Hubble} it was largely ignored for much of the mid-20th century and Einstein, himself, would later term it his "greatest mistake". In 1998 and 1999 two teams studying supernovae were able to show that the universe's expansion continues to accelerate.\cite{Perlmutter:1998np,Riess:1998cb} At this point, discussions of cosmology began reincorporating the cosmological constant, now as an indication for and quantification of our universe's acceleration (should it be constant). Dark Energy, as this phenomenon has come to be called, is the least understood problem in this section. Suggested explanations have varied widely from undiscovered fundamental forces to rapidly shifting dark energy densities.\cite{Caldwell:1997ii,Frampton:2011sp} It remains of crucial interest to physicists as estimates of this dark energy suggest it comprises $69\%$ of the energy in the universe.\cite{Ade:2013xsa}

\subsubsection{Theoretical Inconsistencies}
One of the most obvious inconsistencies in modern theory is the irreconcilability of general relativity and quantum physics. In a way, general relativity is the last vestige of classical physics, that is to say, physics before the adoption of quantum principles. General Relativity creates a smooth geometric interpretation of space and gravity, in seeming contradiction to the bubbling chaos observed on quantum scales. It would make no sense to try and build a classical quantum theory as anything other than a toy model. On the other hand, attempts to create theories of quantum gravity have proven extremely difficult. The graviton was proposed in the 1930s as a spin-2 boson conveying gravity, and one can create a unified theory for empty space without much effort. Sadly, once one begins to bend space, irreconcilable infinities begin to enter the calculations. There have been several attempts to confront these issues, most notably by string theory,\cite{Siegel:1988yz} which posits we live in a 10- or 11-dimensional universe, with the unseen dimensions existing in compact spaces. 

Another inconsistency is the hierarchy problem (sometimes known as the naturalness problem). This is the desire to achieve a working theory that explains fundamental constant values without introducing an arbitrary fine-tuning. Fine-Tuning is the suggestion that sets of constants' values are, without good justification, suspiciously coincidental. It can also mean assuming a model will match experimental data based on the exact and arbitrary placement of constants. There are whole classes of questions about either conveniently placed or coincidentally canceling fundamental constants. One example would be to ask why the gauge couplings are placed as they are. An explanation that has gathered a significant following in the physics community is \gls{SUSY}, which posits that all fermions have a bosonic superpartner and all known bosons have a fermionic superpartner. \gls{SUSY} allows the 3 \gls{MSM} force gauge couplings to unify at high energies,\cite{Dimopoulos:1981yj,Ibanez:1981yh,Marciano:1981un} answering one of these naturalness problems. There are innumerable variations of \gls{SUSY} under investigation, given the numerous mechanisms to limit or organize the additional parameters, with the search for the so-called "Lightest Supersymmetric Particle," a dark matter candidate, making up a great deal of the current generation of \gls{BSM} physics searches. It is noteworthy that string theories typically incorporate \gls{SUSY}, though for slightly different reasons. 

Another problem involving fine-tuning arises in \gls{CP}. It involves two significant physical symmetries: charge conjugation (symmetry under reversal of electric charge) and parity (symmetry under inversion of one spatial direction). These symmetries are not absolute as the \gls{MSM} contains mechanisms to allow \gls{CP}-violation. Notably the weak nuclear force violates parity alone and allows for \gls{CP}-violation in such phenomena as neutral kaon mixing. The problem enters when one notices that the \gls{QCD} contains a term that would allow for \gls{CP}-violation, but that experimentally none has been found as a result of the strong force alone. Meaning that the \gls{QCD} vacuum angle, which acts as a measure of \gls{QCD} \gls{CP}-violation strength, needs to be extremely small, or just zero. As there is no good theory-based reason to do this within the \gls{MSM}, physicists turned to \gls{BSM} theories, such as axions,\cite{Peccei:1977hh} to explain it. Another problem that develops out of \gls{CP}-violation is matter-antimatter asymmetry. If \gls{CP} were perfectly preserved all matter and antimatter would have cancelled out in the early universe. As we live in a matter-dominated universe, we are left with the question of how did matter achieve dominance. Some manner of \gls{CP}-violation seems warranted, but as \gls{QCD} sources remain ill understood, and weak sources seem poorly equipped for the magnitude of the problem, we are left with an unanswered mystery.

Another notable theoretic question is the coincident number of families in both quarks and leptons. Furthermore, each sector of fermions (excluding neutrinos where data remains limited) also exhibits a steeply increasing mass hierarchy. These similarities have given rise to the belief that an unexplained symmetry may be giving rise to the patterned behaviors of the fermions. Coincidentally, as the community discovers how neutrinos break with expectations, they may also serve as the best guide to this family symmetry. Explaining these coincidences also act as primary motivations for our research.

There are many other problems under investigation by the physics community. We have simply tried to sketch a few of the best known and most relevant here in hopes of demonstrating the continuing need for research and experimentation to further our understanding of the universe beyond that offered by the \gls{MSM}.

\section{Group Theory}
\label{groups}
\noindent
{\it This Section is largely based on Ref.~\citen{Ramond2} and Ref.~\citen{Carr:2007qw}}
\subsection{Symmetry Groups}
\subsubsection{Axioms and Examples}
We will begin our discussion of mathematical groups by describing their defining qualities. A group is a collection of operators and a defined operation. Combining those, a group containing elements: $a_1,a_2,\dotso,a_n$, must follow four mathematical rules:

\begin{spacing}{1.0}
\begin{itemize} 
\item $ {\bf Closure}{\rm~The~result~of~operating~a~group~element~on~any~other}$\\
${\rm ~will~result~in~another~element~of~the~group:}$\\
$a_1\times a_2=a_3$

\item $ {\bf Associativity}{\rm~The~group~operation~is~not~dependent~on~order:}$\\
$(a_1\times a_2) \times a_3=a_1\times (a_2 \times a_3)$

\item $ {\bf A~Unit~Element}{\rm~The~group~contains~an~element~that~can~operate~on~any}$\\
${\rm element,~including~itself,and~return~that~other~element:}$\\
$e\times a_1=a_1\times e=a_1,~~e\times e=e$

\item $ {\bf An~Inverse~Element}{\rm~For~each~group~element,}~a_1{\rm~there~exists~a~unique~element,}$\\
$(a_1)^{-1},~{\rm which~yields~the~unit~element~when~the~two~are~operated~together:}$\\
$a_1\times (a_1)^{-1}=e$
\end{itemize} 
\end{spacing}

As a demonstration, we can observe how these principles apply to the \gls{Z2}. In addition to being fairly simple and relevant to our later discussions, this group should be well-known to readers as the symmetry of multiplying positive and negative one ($+1$, $-1$). We will begin by laying out the \gls{Z2} multiplication table.
\begin{table}[ht]
\renewcommand{\arraystretch}{1.75}
\begin{center}
\begin{tabular}{||c||c|c|}
\hline\hline
${\bf Z_2}$ & a & b \\
\hline\hline
a & a & b \\
\hline
b & b & a \\
\hline
\end{tabular}
\label{Z2irreps}
\caption{Multiplication Table for \glsentrytext{Z2}}
\end{center}
\end{table}
Closure is easy to demonstrate given the only two products on the table are clearly elements of the group. As this is the symmetry demonstrated by multiplying positive and negative one, we can take the associativity of arithmetic multiplication to hold for \gls{Z2} as well. Following from that, it is fairly clear that $a$ is the unit element, while, in this case, $a$ and $b$ are each their own inverse element.

\subsubsection{Lie Groups}
While they are not the primary focus for much of our study, Lie groups are perhaps the best-known symmetry groups to most physicists, and warrant remarking on. We have already shown a finite symmetry with \gls{Z2}, but Lie groups are different in that they are continuous symmetries with an infinite number of elements. Some of the easiest examples of continuous symmetries would be rotations about various axes. ${\bf SO(2)}$ and ${\bf SO(3)}$, the special orthogonal groups, are the symmetry transformations for spherically symmetric 2- and 3-dimensional objects rotating about their center.

Also of great significance in particle physics are special unitary groups, ${\bf SU(N)}$. These groups (and their close relatives, the unitary groups, ${\bf U(N)}$) form the (${\bf SU(3)_{\rm C}}\times {\bf SU(2)_{\rm L}}\times {\bf U(1)_{\rm Y}}$) basis of the \gls{MSM}. This is, in part, due to their ability to represent spinors in quantum mechanics. It is also important to point out that ${\bf SU(2)}$ is the double cover of ${\bf SO(3)}$. Defining a double cover is difficult without getting overly technical, but can be roughly illustrated by saying that the double cover of a group will always have two elements representing a single element of the group it covers. 

\begin{figure}[ht]
\centering
\includegraphics[height=70mm]{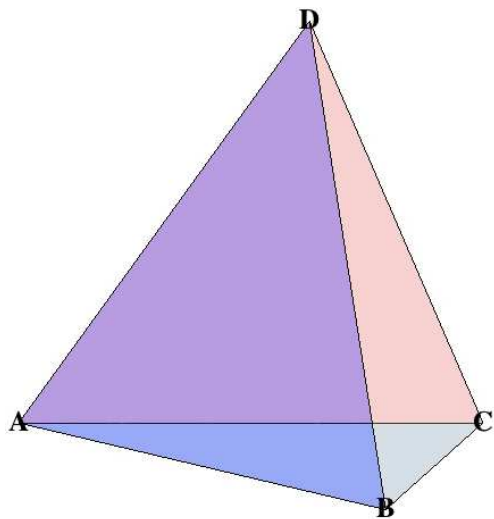}
\caption{A Reference Tetrahedron}
\label{Tfig}
\end{figure}

\subsection{\texorpdfstring{\glsentrytext{A4} and \glsentrytext{T'}}{A4 and T'}}
More pertinent to our research are two related groups, the \gls{A4} and the \gls{T'}. Like \gls{Z2} above, these are both finite non-abelian (the group elements do not necessarily commute) point groups. \gls{A4} is rank 12 and, as the name implies, consists of the elements analogous to the symmetry transformations of a tetrahedron. These transformations fall into three conjugacy classes: 

\begin{spacing}{1.0}
\begin{itemize} 
\item $C_1~{\rm the~unit~element}$

\item $C_2~{\rm a~clockwise~shift~of~vertices~by~120^{\circ}~around~the~center~of~any~of~the~four~faces}$

\item $C_3~{\rm a~counter\mhyphen clockwise~equivalent~of}~C_2$

\item $C_4~{\rm the~three~double~transpositions~of~vertices}$
\end{itemize} 
\end{spacing}

This behavior can be summarized in a group's character table, where the columns are the conjugacy classes (with a number listing its size), and the rows are the \glspl{irrep}. While there are typically several choices for any group's representation, an \gls{irrep} is a representation that cannot be reduced any further (for our purposes, one might think of it as the most efficient packaging of a group's various potential behaviors). It is interesting to note that $C_1$ and $C_4$ have real characters because they act as their own inverse elements, and $C_1$'s characters list the\gls{irrep} dimensions. The \glspl{irrep} of dimension 1 are called singlets, while the \gls{irrep} of dimension 3 is a triplet. The factor of $\omega=\exp{(2\pi i/3)}$ is the complex cube root of unity, and has a notable function once one observes that three repetitions of any single element of either $C_2$ or $C_3$ become a trivial transformation. Following the \gls{A4} character table is a second table for the Kronecker products of \glspl{irrep} operating on each other, and achieves a similar practical use as the \gls{Z2} multiplication table above.

\begin{table}[!h]
\renewcommand{\arraystretch}{1.75}
\begin{center}
\begin{tabular}{||c||c|c|c|c|}
\hline\hline
${\bf A_4}$ & $C_1$ & $4C_2$ & $4C_3$ & $3C_4$ \\
\hline\hline
$1_1$ & 1 & 1 & 1 & 1 \\
\hline
$1_2$ & 1 & $\omega$ & $\omega^2$ & 1 \\
\hline
$1_3$ & 1 & $\omega^2$ & $\omega$ & 1 \\
\hline
3 & 3 & 0 & 0 & -1 \\
\hline
\end{tabular}
\label{A4CharTab}
\caption[Character Table of \glsentrytext{A4}]{Character Table of \glsentrytext{A4} with $\omega = {\rm exp}(2i\pi/ 3)$}
\end{center}
\end{table}

\begin{table}[!h]
\renewcommand{\arraystretch}{1.75}
\begin{center}
\begin{tabular}{||c||c|c|c|c|}
\hline\hline
${\bf A_4}$ & $1_1$ & $1_2$ & $1_3$ & 3 \\
\hline\hline
$1_1$ & $1_1$ & $1_2$ & $1_3$ & $3$ \\
\hline
$1_2$ & $1_2$ & $1_3$ & $1_1$ & $3$ \\
\hline
$1_3$ & $1_3$ & $1_1$ & $1_2$ & $3$ \\
\hline
$3$ & $3$ & $3$ & $3$ & $1_1+1_2+1_3+3+3$ \\
\hline
\end{tabular}
\label{tab:A4irreps}
\caption{Kronecker Products for \glsentryplural{irrep} of \glsentrytext{A4}}
\end{center}
\end{table}

Most relevant to the models we discuss later is \gls{T'}. This group is rank 24 and is the double cover of \gls{A4} (though it is interesting to note \gls{A4} is {\it not} a subgroup of \gls{T'},\cite{Frampton:2009pr} merely its central quotient). As mentioned above, it is difficult to give a nontechnical definition of a double cover, but here it can be taken as the difference between a permutation and an oriented permutation. While a simple illustration of a permutation might be rearranging a set of playing cards, an oriented permutation would also include the possibility that cards shift from face up to face down and back. It contains 7 classes, which, in terms of \glspl{irrep}, translate into the three singlets and single triplet of \gls{A4}, as well as an additional three doublets. 

\begin{table}[!ht]
\renewcommand{\arraystretch}{1.75}
\begin{center}
\begin{tabular}{||c||c|c|c|c|c|c|c|}
\hline\hline
${\bf T^{'}}$ & $C_1$ & $4C_2$ & $4C_3$ & $C_4$ & $4C_5$ & $4C_6$ & $6C_7$ \\
\hline\hline
$1_1$ & $1$ & $1$ & $1$ & $1$ & $1$ & $1$ & $1$ \\
\hline
$1_2$ &$1$ & $\omega$ & $\omega^2$ & $1$ & $\omega$ & $\omega^2$ & $1$ \\
\hline
$1_3$ & $1$ & $\omega^2$ & $\omega$ & $1$ & $\omega^2$ & $\omega$ & $1$ \\
\hline
$2_1$ & $2$ & $1$ & $1$ & $-2$ & $-1$ & $-1$ & $0$ \\
\hline
$2_2$ & $2$ & $\omega$ & $\omega^2$ & $-2$ & $-\omega$ & $-\omega^2$ & $0$ \\
\hline
$2_3$ & $2$ & $\omega^2$ & $\omega$ & $-2$ & $-\omega^2$ & $-\omega$ & $0$ \\
\hline
$3$ & $3$ & $0$ & $0$ & $3$ & $0$ & $0$ & $-1$ \\
\hline
\end{tabular}
\label{T'CharTab}
\caption[Character Table of \glsentrytext{T'}]{Character Table of \glsentrytext{T'} with $\omega = {\rm exp}(2i\pi/3)$}
\end{center}
\end{table}

It is worth pointing out that one of the most significant qualities of \gls{T'} is that the singlet and triplet \glspl{irrep} and their multiplication remain unaltered from \gls{A4}. This, potentially, allows us to expand on ideas originally constructed for \gls{A4} without alteration, while allowing us a greater flexibility in model building due to the doublets.

\begin{table}[!ht]
\renewcommand{\arraystretch}{1.75}
\begin{center}
\begin{tabular}{||c||c|c|c|c|c|c|c|}
\hline\hline
${\bf T^{'}}$ & $1_1$ & $1_2$ & $1_3$ & $2_1$ & $2_2$ & $2_3$ & $3$ \\
\hline\hline
$1_1$ & $1_1$ & $1_2$ & $1_3$ & $2_1$ & $2_2$ & $2_3$ & $3$ \\
\hline
$1_2$ & $1_2$ & $1_3$ & $1_1$ & $2_2$ & $2_3$ & $2_1$ & $3$ \\
\hline
$1_3$ & $1_3$ & $1_1$ & $1_2$ & $2_3$ & $2_1$ & $2_2$ & $3$ \\
\hline
$2_1$ & $2_1$ & $2_2$ & $2_3$ & $1_1 + 3$ & $1_2 + 3$ & $1_3 + 3$ & $2_1 + 2_2 + 2_3$ \\
\hline
$2_2$ & $2_2$ & $2_3$ & $2_1$ & $1_2 + 3$ & $1_3 + 3$ & $1_1 + 3$ & $2_1 + 2_2 + 2_3$ \\
\hline
$2_3$ & $2_3$ & $2_1$ & $2_2$ & $1_3 + 3$ & $1_1 + 3$ & $1_2 + 3$ & $2_1 + 2_2 + 2_3$ \\
\hline
$3$ & $3$ & $3$ & $3$ & $2_1 + 2_2 + 2_3$ & $2_1 + 2_2 + 2_3$ & $2_1 + 2_2 + 2_3$ &
$1_1 + 1_2 + 1_3 + 3 + 3$ \\
\hline
\end{tabular}
\label{T'irreps}
\caption{Kronecker Products for \glsentryplural{irrep} of \glsentrytext{T'}}
\end{center}
\end{table}

\section{Recent Developments in Neutrino Theory}
\label{recent}
\subsection{Neutrino Masses and Mixings}
As previously hinted in Table~\ref{28para} neutrino masses are not quite the same as other known fundamental particles. The alignment between flavor and mass eigenstates is termed, mixing. For quark mixing, flavors and masses are very closely aligned, with only some incidental mixing between flavors. As of 1998, and the discovery of neutrino mass, physicists realized that mixing would also occur in neutrinos. Remarkably, neutrino mixing is not nearly as simple, the matrix which translates between neutrino flavor and mass eigenstates is labeled for its developers, the \gls{PMNS}.\cite{Pontecorvo:1957qd,Maki:1962mu} The matrix allows us to see that each flavor of neutrino exists as a superposition of three mass eigenstates, without any state dominating any particle,
\begin{equation}
\left( \begin{array}{c} \nu_1 \\ \nu_2 \\ \nu_3 \end{array} \right)
= U_{{\rm PMNS}} 
\left( \begin{array}{c} \nu_{\tau} \\ \nu_{\mu} \\ \nu_e \end{array} \right).
\end{equation}

In order to better understand and measure this matrix, we can parametrize the elements of the \gls{PMNS} matrix. This parametrization can be constructed of three (3$\times$3) matrices, each one aligned with a different neutrino mixing angle. The three angles are named for the type of experiment best able to determine their values: $\theta_{12}$ is the solar angle, $\theta_{13}$ is the reactor angle, and $\theta_{23}$ is the atmospheric angle. This parametrization is constructed as follows,
\begin{equation}
U_{{\rm PMNS}} =
\begin{pmatrix}
0 & 0 & 1 \\
-s_{23} & c_{23} & 0 \\
c_{23} & s_{23} & 0
\end{pmatrix}
\begin{pmatrix}
-s_{13} e^{i \delta} & 0 & c_{13} \\
0 & 1 & 0 \\
c_{13} & 0 & s_{13} e^{-i \delta}
\end{pmatrix}
\begin{pmatrix}
0 & -s_{12} & c_{12} \\
0 & c_{12} & s_{12} \\
1 & 0 & 0
\end{pmatrix},
\end{equation}
which yield the form,
\begin{equation}
U_{{\rm PMNS}} = \left( \begin{array}{ccc}
+s_{12}s_{23} - c_{12}c_{23}s_{13} e^{i\delta_{{\rm CP}}} & -s_{12}c_{23} - c_{12}s_{23}s_{13} e^{i\delta_{{\rm CP}}} & +c_{12}c_{13} \\
-c_{12}s_{23} - s_{12}c_{23}s_{13} e^{i\delta_{{\rm CP}}} & +c_{12}c_{23} - s_{12}s_{23}s_{13} e^{i\delta_{{\rm CP}}} & +s_{12}c_{13} \\
+c_{23}c_{13} & +s_{23}c_{13} & +s_{13} e^{-i\delta_{{\rm CP}}}
\end{array}
\right),
\label{PMNS}
\end{equation}
where $c$ and $s$ stand for $\cos$ and $\sin$, respectively; so $c_{12}$ is equivalent to $\cos{\theta_{12}}$. As seen above, the \gls{PMNS} matrix is comprised of the three angles ($\theta_{12}$, $\theta_{13}$, and $\theta_{23}$), as well as a phase ($\delta_{{\rm CP}}$). While each of the angles has been measured, to one degree or another, the \gls{CP}-violating phase has merely been reported at preferred values, and at present even the strictest experimental bounds encompass the entire feasible range from $0\mhyphen2\pi$. For the purposes of this document, and for the sake of simplicity in our algebra, we will assume $\delta_{{\rm CP}}=0$. We freely admit that this questionable assumption may need to be revisited in the face of future evidence to the contrary; as, indeed, current best-fit approximations place the value closer to $\delta_{{\rm CP}}=\pi$.

While non-zero neutrino mass has been clearly demonstrated, individual mass measurements still elude us (though upper bounds to neutrino masses do exist). Instead, experiments have been able to determine difference between the squares of the neutrino mass eigenstates. Consequently, we can see that for $m_2$ and $m_3$, $\Delta m_{32}^2=0.0023 {\rm eV}^2$, and for $m_1$ and $m_2$, $\Delta m_{21}^2=7.5\times 10^{-5} {\rm eV}^2$. As we do not yet know the sign for $\Delta m_{32}^2$, the ordering of the mass states remains unknown. This has resulted in two Mass Hierarchies: The \gls{NH} exhibits a natural ordering of $m_1<m_2<m_3$, while the alternative \gls{IH} places the third mass eigenstate notably lower than the other two, $m_3<m_1<m_2$.\cite{Beringer:1900zz}

\subsection{Majorana Neutrinos}
As our observations of neutrinos have grown more detailed, there have been many unexpected discoveries. Perhaps, the most puzzling to the theory community is neutrino helicity. In all other observed fermions there exist two variations: a left-handed and right-handed variant. These are determined and named for the handedness of the projection of the spin vector onto the momentum. Notably, for massless spin-$\tfrac{1}{2}$ particles, helicity can, then, be interchanged with chirality. However, neutrinos have only ever been observed with left-handedness, while anti-neutrinos are only observed with right-handedness. If we were to assume that neutrinos shared this symmetry with other Dirac fermions, then there should be four forms, not two: right- and left-handed variants of both the neutrino and anti-neutrinos. All of this begs the question ``where are the right-handed neutrinos?"

There have been several suggested solutions to this dilemma. One of the simplest, but least satisfying, is that neutrinos are simply different. Many of our expectations for the behavior of neutrinos come from observing other fermions. This bias is largely due to our proficiency at measuring particles with charges or larger masses than neutrinos. Nonetheless, neutrinos have been confounding our expectations for decades and this may simply be another difference from the rest of the \gls{MSM}. In this case, there simply would be no right-handed neutrino and, disappointingly, no significant new physics to be discovered.

A second possibility is that of sterile neutrinos. This theory holds that right-handed neutrinos exist, but have significantly different properties than their left-handed counterparts. While left-handed neutrinos interact primarily via the weak nuclear force and via gravitation, right-handed neutrinos would interact only via gravitation and, potentially, through some mixing between left- and right-handed types. This idea has some backers both among theorists and experimentalists. A number of theories require either sterile right-handed neutrinos or a 4th generation of sterile neutrinos. Experimental evidence, by comparison, falls into two camps divided over the likely mass of these particles. There have been some indications at terrestrial detectors of sterile neutrinos with mass in the ${\rm eV}$ range, most notable at LSND,\cite{Aguilar:2001ty} and more recently at MiniBooNE.\cite{AguilarArevalo:2012va} The other experimental evidence for sterile neutrinos is largely astrophysical,\cite{Abazajian:2012ys} and suggest a ${\rm keV}$ mass scale would answer questions about primordial element abundances. 

The third, and to our mind, most convincing solution to the mystery of the missing right-handed neutrinos are the so-called Majorana neutrinos. Neutrinos aside, all known fermions fall into the classification of Dirac particles, that is to say they obey the Dirac Equation,\cite{Dirac:1928hu}
\begin{equation}
(i {\not \delta}-m)\psi=0~,
\label{dirac}
\end{equation}
and are not their own anti-particles (by comparison, a photon, a boson, {\it is} its own antiparticle). Another equation and classification of particle is that of Majorana particles from 1937's Ref.~\citen{Majorana:1937vz}:
\begin{equation}
-{\not \delta}\psi +m\psi^{\ast}=0~,
\label{Majorana}
\end{equation} 
where fermions {\it will} act as their own anti-particles. While there are no confirmed examples, many in the community believe neutrinos to be attractive candidates for this designation. Under many of these proposed models, including ours, a set of undiscovered, right-handed Majorana neutrinos also exist. Hints to these particles' existence remain a high priority of neutrino detector searches.

Given the significant interest on the part of the community, an experimental design was proposed to test whether neutrinos are, indeed, Majorana particles titled, Neutrinoless Double-Beta Decay.\cite{Furry:1939qr} In typical beta decay a single neutrino is emitted; thus, in the most common variant of double-beta decay, one would observe two emitted neutrinos. However, if neutrinos are Majorana particles, then they are their own antiparticle and can coannihilate. This would lead to the rare but statistically significant possibility of neutrinos from two proximate beta decays annihilating with one another, giving the test its name. Neutrinos in this case, as in most cases, remain difficult to detect, necessitating expensive materials and long running times in order to build up the statistically necessary evidence. This is further complicated by the enduring vagaries of neutrino behavior including their mass hierarchy and mass scale.

\subsection{Tribimaximal Mixing}
In the first five years following the determination that neutrinos had mass and that their mixing, encapsulated in the \gls{PMNS} matrix, was nontrivial, there were many attempts to introduce a flavor symmetry explaining quark and lepton mixing. Many of these introduced so-called texture zeros. These would be elements of the \gls{PMNS} matrix set to zero by the model. Introducing one, or more, of these zeros created a structural stability that benefited many models by making it simpler to explain. Over the years, there were even several works that attempted to categorize the likelihood of any of each element being a texture zero and their potential to explain the behaviors of flavors and families.\cite{Frampton:1999yn,Frampton:2002yf,Frampton:2002qc}

In 2002, a paper postulated a form for the \gls{PMNS} matrix. This practice, while not unusual following on earlier attempts such as the Bi-Maximal and Tri-Maximal models,\cite{Barger:1998ta,Harrison:1994iv} proved fairly accurate and was summarily dubbed \gls{TBM}.\cite{Harrison:2002er} In it, 
\begin{equation}
\theta_{13}=0^{\circ}, \theta_{23}=45^{\circ},~{\rm and}~\theta_{12}=\sin^{-1} (\tfrac{1}{\sqrt{3}})\simeq 35.3^{\circ}~, 
\label{TBMVal}
\end{equation}
leading to a \gls{PMNS} matrix of the form:

\begin{equation}
U_{\rm TBM} = \left(
\begin{array}{ccc}
-\sqrt{\frac{1}{6}} & -\sqrt{\frac{1}{6}} & \sqrt{\frac{2}{3}} \\
\sqrt{\frac{1}{3}} & \sqrt{\frac{1}{3}} & \sqrt{\frac{1}{3}} \\
\sqrt{\frac{1}{2}} & -\sqrt{\frac{1}{2}} & 0 
\end{array}
\right).
\label{TBM}
\end{equation}
A noted benefit to this depiction is the rational root form that leads to an ease of incorporation into theoretical models, and in particular finite symmetry models (given their Kronecker products). Indeed, many attempts to utilize various finite or flavor symmetries since the initial \gls{TBM} proposal have attempted to show that they are able to incorporate this symmetry structure.

Although theories have successfully demonstrated that there are numerous potential paths that all arrive at the \gls{TBM} form, it should be noted that this form was merely a guess at the actual values of \gls{PMNS} elements. As the years have progressed, we have indeed seen that the initial \gls{TBM} form may not be correct and that either a new form should be adopted or additional mechanisms are needed to shift \gls{TBM} values in line with experimental data.

\subsubsection{\texorpdfstring{$\mu\mhyphen\tau$ Symmetry}{mutau}}
Another notable feature of \gls{TBM} is that it exhibits $\mu\mhyphen\tau$ symmetry.\cite{Fuki:2006ag} This can be easily demonstrated by examining the first and second columns of $U_{{\rm TBM}}^2$ and noting that they are identical. Physically this implies that $\nu_{\mu}$ and $\nu_{\tau}$ have identical superpositions of the three neutrino mass eigenstates. Also of note, is that slight breaking of the $\mu\mhyphen\tau$ symmetry can lead to a similar perturbation as that seen in Chapter~\ref{EFM1}.

\section{\texorpdfstring{The \glsentrytext{T'} Model}{T' Model}}
\label{Original Model}
\noindent
{\it This Section is largely based on the work of Ref.~\citen{Frampton:2008ci} and Ref.~\citen{Frampton:2008bz}}
\subsection{\texorpdfstring{\glsentrytext{A4} and the Lepton Sector}{A4 Model}}
\subsubsection{Model Characteristics}
Having concluded our historical background, we shall proceed to lie out the models that form the basis of our work. In this chapter we show two derivations of the Majorana mass matrix, $M_\nu$, and the conclusions found from relating the two. 

We shall start by crafting an initial \gls{A4} model comprised of (\gls{A4}$\times$\gls{Z2}),\cite{Ma:2001dn,Babu:2002dz,Ma:2005mw,Ma:2005pd,Ma:2005qf,Altarelli:2005yp,Altarelli:2005yx,Altarelli:2006kg,Ma:2006sk,Ma:2006ip,Adhikary:2006wi} where the various particles are assigned to \glspl{irrep} as,
\begin{equation}
\begin{array}{cccc}
\left. \begin{array}{c}
\left( \begin{array}{c} \nu_{\tau} \\ \tau^- \end{array} \right)_{{\rm L}} \\
\left( \begin{array}{c} \nu_{\mu} \\ \mu^- \end{array} \right)_{{\rm L}} \\
\left( \begin{array}{c} \nu_e \\ e^- \end{array} \right)_{{\rm L}}
\end{array} \right\}
L_{\rm L} (3, +1)~, &
\begin{array}{l}
~~ \tau^-_{{\rm R}}~ (1_1, -1) \\
~~ \mu^-_{{\rm R}} ~ (1_2, -1) \\
~~ e^-_{{\rm R}} ~ (1_3, -1)~, \end{array}
&
{\rm and}
&
\begin{array}{l}
N^{(1)}_{\rm R} ~ (1_1, +1) \\
N^{(2)}_{\rm R} ~ (1_2, +1) \\
N^{(3)}_{\rm R} ~ (1_3, +1)~.\\ \end{array}
\end{array}
\label{LepAssign}
\end{equation}
As Eq.~(\ref{LepAssign}) shows, this model, as with all \gls{A4} models, will only include the leptons (including right-handed Majorana neutrinos). Refs.~\citen{Altarelli:2005yp,Altarelli:2005yx,Altarelli:2006kg} have shown, \gls{A4} is not capable of replicating quark mixing${\it -}$a mixing typically encapsulated in the \gls{CKM}.\cite{Cabibbo:1963yz,Kobayashi:1973fv} In parentheses next to every particle are the specifications for how that particle rotates, first under \gls{A4}, then under \gls{Z2}. In this setup, we have placed the left-handed leptons in a triplet, and the right-handed leptons in various singlets.

From here we can proceed to the formation of a Lagrangian. This crucial step must be carefully considered, for while we hope to discover new physics in the course of our investigation, we must tread lightly in order to avoid blatantly contradicting historic experimental particle physics data. This, oddly, leads to a middle ground where some things are new, but not too many. We will also include the constraint of using only renormalizable couplings, and though the \gls{A4} model contains an anomaly, this is subsequently cancelled by the \gls{T'} model discussed later.

The Lagrangian for this model is then,
\begin{eqnarray}
{\cal L}_{\rm Y}
&=&
\frac{1}{2} M_1 N_{\rm R}^{(1)} N_{\rm R}^{(1)} + M_{23} N_{\rm R}^{(2)} N_{\rm R}^{(3)} \nonumber \\
& & + \Bigg\{
Y_{1} \left( L_{\rm L} N_{\rm R}^{(1)} H_3 \right) + Y_{2} \left( L_{\rm L} N_{\rm R}^{(2)} H_3
\right) + Y_{3} 
\left( L_{\rm L} N_{\rm R}^{(3)} H_3 \right) \nonumber \\ 
&& +
Y_\tau \left( L_{\rm L} \tau_{\rm R} H'_3 \right)
+ Y_\mu \left( L_{\rm L} \mu_{\rm R} H'_3 \right) + 
Y_e \left( L_{\rm L} e_{\rm R} H'_3 \right) 
\Bigg\}
+ 
{\rm h.c.~.} 
\label{Leplagrangian}
\end{eqnarray}
In this form it is clear that that many of the classic features of the \gls{MSM} remain. We have also added 2 triplet Higgs scalars (6 doublets under ${\bf SU(2)_{\rm L}}$, 2 triplets under \gls{A4}) where needed of $H_3(3, +1)$ and $H_3^{'}(3, -1)$. These additions are needed in order to ensure that each Lagrangian term rotates as a singlet under \gls{A4}. By referencing this Lagrangian, the assignments in Eq.~(\ref{LepAssign}), and the Kronecker product table in Sec.~\ref{groups}, one can see this approach is consistently applied. The factors of $\tfrac{1}{2}$ have been added in order to mitigate the identical hermitian conjugates of Majorana mass terms.

Our model maintains that the masses of the charged leptons ($e, \mu, \tau$) emerge from the \glspl{VEV} of 
\begin{equation}
<H_3^{'}> = (\frac{m_{\tau}}{Y_{\tau}},\frac{m_{\mu}}{Y_{\mu}},\frac{m_{e}}{Y_{e}})=(M_{\tau},M_{\mu},M_{e})~. 
\end{equation}
If, largely for the sake of simplicity, we then choose a flavor basis where the charged leptons act as mass eigenstates, we can then separate, at leading order, charged lepton and neutrino masses. We also note that the $N_{{\rm R}}^{(i)}$ masses break $L_{\tau} \times L_{\mu} \times L_e$ symmetry, but will alter the charged lepton masses at the one-loop level only by a factor $\propto Y^2m_i/M_{\rm R}$.

One of the most notable features of this model are the Majorana neutrinos, whose benefits were detailed in Sec.~\ref{recent}. Given this, we must now further specify the form of the neutrino mass matrices. First the Majorana mass matrix in typical form,
\begin{equation}
M_{\rm R} =
\left(
\begin{array}{ccc}
M_1 & 0 & 0 \\
0 & 0 & M_{23} \\
0 & M_{23} & 0
\end{array}
\right).
\label{MN}
\end{equation}
Next is the Dirac mass matrix formed from the Lagrangian Yukawa couplings and a generic set of \glspl{VEV} from the other Higgs triplet,
\begin{equation}
<H_3> = (V_1, V_2, V_3)~,
\label{VEV}
\end{equation}
\begin{equation}
M_{\nu}^D = 
\left(
\begin{array}{ccc}
Y_{1} V_1 & ~~~ Y_{2} V_3 & ~~~ Y_3 V_2 \\
Y_1 V_3 & ~~~ Y_2 V_2 & ~~~ Y_3 V_1 \\
Y_1 V_2 & ~~~ Y_2 V_1 & ~~~ Y_3 V_3
\end{array}
\right).
\label{MD}
\end{equation}

The Majorana mass matrix, $M_{\nu}$, is then given by
\begin{equation}
M_{\nu} = M_{\nu}^D M_{\rm R}^{-1} (M_{\nu}^D)^{T}~.
\label{seesaw1}
\end{equation}
Defining $x_1 \equiv Y_1^2/M_1$ and $x_{23} \equiv Y_2Y_3/M_{23}$ yields the symmetric form of,
\begin{equation}
M_{\nu}=\left(
\begin{array}{ccc}
x_1 V_1^2 + 2 x_{23} V_2V_3 & 
~~~ x_1V_1V_3 + x_{23} (V_2^2+V_1V_3) & 
~~~ x_1V_1V_2 + x_{23} (V_3^2+V_1V_2) \\
 & 
~~~ x_1V_3^2 + 2 x_{23} V_1V_2 & 
~~~ x_1V_2V_3 + x_{23} (V_1^2+V_2V_3) \\ 
 & 
 & 
~~~ x_1V_2^2 + 2 x_{23} V_1V_3
\end{array}
\right).
\label{Mnu2}
\end{equation}

\subsubsection{\texorpdfstring{Incorporating \glsentrytext{TBM} and Majorana Neutrinos}{TBM Majorana}}
We shall now attempt to approach the same matrix from a different direction and determine what limits can be placed as a result of assuming the symmetries of \gls{TBM}. 

As seen in Eq.~(\ref{TBM}), this proposed mixing structure takes the form,
\begin{equation}
U_{{\rm TBM}} = \left(
\begin{array}{ccc}
-\sqrt{\frac{1}{6}} & -\sqrt{\frac{1}{6}} & \sqrt{\frac{2}{3}} \\
\sqrt{\frac{1}{3}} & \sqrt{\frac{1}{3}} & \sqrt{\frac{1}{3}} \\
\sqrt{\frac{1}{2}} & -\sqrt{\frac{1}{2}} & 0 
\end{array}
\right),
\label{TBM2}
\end{equation}
and acts to delineate the relation between flavor and mass eigenstates,
\begin{equation}
\left( \begin{array}{c} \nu_1 \\ \nu_2 \\ \nu_3 \end{array} \right)
= U_{{\rm TBM}} 
\left( \begin{array}{c} \nu_{\tau} \\ \nu_{\mu} \\ \nu_e \end{array} \right).
\end{equation}

Assuming no \gls{CP}-violation, the Majorana matrix $M_{\nu}$ is real and symmetric, and has the general form of
\begin{equation}
M_{\nu} = \left(
\begin{array}{ccc}
A & B & C \\
B & D & E \\
C & E & F 
\end{array}
\right),
\label{ABCDEF}
\end{equation}
which, in general can be diagonalized by the \gls{PMNS} matrix. In order to incorporate symmetries present in \gls{TBM}, specifically, one can use $U_{{\rm TBM}}$ to diagonalize:
\begin{equation}
M_{diag} = \left( \begin{array}{ccc}
m_1 & 0 & 0 \\
0 & m_2 & 0 \\
0 & 0 & m_3
\end{array}
\right) = U_{{\rm TBM}} M_{\nu} U_{{\rm TBM}}^{T}~.
\label{diag}
\end{equation}
Substituting Eq.~(\ref{TBM2}) into Eq.~(\ref{diag}) and solving for $M_{\nu}$ leads to a further reduction of Eq.~(\ref{ABCDEF}) to the three real parameters $A, B, C$:
\begin{equation}
M_{\nu} = \left(
\begin{array}{ccc}
A ~~~ & B & ~~~ C \\
B & A & ~~~ C \\
C & C & ~~~ A+B-C
\end{array}
\right),
\label{ABC}
\end{equation}
with eigenvalues,
\begin{eqnarray}
m_1 &=& (A + B - 2C)~, \nonumber \\
m_2 &=& (A + B + C)~, \nonumber \\
m_3 &=& (A - B)~,
\label{masses}
\end{eqnarray}
whose individual assignments can be found by a substitution of Eq.~(\ref{ABC}) back into Eq.~(\ref{diag}) and multiplying out the right-hand side.

Now, by relating the two forms of $M_{\nu}$, Eq.~(\ref{Mnu2}) and Eq.~(\ref{ABC}), we find three equations,
\begin{equation}
x_1V_1^2 + 2x_{23}V_2V_3 = x_1V_3^2 + 2x_{23}V_1V_2~,
\label{A}
\end{equation}
\begin{equation}
x_1V_1V_2 + x_{23}(V_3^2+V_1V_2) = x_1V_2V_3 +x_{23}(V_1^2+V_2V_3)~,
\label{B}
\end{equation}
\begin{equation}
x_1(V_1^2+V_1V_3-V_1V_2) + x_{23}(2V_2V_3 +V_2^2 +V_1V_3 -V_3^2 - V_1V_2)
 = x_1V_2^2 + 2 x_{23} V_1V_3~,
\label{C}
\end{equation}
corresponding to $A$, $C$, and $A+B-C$, respectively.

We find no solutions of Eqs.~(\ref{A}, \ref{B}, \ref{C}) with any of $x_1, x_{23}, V_1, V_2, V_3$ vanishing. It is straightforward to note that Eq.~(\ref{A}) and Eq.~(\ref{B}) can only both be satisfied if $V_1=V_3$. Solving Eq.~(\ref{C}) further requires $(2V_1+V_2)(V_1-V_2) = 0$, since it can be shown that $x_1=x_{23}$ is not possible for any hierarchy consistent with experiment. In this \gls{A4} model, therefore, only two \glspl{VEV} of $H_3$ give \gls{TBM}.

The first is,\footnote{This \gls{VEV}, $<H_3> \propto (1, 1, 1)$, can be transformed to $<H_3> \propto (0, 0, 1)$ by an \gls{A4} transformation. The literature distinguishes these designations as the Ma-Rajaskaran and Altarelli-Feruglio bases respectively.}
\begin{equation}
<H_3> = (V, V, V)~,
\label{VVV}
\end{equation}
which is studied in Ref.~\citen{Lee:2006pr}. Now by equating Eq.~(\ref{Mnu2}) and Eq.~(\ref{ABC}) one can find relations for $A$, $B$, and $C$, from there one can use Eq.~(\ref{masses}) and the relative values of Eq.~(\ref{VVV}) to find expressions for the neutrino mass eigenstates,
\begin{equation}
\begin{array}{cc}
\begin{array}{c}
A = V^2(x_1 +2x_{23})~, \\
B = V^2(x_1 +2x_{23})~, \\
C = V^2(x_1 +2x_{23})~, 
\end{array}
&
\begin{array}{l}
m_1 = 0~, \\
m_2 = V^2(3x_1 + 6x_{23})~, \\
m_3 = 0.
\end{array}
\end{array}
\label{brokenhierarchy}
\end{equation}
Clearly this implies $m_2 \gg m_1=m_3=0$, an inappropriate hierarchy being neither \gls{NH} or \gls{IH}, thus Eq.~(\ref{VVV}) is an unacceptable \gls{VEV} for $<H_3>$ in our model.

The only other \gls{VEV} for \gls{A4} is therefore,\footnote{Because $<H_3> \propto (1, 1, 1)$ could be made consistent with the neutrino masses in most previous \gls{A4} models, due to additional parameters, the alternative of $<H_3> \propto (-2, 1, 1)$ seems not to have been previously studied.}
\begin{equation}
<H_3> = (V, -2V, V)~.
\label{V2VV}
\end{equation}
As before, the forms of $A$, $B$, and $C$ as well as the masses can be found from the combination of Eqs.~(\ref{Mnu2}, \ref{ABC}, \ref{masses}) with the relative values of Eq.~(\ref{V2VV}),
\begin{equation}
\begin{array}{cc}
\begin{array}{l}
A = V^2(x_1 -4x_{23})~, \\
B = V^2(x_1 +5x_{23})~, \\
C = V^2(-2x_1 -1x_{23})~, 
\end{array}
&
\begin{array}{l}
m_1 = x_1 V^2(6 + 3y)~, \\
m_2 = 0~,\\
m_3 = x_1 V^2(-9y)~,
\end{array}
\end{array}
\label{mendedhierarchy}
\end{equation}
where $y=x_{23}/x_1$. If we continue in our assumption that $m_2\simeq m_1$ (an appealing choice given the known sign of $\Delta m_{21}$ invalidates $m_2<m_1$), then $y$ is constrained to a value $y=-2$. 

We are then left with a strong model preference for the \gls{NH},\footnote{Previously, other \gls{A4} models have shown more flexible in choosing between \gls{IH} and \gls{NH}, they also include more parameters, and, as a result, incur diminished predictivity.} with $m_3\gg m_2\simeq m_1$, making Eq.~(\ref{V2VV}) our only feasible \gls{VEV} for $<H_3>$. 

\subsection{\texorpdfstring{The Minimal \glsentrytext{T'} Model}{MT'M}}
\subsubsection{History}
In 1994, Ref.~\citen{Frampton:1994rk} investigated the simple non-abelian discrete groups up to order 31 (they stop before order 32 because of the large number of additional groups at every power of 2) as potential family symmetries. They began by laying out a set of model-building guidelines, and proposed model assignments for each symmetry group. One such group, labeled at the time the Double Tetrahedral Group, was detailed as a subset of ${\bf SU(2)}$. There, they detailed a particle assignment set for the six \gls{MSM} quarks and the leptons known at the time: 

\begin{equation}
\begin{array}{cccc}
\begin{array}{c}
\left( \begin{array}{c} t \\ b \end{array} \right)_{{\rm L}}~~~1 \\
\left. \begin{array}{c} \left( \begin{array}{c} c \\ s \end{array} \right)_{{\rm L}} \\
\left( \begin{array}{c} u \\ d \end{array} \right)_{{\rm L}} \end{array} \right\} 2 
\end{array}
 &
\begin{array}{c}
~~ t_{{\rm R}}~~~~ 1 \\
~~ c_{{\rm R}} ~~~~ 1 \\
~~ u_{{\rm R}} ~~~~ 1 \\
\left. \begin{array}{c}
~~ b_{{\rm R}} \\
~~ s_{{\rm R}} \\
~~ d_{{\rm R}} \end{array} \right\} 3 \end{array}
~~~&~~~
\begin{array}{c}
\left( \begin{array}{c} \nu_{\tau} \\ \tau^- \end{array} \right)_{{\rm L}}~~~1 \\
\left. \begin{array}{c} \left( \begin{array}{c} \nu_{\mu} \\ \mu^- \end{array} \right)_{{\rm L}} \\
\left( \begin{array}{c} \nu_e \\ e^- \end{array} \right)_{{\rm L}} \end{array} \right\} 2 
\end{array}
 &
\left. \begin{array}{c}
~~ \tau^-_{{\rm R}} \\
~~ \mu^-_{{\rm R}} \\
~~ e^-_{{\rm R}} \end{array} \right\} 3
\end{array}
\end{equation}

This particle assignment had the benefits of containing all of the, then known, fundamental fermions. In addition, it labeled the top quark mass as a singlet of the group symmetry, better explaining its extremely high mass of $m_t=173.5~{\rm GeV}$.\cite{Beringer:1900zz} This model is not without problems in light of later discoveries and data, indeed neutrino masses and non-trivial mixing came as a surprise to much of the community. It would be over a decade before this basic model was fully overhauled to comport with our modern understanding.

\subsubsection{\texorpdfstring{\glsentrytext{T'} as Flavor Symmetry}{T' Flavor}}
Following their articulation of a minimal \gls{A4} model, as detailed in earlier, the authors of Ref.~\citen{Frampton:2008ci} sought to expand their treatment to include quark mixing by converting to the symmetry now titled, the Binary Tetrahedral Group (\gls{T'}).\cite{Frampton:2000mq,Chen:2007afa,Feruglio:2007uu,Frampton:2007et} As detailed in~\ref{groups} the connection between \gls{A4} and \gls{T'} is quite special. Since they share the singlet and triplet elements and multiplications, the prior \gls{A4} minimal model can easily be converted to \gls{T'}. Thus, the assignments of the leptons will remain the same in the new model, as will the treatment of the \gls{TBM} angles and Higgs \glspl{VEV}. Now, though, \gls{T'} affords the advantage of doublet symmetry elements, which we can use to accommodate the, mostly diagonal, \gls{CKM} matrix.

As before, it is best to keep the top quark in its own singlet as a way of motivating a higher mass. So, the lighter two families will be placed in doublets: one doublet for both left handed families labeled $Q_{\rm L}$, one doublet for the two lighter down-type right-handed quarks labeled ${\cal S}_{\rm R}$, and one doublet for the two lighter up-type right-handed quarks labeled ${\cal C}_{\rm R}$. The left-handed third family will be in a singlet ${\cal Q}_{\rm L}$, while the right-handed third family will be in two self-named singlets. These new assignments are summarized below,
\begin{equation}
\begin{array}{ccc}
\begin{array}{c}
\left( \begin{array}{c} t \\ b \end{array} \right)_{{\rm L}}~ {\cal Q}_{\rm L} ~~~~~~~~~~~ (1_1, +1)~ \\
\left. \begin{array}{c} \left( \begin{array}{c} c \\ s \end{array} \right)_{{\rm L}} \\
\left( \begin{array}{c} u \\ d \end{array} \right)_{{\rm L}} \end{array} 
\right\} Q_{\rm L} ~~~~~~~~ (2_1, +1)~, \end{array}
&
{\rm and}
&
\begin{array}{l}
~~t_{{\rm R}} ~~~~~~~~~~~~~~ (1_1, +1)~~ \\
~~b_{{\rm R}} ~~~~~~~~~~~~~~ (1_2, -1)~~ \\
\left. \begin{array}{c} c_{{\rm R}} \\ u_{{\rm R}} \end{array} \right\}
{\cal C}_{\rm R} ~~~~~~~~ (2_3, -1)~ \\
\left. \begin{array}{c} s_{{\rm R}} \\ d_{{\rm R}} \end{array} \right\}
{\cal S}_{\rm R} ~~~~~~~~ (2_2, +1)~. \end{array}
\end{array}
\label{QuarkAssign}
\end{equation}
As before, the parenthetical numbers state the element assignment under the main group algebra first (\gls{T'} this time) followed by the auxiliary \gls{Z2}. 

Now, as we prepare to state a Lagrangian for both lepton and quark sectors, we should note it remains the intent to craft a minimal model using finite symmetry. When the Higgs were chosen for the leptons, they were limited by the particle assignments and the elements of \gls{A4}. Now that we have progressed to \gls{T'}, doublets are available as well. However, in this \glsreset{MRTM}\gls{MRTM} only Higgs singlets and triplets (under \gls{T'}, they all remain doublets under ${\bf SU(2)_{\rm L}}$) will be used. With these stipulations in place, we can write down the \gls{MRTM} Lagrangian:

\begin{align}
{\cal{L}}_{\rm Y} = \frac{1}{2} M_1 N_{\rm R}^{(1)} N_{\rm R}^{(1)} +
 M_{23} N_{\rm R}^{(2)} N_{\rm R}^{(3)} + \nonumber \\
Y_{e} L_{\rm L} e_{\rm R} H_3^{'} + Y_{\mu} L_{\rm L} \mu_{\rm R} H_3^{'} + 
Y_{\tau} L_{\rm L} \tau_{\rm R} H_3^{'} + \nonumber \\
Y_1 L_{\rm L} N_{\rm R}^{(1)} H_3 + Y_2 L_{\rm L} N_{\rm R}^{(2)} H_3 + 
Y_3 L_{\rm L} N_{\rm R}^{(3)} H_3 + \nonumber \\
Y_t ({\cal{Q}}_{\rm L} t_{\rm R} H_{1_1}) + Y_b ({\cal{Q}}_{\rm L} b_{\rm R} H_{1_3}) + \nonumber \\
Y_{\cal C} (Q_{\rm L} {\cal{C}}_{\rm R} H^{'}_{3}) + Y_{\cal S} (Q_{\rm L} {\cal{S}}_{\rm R} H_{3}) + h.c.~.
\end{align}
This equation, naturally, now includes several new terms. To complement and complete these terms, two new Higgs scalars are required, $H_{1_{1}}(1_1,+1)$ and $H_{1_{3}}(1_3,-1)$, with \glspl{VEV} of:
\begin{equation}
<H_{1_1}> = m_t/Y_t~, ~~~~ <H_{1_3}> = m_b/Y_b~,
\label{H13VEV}
\end{equation}
to provide the masses of the third family. Taken together this allows for the established quark mass hierarchy of $m_t \gg m_b > m_{c,s,d,u}$. 

Also notable is the fact that the quark and lepton sectors reuse the same two Higgs triplets. This can serve as the basis for a connection between the families and a unifying foundation among all fermions.

\subsection{Cabibbo Angle Prediction}
\noindent
{\it A note on formalism: We will be parametrizing the \gls{CKM} matrix in an identical way to our previously described depiction of the \gls{PMNS} matrix (Sec.~\ref{recent}) as is customary given the parametrization we use originated with \gls{CKM}, and distinguish the angles of the two by using $\Theta_{ij}$ for \gls{CKM}, and $\theta_{ij}$ for \gls{PMNS}. Additionally, we will distinguish between the two \gls{CP}-violating phases with $\delta_{{\rm CP}}$ for \gls{PMNS} and $\delta_{{\rm KM}}$ for \gls{CKM}.}

Due to our choice to avoid \gls{T'} doublet Higgs terms in this \gls{MRTM}, an assessment of the \gls{CKM} matrix will need to be reduced down to the ($2\times2$) quark mixing matrices, which assume $\Theta_{13}=\Theta_{23}=0$. While this is demonstrably inaccurate, it remains a decent approximation given that both angles are quite small, $\Theta_{13} = 0.201^{\circ}$ and $\Theta_{23} = 2.38^{\circ}$. 

The two remaining nontrivial ($2\times2$) matrices for $(c,~u)$ and $(s,~d)$ will hereafter be denoted $U^{'}$ and $D^{'}$, respectively, and calculated using the \gls{T'} complex Clebsch-Gordan coefficients illustrated in Ref.~\citen{Fairbairn:1963}. If we divide $U^{'}$ by $Y_{\cal C}$ we find,
\begin{equation}
\begin{array}{cc}
U \equiv \left( \frac{1}{Y_{{\cal C}}}\right) U^{'} = \left(\begin{array}{cc}
\sqrt{\frac{2}{3}} \omega M_{\tau} & \frac{1}{\sqrt{3}} M_e \\
- \frac{1}{\sqrt{3}} \omega M_e & \sqrt{\frac{2}{3}} M_{\mu}
\end{array}
\right),
&
{\rm where}~~\omega=e^{2i\pi/3}~.
\end{array}
\label{Umatrix}
\end{equation}

If we take the additional step of setting the electron mass, $M_e$, to zero, $U$ becomes immediately diagonal and leaves $m_u$, $m_c$, $m_{\mu}$, and $m_{\tau}$ free. 

Next we shall take a look at the ($2\times2$) Cabibbo Matrix. In its general parametrized form it appears as
\begin{equation}
P \equiv \left( \begin{array}{cc}
\cos \Theta_{12} & -\sin \Theta_{12} \\
\sin \Theta_{12} & \cos \Theta_{12} 
\end{array}
\right).
\label{CabibboM}
\end{equation}
We can use this form to diagonalize the hermetian square of $D^{'}$, after dividing by $Y_{\cal S}$,
\begin{equation}
D \equiv \left( \frac{1}{V Y_{{\cal S}}} \right)
D^{'} = \left( \begin{array}{cc}
\frac{1}{\sqrt{3}} & 2 \sqrt{\frac{2}{3}} \omega^2 \\
\sqrt{\frac{2}{3}} & -\frac{1}{\sqrt{3}} \omega^2
\end{array}
\right),
\label{Dmatrix}
\end{equation}
\begin{equation}
{\bf {\cal D}} \equiv D D^{\dagger} = \left(
\frac{1}{3} \right) \left(
\begin{array}{cc} 
9 & - \sqrt{2} \\
- \sqrt{2} & 3 
\end{array}
\right).
\label{DDdagger}
\end{equation}
We now have developed the tools needed to solve
\begin{equation}
\left( \begin{array}{cc}
m_d^2 & 0 \\
0 & m_s^2 
\end{array}
\right)
~=~
P^T {\cal D} P~.
\label{diagCab}
\end{equation}
for the two remaining unknowns, $\Theta_{12}$ and $(m_d^2/m_s^2)$.

The first result, that of the Cabibbo Angle, yields:

\begin{equation}
\tan 2\Theta_{12} = \left( \frac{\sqrt{2}}{3} \right),
\label{Cabibbo}
\end{equation}
which converts to a decimal prediction of $\Theta_{12} = 12.6^{\circ}$ (by comparison its experimental value is $\Theta_{12} = 13.04^{\circ}\pm0.05^{\circ}$). While this is $9\sigma$ away from the measured value, this is largely due disparity in precision between quark and neutrino data, and it remains an adequate first-order prediction. As we shall see, later attempts to adjust the theory achieve better agreement. 

As for $(m_d^2/m_s^2)$, the solution to Eq.~\ref{diagCab} yields a predicted value of $\approx 0.288$, compared with experimental findings of $(m_d^2/m_s^2)\backsimeq 0.003$. While admittedly a poor initial guess, one might suppose that this is due to the assumption of $\Theta_{23} = \Theta_{13} = 0$, and that incorporation of mixing between $(d,~s)$ and $b$ would help matters.

This model proved to be an important step, both in demonstrating the viability of \gls{T'} as a suitable basis for neutrino mixing models, and its superiority over \gls{A4} given it has the ability to address the quark sector. Having achieved a semi-reliable framework to connect the mixing angles of quarks and leptons, we can begin to ask questions of the underlying hypotheses. The \gls{A4} model was constructed to handle neutrinos, and was incapable of addressing quarks. The \gls{MRTM} model made strong assumptions about the neutrino mixing angles and used them to make moderately close predictions of the well-measured quark angles. As Ch.~\ref{EFM1} will show, it bears some investigation to see if we can use quark mixing to learn about neutrinos.

\chapter{\texorpdfstring{\glsentrytext{T'} Model Perturbations and Neutrino Mixing}{Perturbations on the T' Model}}
\label{EFM1}
\noindent
{\it This Chapter is largely based on the work of Ref.~\citen{Eby:2008uc}}
\section{Revised Assumptions}

So far, the \gls{T'} model has shown itself capable of creating a successful mechanism to link the mixing angles of quarks and neutrinos. Yet, thus far, we have assumed the exact accuracy of the \gls{TBM} angles, and their underlying symmetry, in order to develop a prediction for quark mixing. This was largely due to the historical development of the theory, from \gls{A4} (which can only function for leptons), to \gls{T'} (which can accommodate both). But given the fact that \gls{T'} can accommodate both and the measured value of the Cabibbo angle is both highly precise, and not easily reducible to a rational form, it may make more sense to assume the measured value of the Cabibbo angle and use our framework to develop a prediction for neutrino mixing. When this model was initially developed the neutrino mixing angles had as much as $10^{\circ}$ of uncertainty around the \gls{TBM} values. Consequently, we will keep the mechanism discussed in~\ref{Original Model}, but introduce perturbations.

We will begin by redefining the neutrino mixing angles as, 
\begin{equation}
\theta_{ij} = (\theta_{ij})_{\rm TBM} + \epsilon_k~,
\label{PerturbedTBM}
\end{equation}
(where $\epsilon_3$ is used for $\theta_{12}$, etc.), and proceed to perturb around Eq.~(\ref{Cabibbo}).

First, we recall a few salient points about the model in Sec.~\ref{Original Model} based on \gls{A4} symmetry. The only important scalar for the present analysis is the triplet $H_3(3, +1)$ whose vacuum expectation value in Sec.~\ref{Original Model} was taken as
\begin{equation}
<H_3> = (V_1, V_2, V_3) = V(1, -2, 1)~,
\label{H3TBM}
\end{equation}
which linked to the \gls{TBM} form seen in Eq.~(\ref{TBM}). We shall consider the perturbation 
\begin{equation}
<H_3> = (V^{'}_1, V^{'}_2, V^{'}_3) = V^{'}(1, -2+b, 1+a)~,
\label{H3new}
\end{equation}
where $|a|, |b| \ll 1$.

\section{Perturbations}
We shall first consider the calculation of a perturbation around the earlier work in Sec.~\ref{Original Model} by using Eq.~(\ref{H3new}) in place of Eq.~(\ref{H3TBM}). The down-type quark ($2\times 2$) mass matrix for the first two families $(s, d)$ is perturbed to
\begin{equation}
D \equiv \left( \frac{1}{V^{'} Y_{{\cal S}}} \right)
D^{'} = \left( \begin{array}{cc}
\frac{1}{\sqrt{3}} & (2-b) \sqrt{\frac{2}{3}} \omega^2 \\
\sqrt{\frac{2}{3}}(1+a) & -\frac{1}{\sqrt{3}} \omega^2
\end{array}
\right),
\label{DmatrixAlt}
\end{equation}
where, again, $\omega=\exp{2i\pi/3}$. The hermitian square ${\bf {\cal D}} \equiv D D^{\dagger}$ is, to first order in $a$ and $b$,
\begin{equation}
{\bf {\cal D}} \equiv D D^{\dagger} \simeq \left(
\frac{1}{3} \right) \left(
\begin{array}{cc}
9 - 8b & \sqrt{2} (-1+a+b) \\
\sqrt{2} (-1+a+b) & 3+4a 
\end{array}
\right).
\label{DDdagger2}
\end{equation}
The eigenvalues of Eq.~(\ref{DDdagger2}) satisfy the quadratic equation
\begin{equation}
(9 - 8b-\lambda_e)(3+4a-\lambda_e) - 2 (1-a-b)^2 = 0~,
\label{quadratic}
\end{equation}
with solutions of
\begin{equation}
\lambda_e{\pm} = (6 \pm \sqrt{11}) + 2a\left( 1 \mp \frac{4}{\sqrt{11}} \right)
-2b \left( 2 \pm \frac{7}{\sqrt{11}} \right)~.
\label{eigenvalues}
\end{equation}
An eigenvector $(\alpha, \beta)$ has components satisfying
\begin{equation}
\left( \frac{\beta}{\alpha} \right) = \left(
\frac{3 - \sqrt{11}}{\sqrt{2}} \right) \left[ 1 - \frac{a}{\sqrt{11}}
+ \frac{b}{\sqrt{11}} \right],
\label{components}
\end{equation}
whose normalization $N(\alpha, \beta)$ satisfies
\begin{equation}
N^{-2} = 1 + \beta^2/\alpha^2,
\label{normalization}
\end{equation}
from which the Cabibbo angle $\sin \Theta_{12} = N \beta/\alpha$ is
\begin{equation}
\sin \Theta_{12} = \sqrt{ \left(\frac{1}{2} - \frac{3}{2 \sqrt{11}} \right) }
\left(1 - \frac{3 + \sqrt{11}}{22} (a - b) \right).
\label{sin}
\end{equation}
From this one finds at leading order 
\begin{equation}
\cos 2 \Theta_{12} \simeq \left(\frac{3}{\sqrt{11}} \right)
\left( 1+ \frac{2}{33} (a - b) \right),
\label{Cos}
\end{equation}
and
\begin{equation}
\sin 2\Theta_{12} \simeq \left(\frac{\sqrt{2}}{\sqrt{11}} \right) 
\left( 1 - \frac{3}{11} (a - b) \right),
\label{Sin}
\end{equation}
where
\begin{equation}
\tan 2 \Theta_{12} \simeq (\sqrt{2})/3 \left(1 - \frac{1}{3} (a - b) \right),
\label{Tan}
\end{equation}
which, of course, reduces back to \gls{TBM} values (and the Sec.~\ref{Original Model} Cabibbo prediction) for $a=b=0$.

Next, we will relate the $\epsilon_i$ neutrino angle perturbations of Eq.~(\ref{PerturbedTBM}) to the vacuum alignment perturbations $a$ and $b$ of Eq.~(\ref{H3new}).

As before we will start with the basic \gls{TBM} form as seen in Eq.~(\ref{TBM}, \ref{TBM2}),
\begin{equation}
U_{\rm TBM} = \left(
\begin{array}{ccc}
-\sqrt{\frac{1}{6}} & -\sqrt{\frac{1}{6}} & \sqrt{\frac{2}{3}} \\
\sqrt{\frac{1}{3}} & \sqrt{\frac{1}{3}} & \sqrt{\frac{1}{3}} \\
\sqrt{\frac{1}{2}} & -\sqrt{\frac{1}{2}} & 0 
\end{array}
\right).
\label{TBM3}
\end{equation}
using the neutrino mixing angle values given in Eq.~(\ref{TBMVal}).

By utilizing the small-angle approximations and the added-angle trigonometric identities, one can fashion a practical form of Eq.~(\ref{PerturbedTBM}), which at first order comes to,
\begin{spacing}{1.0}
\begin{itemize} 
\item $s_{12} \simeq \sqrt{\frac{1}{3}}(1+\sqrt{2}\epsilon_3)$
\item $c_{12} \simeq \sqrt{\frac{2}{3}}(1-\epsilon_3/\sqrt{2})$
\item $s_{23} \simeq \sqrt{\frac{1}{2}}(1+\epsilon_1)$
\item $c_{23} \simeq \sqrt{\frac{1}{2}}(1-\epsilon_1)$
\item $s_{13} \simeq \epsilon_2$
\item $c_{13} \simeq 1$
\end{itemize} 
\end{spacing}

Consequently, one may write
\begin{equation}
U \simeq U_{{\rm TBM}} + \delta U = U_{{\rm TBM}} + 
\delta U_1\epsilon_1 + \delta U_2 \epsilon_2 + \delta U_3 \epsilon_3~,
\label{deltaU}
\end{equation}
where
\begin{equation}
\delta U_1 = \left(\begin{array}{ccc}
-\sqrt{\frac{1}{6}} & +\sqrt{\frac{1}{6}} & 0 \\
+\sqrt{\frac{1}{3}} & -\sqrt{\frac{1}{3}} & 0 \\
-\sqrt{\frac{1}{2}} & -\sqrt{\frac{1}{2}} & 0
\end{array}
\right),
\label{deltaU1}
\end{equation}

\begin{equation}
\delta U_2 = \left(\begin{array}{ccc}
-\sqrt{\frac{1}{3}} & +\sqrt{\frac{1}{3}} & 0 \\
-\sqrt{\frac{1}{6}} & +\sqrt{\frac{1}{6}} & 0 \\
0 & 0 & 1 
\end{array}
\right),
\label{deltaU2}
\end{equation}

\begin{equation}
\delta U_3 = \left(\begin{array}{ccc}
-\sqrt{\frac{1}{3}} & -\sqrt{\frac{1}{3}} & -\sqrt{\frac{1}{3}} \\
-\sqrt{\frac{1}{6}} & -\sqrt{\frac{1}{6}} & +\sqrt{\frac{2}{3}} \\
0 & 0 & 0 
\end{array}
\right),
\label{deltaU3}
\end{equation}
combine to form
\begin{equation}
\delta U = \frac{1}{\sqrt{6}}\left(\begin{array}{ccc}
-\epsilon_1-\sqrt{2}(\epsilon_2+\epsilon_3) & \epsilon_1+\sqrt{2}(\epsilon_2-\epsilon_3) & -\sqrt{2}\epsilon_3 \\
\sqrt{2}\epsilon_1-(\epsilon_2+\epsilon_3) & -\sqrt{2}\epsilon_1+(\epsilon_2-\epsilon_3) & 2\epsilon_3 \\
-\sqrt{3}\epsilon_1 & -\sqrt{3}\epsilon_1 & \sqrt{6}\epsilon_3 
\end{array}
\right).
\label{deltaUC}
\end{equation}

By inserting Eq.~(\ref{TBM3}) into Eq.~(\ref{diag}), we arrive at:
\begin{equation}
(M_{\nu})_{{\rm TBM}}
= 
\left( \frac{1}{6} \right) 
\left( \begin{array}{ccc}
m_1 + 2m_2 + 3m_3 & m_1 + 2m_2 -3m_3 & -2m_1+2m_2 \\
& m_1 + 2m_2 + 3 m_3 & -2m_1+2m_2 \\
& & 4 m_1 + 2m_2
\end{array}
\right).
\label{MnuTBM}
\end{equation}

Analysis of Eq.~(\ref{diag}) leads to the full perturbation of,
\begin{eqnarray}
\delta (M_{\nu})_{{\rm diag}} & = & \left( 
\begin{array}{ccc}
\delta m_1 & 0 & 0 \\
0 & \delta m_2 & 0 \\
0 & 0 & \delta m_3 
\end{array}
\right) \nonumber \\
& = & \delta U (M_{\nu})_{{\rm TBM}} U_{{\rm TBM}}^T \nonumber \\
&+& U_{{\rm TBM}} \delta M_{\nu} U_{{\rm TBM}}^T \nonumber \\
&+& U_{{\rm TBM}} (M_{\nu})_{{\rm TBM}} \delta U^T~,
\label{deltaMdiag}
\end{eqnarray}
in which $U_{{\rm TBM}}$ is known from Eq.~(\ref{TBM3}) and $\delta U$ from Eq.~(\ref{deltaUC}). Since the derivation of $(M_{\nu})_{{\rm TBM}}$ contains further multiplications by the unitary matrix $U_{{\rm TBM}}$, we can further simplify by eliminating factors of $U_{{\rm TBM}}U_{{\rm TBM}}^T=U_{{\rm TBM}}^T U_{{\rm TBM}}={\bf 1}$. This abbreviated form is simply:
\begin{eqnarray}
\left( 
\begin{array}{ccc}
\delta m_1 & 0 & 0 \\
0 & \delta m_2 & 0 \\
0 & 0 & \delta m_3 
\end{array}
\right) 
 = \delta U~U_{{\rm TBM}}^T M_{{\rm diag}} 
+ U_{{\rm TBM}} \delta M_{\nu} U_{{\rm TBM}}^T 
+ M_{{\rm diag}} U_{{\rm TBM}} \delta U^T~.
\label{abbrdeltaMdiag}
\end{eqnarray}

To compute $\delta M_{\nu}$ in Eq.~(\ref{abbrdeltaMdiag}) we start from Eq.~(\ref{Mnu2}),
\begin{equation}
\setlength{\arraycolsep}{2pt}
(M_{\nu})_{\rm TBM} =
\begin{pmatrix}
x_1 V_1^2 + 2 x_{23} V_2V_3 &
~~ x_1V_1V_3 + x_{23} (V_2^2+V_1V_3) &
~~ x_1V_1V_2 + x_{23} (V_3^2+V_1V_2) \\
& 
~~ x_1V_3^2 + 2 x_{23} V_1V_2 & 
~~ x_1V_2V_3 + x_{23} (V_1^2+V_2V_3) \\
&
& 
~~ x_1V_2^2 + 2 x_{23} V_1V_3
\end{pmatrix}.
\label{MnuA4}
\end{equation}
where $<H_3> = (V_1, V_2, V_3)$, $x_1=Y_1^2/M_1$ and $x_{23} = Y_2Y_3/M_{23}$. These variables, including Yukawa couplings and right-handed neutrino masses, all remain empirically unknown. As all terms include either $x_1$ or $x_{23}$, Eq.~(\ref{MnuA4}) can be further simplified by combining these factors into $y=x_{23}/x_1$, leaving us to obtain predictions by determining this unknown.

We shall now introduce our perturbation of the vacuum alignment, Eq.~(\ref{H3new}), to Eq.~(\ref{MnuA4}), again at first-order only in $a$ and $b$, to find,
\begin{equation}
\delta M_{\nu} =
x_1 (V^{'})^{2} \left(
\begin{array}{ccc}
~~ 2(-2a+b)y & a + (a-4b)y & b+(2a+b)y \\
 & 2(a+by) & (-2a+b)(1+y) \\
 & & -4b +2ay 
\end{array}
\right).
\label{MA4}
\end{equation}

By inserting this $\delta M_{\nu}$ into Eq.~(\ref{abbrdeltaMdiag}) we obtain six equations from the ($3 \times 3$) symmetric matrix. In the $\delta m$ of (I) - (III) a common (but unpredicted) normalization factor has been omitted.

\begin{itemize}
\item (I) $\delta m_1 = (2+y) (a -2b)$

\item (II) $\delta m_2 = 0$

\item (III) $\delta m_3 = -3y (a-2b)$

\item (IV) $\epsilon_2 = \sqrt{2} \epsilon_1$

\item (V) $a = \epsilon_1\frac{m3-m1}{1+2y} $

\item (VI) $(a+b) = \left( \frac{1}{\sqrt{2}} \frac{m2-m1}{y-1} \right)\epsilon_3$

\end{itemize}

\section{Direct Predictions}
\label{neutmass}
Result (IV) is significant and contains two interpretations. The first is as written, implying that a $\theta_{13}>0$ results in $\theta_{23}>45^{\circ}$. The second interpretation is to redefine the angle $\theta_{13}$ with the transformation $\theta_{13}\Rightarrow -\theta_{13}$ (note that until now, we had assumed $\theta_{13}=0$, meaning this transformation has no phenomenological affect), which leads to a $\theta_{23}$ in the first quadrant. Summarizing these possibilities is analogous to stating that
\begin{equation}
\theta_{13} = \lvert \sqrt{2} \rvert \left( \frac{\pi}{4} - \theta_{23} \right),
\label{root2}
\end{equation}
which, interestingly, links any non-zero value for $\theta_{13}$ to the departure of the atmospheric neutrino mixing angle $\theta_{23}$ from maximal mixing at $\theta_{23} = \pi/4$. This is our most definite prediction from \gls{T'}, and is independent of phenomenological input.

To arrive at further \gls{T'} predictions for the neutrino mixings, $\theta_{13}$ and $\theta_{23}$, we shall require additional input.

The equation (I) through (III) must be combined with the zeroth-order values
\begin{equation}
\begin{array}{ccc}
\begin{array}{l}
m_1^0 = 3 (y+2)~, \\
m_2^0 = 0~, \\
m_3^0 = - 9y~,
\end{array}
&
\Rrightarrow
&
\begin{array}{l}
m_1 = 3 (y+2) +(a-2b)(2+y)~, \\
m_2 = 0~, \\
m_3 = - 9y-3y(a-2b)~.
\end{array}
\end{array}
\label{mpertfin}
\end{equation}

It is notable that $m_2 = 0$ remains even at first order. This arises from the zero structures in the terms of Eq.~(\ref{deltaMdiag}). They are
\begin{equation}
\delta U (M_{\nu})_{{\rm TBM}} U_{{\rm TBM}}^T 
~~~~~~~~~~ \left(\begin{array}{ccc}
0 & 0 & \\
& 0 & \\
& 0 & 0
\end{array}
\right),
\label{term1}
\end{equation}
\begin{equation}
U_{{\rm TBM}} \delta M_{\nu} U_{{\rm TBM}}^T 
~~~~~~~~~~ \left(\begin{array}{ccc}
& & \\
& 0 & 0 \\
 & 0 & 
\end{array}
\right),
\label{term2}
\end{equation}
\begin{equation}
U_{{\rm TBM}} (M_{\nu})_{{\rm TBM}} \delta U^T 
~~~~~~~~~~ \left(\begin{array}{ccc}
0 & & \\
0 & 0 & 0 \\
 & & 0
\end{array}
\right).
\label{term3}
\end{equation}

In order to satisfy the criterion that $m_1\leq m_2$, there are two possibilities, neither of which is particularly satisfying. 

The first is setting $a=-3+2b$, though upon closer examination, this fails to meet the criteria that $a,b\ll1$, and therefore we discard it.

Second is the phenomenological input, originating in Sec.~\ref{Original Model}, that set $y=-2$. This, combined with $m_1\backsim m_2$, gives $(a+b)=0$ and Eq.~(\ref{Tan}) becomes simply
\begin{equation}
\begin{array}{cccc}
\tan 2 \Theta_{12} = \left( \frac{\sqrt{2}}{3} \right) \left(1 - \frac{2}{3} a \right)
&
{\rm with}
&
a=\epsilon_1 \frac{\Delta m_{31}}{3},
&
\Delta m_{31} \approx \sqrt{\Delta m_{32}^2}.
\end{array}
\label{TanNew}
\end{equation}
 
Eq.~(\ref{TanNew}) allows us to approximate the size of the perturbation from the experimental value reported in Ref.~\citen{preMoriond}, $(\Theta_{12})_{{\rm experiment}} = 13.03 \pm 0.04^{\circ}$, to identify the limits
\begin{equation} 
0.306 < \epsilon_1 < 0.382~,
\label{eps1}
\end{equation}
and by using Eq.~(\ref{root2}),
\begin{equation}
0.433 < \epsilon_2 < .540~.
\label{eps2}
\end{equation}

The values of Eqs.~(\ref{eps1}, \ref{eps2}) lead directly to predictions for the neutrino mixing angles. Substitution of Eqs.~(\ref{eps1}, \ref{eps2}) into Eq.~(\ref{PerturbedTBM}) gives 
\begin{equation}
24.8^{\circ} \leq \theta_{13} \leq 30.9^{\circ}~,
\label{predict13}
\end{equation}
and
\begin{equation}
23.1^{\circ} \leq \theta_{23} \leq 27.4^{\circ}~,
\label{predict23}
\end{equation}

It is notable that this creates a theoretically motivated deviation from \gls{TBM} values, if far larger than experiments indicate. We are inclined to believe that when this \gls{MRTM} is fully expanded to account for full 3-family quark mixing, these projections will better accommodate experimental data.

On the topic of quark and lepton masses, too, we are disappointed with the lack of progress. Although we understand why $m_t \gg m_b > m_{c,s,d,u}$ for quarks and why $m_3 \gg m_1=m_2$ for neutrinos, when we look more closely at the details we find that masses are not quantitatively explained. It is not clear to us whether this will be corrected in the (\gls{T'}$\times$\gls{Z2}) model by higher order corrections, or by adding \gls{T'} doublet \glspl{VEV}. In the present work, we take the view that our model has made reliable predictions about mixing angles even when details of the mass spectra are incomplete.

\section{Correlated Projections}
\noindent
{\it This Section is largely based on the work of Ref.~\citen{Eby:2011aa}}

Recalling the values of the angles $\theta_{13}$ and $\theta_{23}$ listed in the 2010 Review of Particle Physics,\cite{Nakamura:2010zzi} as they help to illustrate the recent leap in experimental precision for \gls{PMNS} parameters,
\begin{equation}
36.8^{\circ} \lesssim \theta_{23} \leq 45.0^{\circ}~,~~~~~~~~~~~~~
0.0^{\circ} \leq \theta_{13} \lesssim 11.4^{\circ}~,
\label{2313}
\end{equation}
consistent with vanishing $\theta_{13}$ and maximal $\theta_{23}$.

Up to 2011, neutrino mixing angles were all empirically consistent with \gls{TBM} values. However, as the experimental precision has now improved due to recent data from T2K,\cite{Abe:2011sj, Dufour:2011zz, Hartz:2012np, Frank:2011zz, Izmaylov:2011np, T2Ktalk} MINOS,\cite{MINOStalk, Adamson:2012rm, Adamson:2011qu, Holin:2012np, Habig:2011zz, Orchanian:2011qq, Evans:2010zza} Double Chooz,\cite{DCtalk, DC, Palomares:2011zz, Abe:2011fz, Palomares:2009wz} Daya Bay,\cite{An:2012eh, DBtalk} and RENO,\cite{Ahn:2012nd, RENOtalk} this situation has changed dramatically. This is clearly seen in the global fits of Refs.~\citen{Fogli:2012ua, Tortola:2012te, GonzalezGarcia:2012sz}; of these we shall primarily use Fogli {\it et al}.,\cite{Fogli:2012ua} but will also include a limited analysis of Tortola {\it et al}.,\cite{Tortola:2012te} given its preference for a $\theta_{23}>45^{\circ}$. These five remarkable experiments have provided us with a rich new perspective on the mixing angles. From flavor symmetry, it is then possible to predict quantitatively how departures from the \gls{TBM} values are related. 

In this section, we intend to thoroughly investigate the ramifications of the most powerful prediction made by the $T^{'}$ model, that deviations from the \gls{TBM} matrix in Eq.~(\ref{TBM3}) in $\theta_{13}$ and $\theta_{23}$ are correlated and independent of the solar neutrino mixing angle $\theta_{12}$. To do this we shall consider only the projection on the two-dimensional $\theta_{23}\mhyphen\theta_{13}$ plane of the three-dimensional $\theta_{12}\mhyphen\theta_{23}\mhyphen\theta_{13}$ space. As a reminder, these perturbations stem from the small angle approximation, requiring $\sin{\alpha} \sim \alpha$ for $\theta_{13}$ and $(\frac{\pi}{4} - \theta_{23})$.\footnote{This is a $ < 1\%$ approximation for $\theta_{13}$ and $(\frac{\pi}{4} - \theta_{23})$ since both angles are less than $\alpha = 12^{\circ} = 0.2094$ radians with $\sin{\alpha}~=~0.2079$.} 
 
The data from KamLAND, LBL accelerators (like T2K and MINOS), solar experiments, SBL accelerators (such as Double Chooz, Daya Bay, and RENO), and Super-Kamiokande, as combined in Ref.~\citen{Fogli:2012ua} currently indicate (accounting for \gls{CP}-violation)
\begin{equation}
\sin^2{\theta_{13}}=0.0241_{~-0.0048}^{~+0.0049}~~~~~
{\rm with}~95\%~{\rm C.L.}~,
\label{t2k}
\end{equation}
for a \gls{NH}, as favored by \gls{T'}. 

As noted in Sec.~\ref{neutmass} our perturbed model leads to the linear relationship,\footnote{\gls{A4} is also capable of producing Eq.~(\ref{eta}) with $\eta = \sqrt{2}$, though we give preference in this analysis to \gls{T'} for its capacity to explain \gls{CKM} mixing.}$^{,}$\footnote{It is notable that Eq.~(\ref{eta}) with $\eta \simeq \sqrt{2}$ appears {\it en passant} in Ref.~\citen{Harrison:2005dj}; see also Ref.~\citen{Fuki:2006xw} which implies that $\eta \sim 2$. Another, model-independent correlation was developed in Ref.~\citen{Ge:2011qn}, including the three \gls{PMNS} mixing angles and the \gls{CP}-violating phase.}
\begin{equation}
\theta_{13} = \lvert \eta \rvert \left( \frac{\pi}{4} - \theta_{23} \right),
\label{eta}
\end{equation} 
with a sharp prediction, from Eq.~(\ref{root2}), of $\eta = \sqrt{2}$. Thus resulting in
\begin{equation}
\theta_{13} = \lvert \sqrt{2} \rvert \left( \frac{\pi}{4} - \theta_{23} \right)~.
\label{eta2}
\end{equation}

\begin{figure*}[t]
\includegraphics[width=\textwidth]{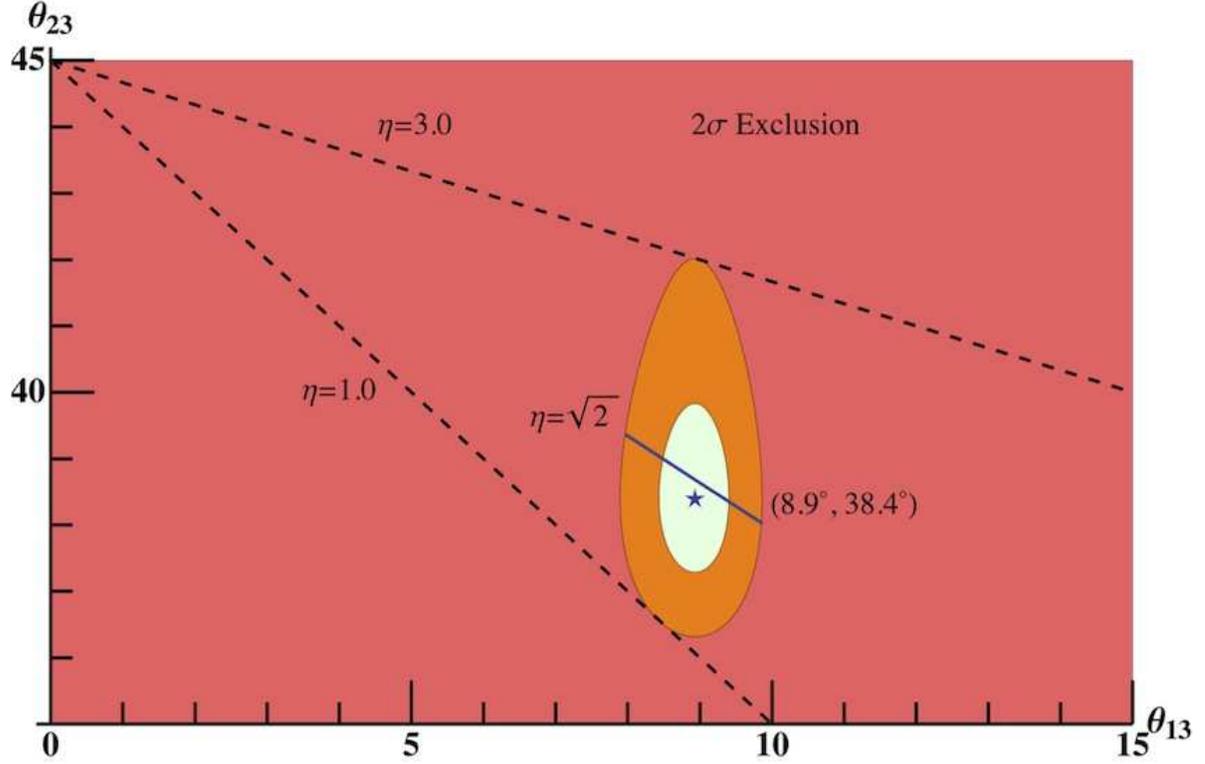}
\caption[Data comparison of $\theta_{13}{\it -}\theta_{23}$ neutrino mixing plane, assuming $0<\theta_{23}<45^{\circ}$]
{The global analysis of Ref.~\citen{Fogli:2012ua}, incorporating SBL, LBL, solar, and atmospheric neutrino observations, excludes the red-shaded region at $2\sigma$. The same assessment excludes the orange-shaded region at $1\sigma$. The best fit value for ($\theta_{13},\theta_{23}$) is indicated by the star at ($8.9^{\circ},38.4^{\circ}$). Extreme values of the linear correlation coefficient, $\eta$, are indicated by dashed lines at $\eta=1.0$ and $\eta=3.0$, while our predicted correlation of $\eta=\sqrt{2}$ is indicated by the solid dark blue line. The combination of our correlation and the experimental value of $\theta_{13}$ result in a prediction of $\theta_{23}=38.7$, a close match to its shown best fit value.}
\label{figure1}
\end{figure*}

\begin{figure*}[t]
\includegraphics[width=\textwidth]{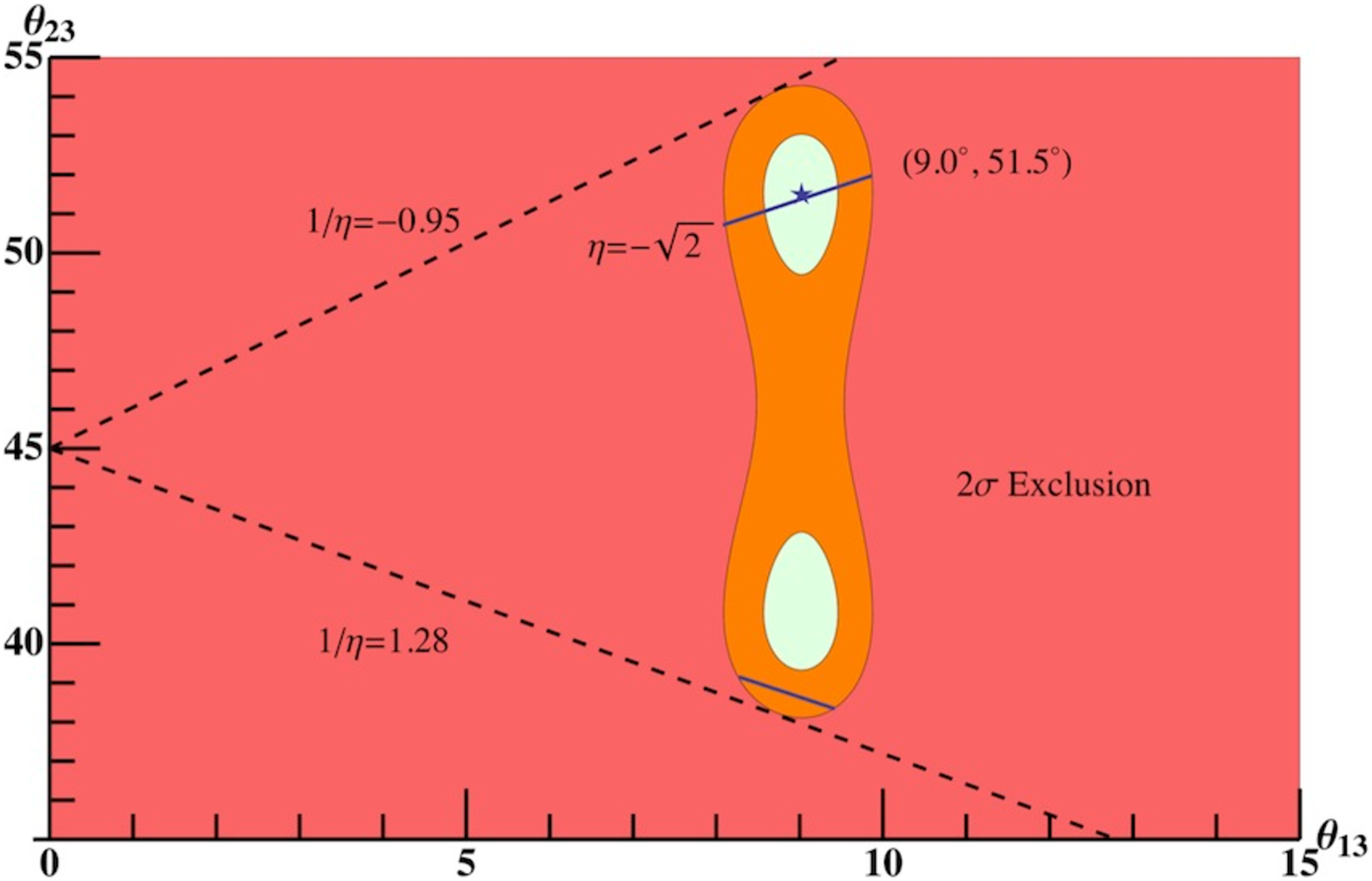}
\caption[Data comparison of $\theta_{13}{\it -}\theta_{23}$ neutrino mixing plane, assuming $0<\theta_{23}<90^{\circ}$]
{This figure shows a second global analysis by Ref.~\citen{Tortola:2012te}, including many of the same sources. The red-shaded region remains excluded at $2\sigma$, with $1\sigma$ exclusion for orange. The difference in this figure is the possibility that $\theta_{23}>45^{\circ}$. Since many experiments are only sensitive to the $\sin^2{2\theta_{23}}$, thus leaving the two octants degenerate, there have been some indications that the assumption $\theta_{23}<45^{\circ}$ is untrue. As it happens, our prediction does not distinguish between the octants and gives a best fit at $\theta_{13} = 9.0^{\circ}$ and $\theta_{23} = 51.4^{\circ}$, extremely close to the experimental best fit at $\theta_{23}=51.1^{\circ}$. In this case, it makes more sense to frame $\eta$ as $1/\eta$ to avoid running through $\infty$. Thus, the allowed range for this global fit exist from $1/\eta=1.28$ to $1/\eta=-0.95$.}
\label{figure2}
\end{figure*}

Several years ago Super-Kamiokande showed $\theta_{23} > 36.8^{\circ}$,\cite{Ashie:2005ik} and current single measurements place it at $\theta_{23} \simeq 40.7^{\circ}$.\cite{SKtalk} Once combined in a global fit of $3 \nu$ oscillation, Ref.~\citen{Fogli:2012ua} states the best fit of $\theta_{23}=38.4^{\circ}$, tantalizingly close to our central value of $\theta_{23} = 38.7^{\circ}$ (or, alternatively, in Ref.~\citen{Tortola:2012te} a best fit of $\theta_{23}=51.5^{\circ}$, compared with our value of $\theta_{23}=51.4^{\circ}$).

As shown in Fig.~\ref{figure1}, the recent experimental data,\cite{Fogli:2012ua} combined with theory, suggest that ($\theta_{13},~\theta_{23}$) are respectively closer to ($8.9^{\circ},~38.7^{\circ}$) than to ($0.0^{\circ},~45.0^{\circ}$). Before the surge of new data $\eta$ was unconstrained, $0 \leq \eta \leq \infty$; with the current global fit data, we find $1.0 \leq \eta \leq 3.0$.

Fig.~\ref{figure2}, using a different global analysis created from an alternate weighting of much of the same data,\cite{Tortola:2012te} suggests that $\theta_{23}$ does not lie in the first octant (i.e. that $\theta_{23}>45^{\circ}$). Because our derivation of Eq.~(\ref{eta}) is not sign dependent, we can alter our projection of $\theta_{23}$ accordingly. Based on this global fit and Eq.~(\ref{eta2}), ($\theta_{13},~\theta_{23}$) are approximately ($9.0^{\circ},~51.3^{\circ}$). Since this analysis still allows $\theta_{23}=45^{\circ}$, albeit at $1\sigma$ exclusion, which remains analogous to an $\eta$ of $\infty$, it makes more sense, for our second analysis, to state limits on $1/\eta$. As such, $1/\eta$ is here constrained to $-0.95 \leq 1/\eta \leq 1.28$.

This is in sharp contrast to the previously widespread acceptance of a maximal $\theta_{23} = \pi/4$, which fitted so well with vanishing $\theta_{13} = 0$ as in \gls{TBM}.

As the measurement of $\theta_{13}$ sharpens experimentally, so will our prediction for $\theta_{23}$ from Eq.~(\ref{eta2}), and an accurate measurement of the atmospheric neutrino mixing's departure from a maximum value will provide an interesting test of Binary Tetrahedral Flavor Symmetry.

While several paper have suggested links between these angles, ours is singular in tying the cause of this exact correlation to the Cabibbo angle's deviation from the rational form of Eq.~(\ref{Cabibbo}). This suggests to us that the \gls{T'} flavor symmetry, introduced in Ref.~\citen{Frampton:1994rk}, should be taken quite seriously. As errors in $\theta_{13}$ and $\theta_{23}$ diminish even further, it will be interesting to see how the \gls{T'} prediction of Eq.~(\ref{eta2}) perseveres, as it would inspire further investigation into other mixing angles for quarks and leptons. This, in turn, may show that \gls{T'}, first mentioned in physics as an example of an ${\bf SU(2)}$ subgroup,\cite{Case:1956zz} is actually a useful approximate symmetry in the physical application of quark and lepton flavors.

\chapter{\texorpdfstring{An Expanded \glsentrytext{T'} Model and Quark Mixing}{NMT'M}}
\label{EFM2}
\noindent
{\it This Chapter is largely based on the work of Ref.~\citen{Eby:2009ii}}
\section{Model Extension}

In the present chapter, we will examine the addition of \gls{T'} Higgs doublet scalars. As anticipated in Sec.~\ref{Original Model}, this allows more possibilities of \gls{T'} symmetry breaking and permits non-zero values for $\Theta_{23}$, $\Theta_{13}$ and $\delta_{KM}$. We present an explicit (\gls{T'}$\times$\gls{Z2}) model and investigate the \gls{CKM} angles.

Note that we continue to focus on a renormalizable model with few, if any, free parameters, and prioritize the mixing matrixes rather than the masses, as the former are more likely to have a geometrical interpretation without adding a surfeit of extra parameters, sadly leaving the masses unpredicted. With that said, the placement of the top quark in a singlet does allow it a much heavier mass in accordance with experiments.

Thus, we shall proceed to develop the \glsreset{NMRTM}\gls{NMRTM}. To do this we will introduce one \gls{T'} doublet scalar in an explicit model. This addition, then, allows non-vanishing $\Theta_{23}$ and $\Theta_{13}$ to be induced by symmetry breaking.

The possible choices under (\gls{T'}$\times$\gls{Z2}) for the new scalar field are:
\begin{equation}
{\bf A} ~~~ H_{2_1}(2_1, +1)~,
\label{newscalarA}
\end{equation}

\begin{equation}
{\bf B} ~~~ H^{'}_{2_3}(2_3, -1)~,
\label{newscalarB}
\end{equation}

\begin{equation}
{\bf C} ~~~ H^{'}_{2_2}(2_2, -1)~,
\label{newscalarC}
\end{equation}

\begin{equation}
{\bf D} ~~~ H_{2_3}(2_3, +1)~,
\label{newscalarD}
\end{equation}
allowing the following Yukawa couplings, respectively,
\begin{equation}
{\bf A} ~~~ Y_{Qt} Q_{\rm L} t_{\rm R} H_{2_1} + h.c.~,
\label{newYukawaA}
\end{equation}

\begin{equation}
{\bf B} ~~~ Y_{Qb} Q_{\rm L} b_{\rm R} H^{'}_{2_3} + h.c.~,
\label{newYukawaB}
\end{equation}

\begin{equation}
{\bf C} ~~~ Y_{{\cal Q}{\cal C}} {\cal Q}_{\rm L} {\cal C}_{\rm R} H^{'}_{2_2} + h.c.~,
\label{newYukawaC}
\end{equation}

\begin{equation}
{\bf D} ~~~ Y_{{\cal Q}{\cal S}} {\cal Q}_{\rm L} {\cal S}_{\rm R} H_{2_3} + h.c.~.
\label{newYukawaD}
\end{equation}

This leaves us to choose between multiple candidates for the \gls{NMRTM}. Largely to ensure computational simplicity, we opt for the single additional term, {\bf D}, inspired by the Chen-Mahanthappa mechanism for \gls{CP}-violation.\cite{Chen:2009gf} We shall choose to keep $Y_{{\cal Q}{\cal S}}$ real, allowing \gls{CP}-violation to arise from the imaginary part of \gls{T'} Clebsch-Gordan coefficients.

The \gls{VEV} for $H_{2_3}$ is taken with the alignment
\begin{equation}
<H_{2_3}> = V_{2_3} (1, 1)~.
\label{newvev}
\end{equation}

\section{\texorpdfstring{\glsentrytext{NMRTM} (D) Predictions}{Predictions of NMRT'M}}
From the Yukawa term, {\bf D}, and the vacuum alignment, we can derive the down-quark mass matrix:
\begin{equation}
D = \left( \begin{array}{ccc}
M_b & 
\frac{1}{\sqrt{2}} Y_{{\cal Q}{\cal S}} V_{2_3} & 
\frac{1}{\sqrt{2}} Y_{{\cal Q}{\cal S}} V_{2_3} \\
0 & 
\frac{1}{\sqrt{3}} Y_{{\cal S}} V & 
2 \sqrt{\frac{2}{3}} \omega^2 Y_{{\cal S}} V \\
0 &
\sqrt{\frac{2}{3}} Y_{{\cal S}} V &
-\frac{1}{\sqrt{3}} \omega^2 Y_{{\cal S}} V
\end{array}
\right),
\label{D}
\end{equation}
where $\omega=e^{2i\pi/3}$, and $M_b=Y_bV_{1_3}$.

The hermitian squared mass matrix ${\cal D} \equiv D D^{\dagger}$ for the down-type quarks is then
\begin{equation}
\resizebox{\linewidth}{!}{$
{\cal D} =
\begin{pmatrix}
M^{'2}_b &
\frac{1}{\sqrt{6}} Y_{{\cal S}} Y_{{\cal Q}{\cal S}} V V_{2_3} (1 - 2\sqrt{2}\omega) &
\frac{1}{\sqrt{6}} Y_{{\cal S}} Y_{{\cal Q}{\cal S}} V V_{2_3} (\omega + \sqrt{2}) \\
\frac{1}{\sqrt{6}} Y_{{\cal S}} Y_{{\cal Q}{\cal S}} V V_{2_3} (1-2\sqrt{2}\omega^{2}) &
3 (Y_{{\cal S}} V)^2 &
-\frac{\sqrt{2}}{3} (Y_{{\cal S}} V)^2 \\
\frac{1}{\sqrt{6}} Y_{{\cal S}} Y_{{\cal Q}{\cal S}} V V_{2_3} (\omega^{2} + \sqrt{2}) &
-\frac{\sqrt{2}}{3} (Y_{{\cal S}} V)^2 &
(Y_{{\cal S}} V)^2
\end{pmatrix},
\label{calD}
$}
\end{equation}
where $M^{'2}_b = M_b^2 + (Y_{{\cal Q}{\cal S}} V_{2_3})^2$.

Note that in this model the mass matrix for the up-type quarks is diagonal,\footnote{This uses the approximation that the electron mass is $m_e=0$; {\it c.f.} Ref.~\citen{Frampton:2008bz}.} so the \gls{CKM} mixing matrix arises purely from diagonalization of ${\cal D}$ in Eq.~(\ref{calD}). The presence of the complex \gls{T'} Clebsch-Gordan in Eq.~(\ref{calD}) acting as the source of the \gls{CP}-violating phase, $\delta_{{\rm KM}}$ (Chen-Mahanthappa mechanism).

In Eq.~(\ref{calD}) the ($2\times2$) sub-matrix for the first two families coincides with the result discussed in Sec.~\ref{Original Model}, thereby preserving the successful Cabibbo Angle formula $\tan 2\Theta_{12} = (\sqrt{3})/2$.

For $m_b^2$ the experimental value is $17.5~{\rm GeV}^2$,\cite{Beringer:1900zz} although the \gls{CKM} angles and phase do not depend on this overall normalization.

Actually our results depend only on assuming that the ratio $(Y_{{\cal Q}{\cal S}} V_{2_3}/ Y_{{\cal S}} V)\ll1$ is much smaller than one.

Defining
\begin{equation}
{\cal D}^{'} = 3 {\cal D}/(Y_{\cal S} V)^2~,
\label{calDprime}
\end{equation}
we find
\begin{equation}
{\cal D}^{'}=
\left( \begin{array}{ccc}
{\cal D}^{'}_{11} &
A e^{-i\psi_1} & 
A \xi e^{-i\psi_2} \\
A e^{i\psi_1} &
9 & 
-\sqrt{2} \\
A \xi e^{i\psi_2} & 
-\sqrt{2} & 
3
\end{array}
\right),
\label{Dprime}
\end{equation}
in which we defined the following:
\begin{equation}
{\cal D}^{'}_{11} = 3 M_b^{'2} /(Y_{\cal S}V)^2~,
\label{calD11}
\end{equation}

\begin{equation}
A = \left( \sqrt{\frac{3}{2}} \right) \left( \frac{Y_{\cal Q \cal S} V_{2_3}}{Y_{\cal S} V} \right)
|1-2 \sqrt{2} \omega|~,
\label{Aeq}
\end{equation}

\begin{equation}
\xi = \left| \frac{\omega + \sqrt{2}}{1-2\sqrt{2}\omega} \right| = 0.36615...~,
\label{xi}
\end{equation}

\begin{equation}
\tan \psi_1 = \frac{-\sqrt{6}}{1+\sqrt{2}}
= - 1.01461...~,
\label{psi1}
\end{equation}

\begin{equation}
\tan \psi_2 = \frac{\sqrt{3}}{2\sqrt{2}-1}
 = 0.94729...~,
 \label{psi2}
\end{equation}
$\psi_1$ and $\psi_2$ have been included to consolidate the imaginary portion of ${\cal D}^{'}$ elements removed by the absolute values in $A$ and $\xi$.

To arrive at predictions for the other \gls{CKM} mixing elements other than the Cabibbo angle ({\it i.e.} $\Theta_{13},\Theta_{23}, \delta_{{\rm KM}}$) one only needs to diagonalize the matrix ${\cal D}^{'}$ in Eq.~(\ref{Dprime}) by 
\begin{equation}
{\cal D}^{'}_{{\rm diag}} = V_{CKM}^{\dagger} {\cal D}^{'} V_{CKM}~.
\label{diagonal}
\end{equation}

We now write the mixing matrix as
\begin{equation}
V_{{\rm CKM}} = \left( \begin{array}{ccc}
1 & V_{ts} & V_{td} \\
V_{cb} & \cos\Theta_{12} & \sin\Theta_{12} \\
V_{ub} & -\sin\Theta_{12} & \cos\Theta_{12}
\end{array}
\right),
\label{fullCKM}
\end{equation}
which, with Eq.~(\ref{Dprime}), can be substituted into Eq.~(\ref{diagonal}), becoming
\begin{equation}
\left( \begin{array}{c}
V_{cb} \\
V_{ub}
\end{array} \right)
= \frac{1}{\hat{{\cal D}}_{11}^{'}}
\left( \begin{array}{cc} 
{\cal D}^{'}_{11} - 3 & -\sqrt{2} \\
-\sqrt{2} & {\cal D}^{'}_{11} - 9
\end{array}
\right)
\left( \begin{array}{c}
A e^{-i\psi_1} \\
A \xi e^{-i\psi_2}
\end{array} \right),
\label{VV1}
\end{equation}
where $\hat{{\cal D}}_{11}^{'} = ({\cal D}^{'}_{11} -6 - \sqrt{11})({\cal D}^{'}_{11} - 6 + \sqrt{11})$, while from unitarity it follows that
\begin{equation}
\left( \begin{array}{c}
V_{ts} \\
V_{td}
\end{array} \right)
= -
\left( \begin{array}{cc}
\cos\Theta_{12} & - \sin\Theta_{12} \\
\sin\Theta_{12} & \cos\Theta_{12} 
\end{array}
\right)
\left( \begin{array}{c}
V_{cb}^{*} \\
V_{ub}^{*}
\end{array} \right).
\label{VV2}
\end{equation}

Our strategy is to now calculate the \gls{CP}-violating Kobayashi-Maskawa phase,
\begin{equation}
\delta_{{\rm KM}} = \gamma = \arg \left( - \frac{V_{ud}V_{ub}^{*}}{V_{cd}V_{cb}^{*}}
\right),
\label{KM}
\end{equation}
and, by using Eqs.~(\ref{fullCKM}, \ref{VV1}), we arrive at the formula in terms of ${\cal D}_{11}$
\begin{equation}
\delta_{KM} = \gamma_{ T^{'}}
= \arg \left[ \frac{-\sqrt{2} + ({\cal D}^{'}_{11}-9) \xi e^{-i(\psi_1 - \psi_2)}}
{({\cal D}^{'}_{11} - 3) - \sqrt{2} \xi e^{-i(\psi_1 -\psi_2)}} \right]
=
\arg [\Gamma({\cal D}_{11}^{'})]~.
\label{gammaTprime}
\end{equation}

From the preceding equations (\ref{fullCKM}, \ref{VV1}) we find a formula for
\begin{equation}
|V_{ub}/V_{cb}| = |\tan\Theta_{13}\csc\Theta_{23}| ~,
\end{equation}
using unitarity, Eq.~(\ref{VV2}), and the form for the ratios of \gls{CKM} matrix elements
\begin{equation}
|V_{td}/V_{ts}| = \left| \frac{ \sin\Theta_{12} + \Gamma({\cal D}_{11}^{'}) \cos\Theta_{12} }
{\cos\Theta_{12} - \Gamma({\cal D}_{11}^{'}) \sin\Theta_{12}} \right|~.
\end{equation}

\begin{figure}[t]
\begin{center}
\includegraphics[height=80mm]{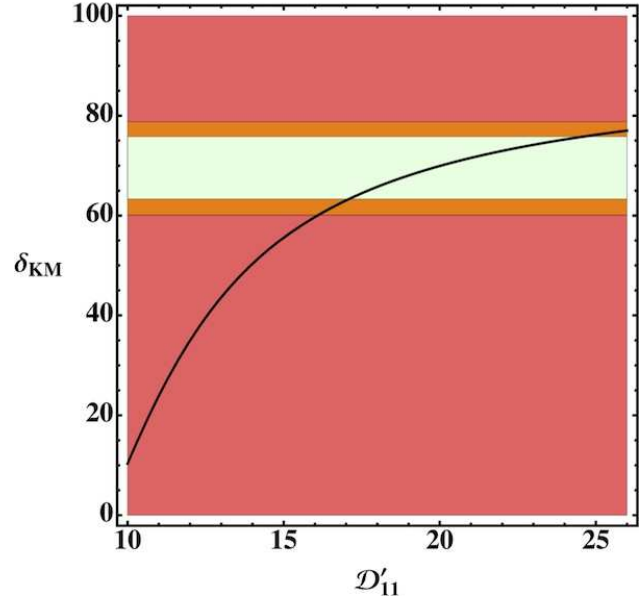}
\caption[Data comparison of quark mixing parameter $\delta_{KM}$]
{\normalsize
The vertical axis is the value of $\delta_{{\rm KM}} \equiv \gamma_{T^{'}}$ in degrees and the horizontal axis is the value of ${\cal D}^{'}_{11}$ defined in the text. The dashed horizontal lines give the $1\sigma$ range for $\delta_{{\rm KM}}$ allowed by the global fit of Ref.~\citen{preMoriond}. The orange area is excluded by $2\sigma$ confidence, while the red region is excluded by $3\sigma$ confidence.}
\label{FigureDprime}
\end{center}
\end{figure}

\section{\texorpdfstring{Comparison with \glsentrytext{CKM} Data}{CKM Data}}
In Fig.~\ref{FigureDprime}, we show a plot of $\gamma_{T^{'}}$ versus ${\cal D}_{11}^{'}$ using Eq.~(\ref{gammaTprime}) and taking the range of experimentally allowed $\gamma \equiv \delta_{{\rm KM}}$ from the global fit of Ref.~\citen{preMoriond} prompts us to use a value ${\cal D}_{11}^{'} \backsim 20\pm 4$ in the subsequent analysis.

\begin{figure}[!ht]
\begin{center}
\includegraphics[height=80mm]{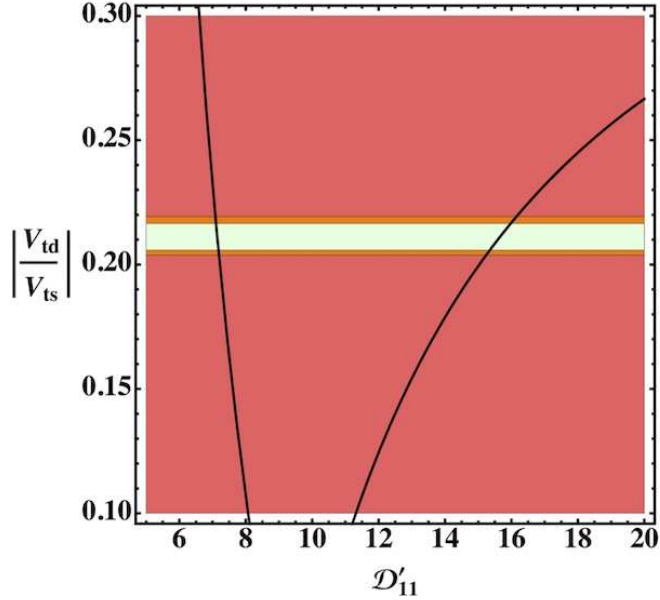}
\caption[Data comparison of quark mixing parameter $| V_{td}/V_{ts} |$]
{\normalsize
The vertical axis is the value of $|V_{td}/V_{ts}|$ and the horizontal axis is the value of ${\cal D}^{'}_{11}$ defined in the text. The dashed horizontal lines give the value with small error allowed by the global fit of Ref.~\citen{preMoriond}. The orange area is excluded by $2\sigma$ confidence, while the red region is excluded by $3\sigma$ confidence.}
\label{FigureVV}
\end{center}
\end{figure}

\begin{figure}[!ht]
\begin{center}
\includegraphics[height=80mm]{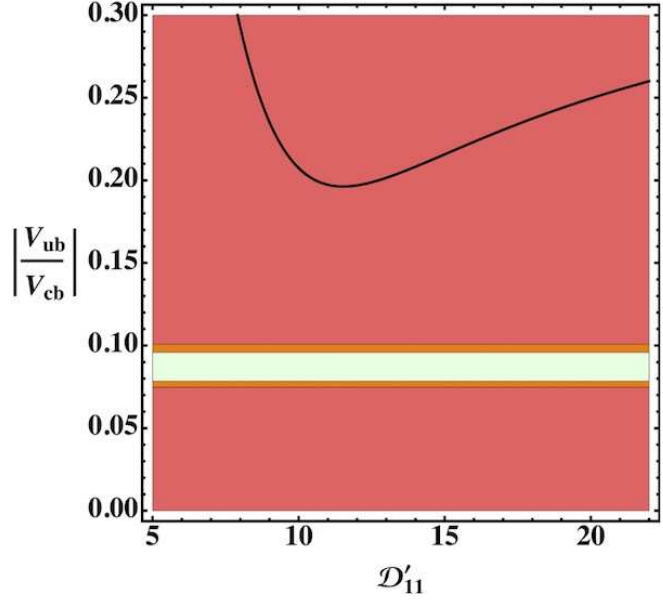}
\caption[Data comparison of quark mixing parameter $| V_{ub}/V_{cb} |$]
{\normalsize
The vertical axis is the value of $|V_{ub}/V_{cb}|$ and the horizontal axis is the value of ${\cal D}^{'}_{11}$ defined in the text. The dashed horizontal lines give the preferred experimental values allowed by the global fit of Ref.~\citen{preMoriond}. The orange area is excluded by $2\sigma$ confidence, while the red region is excluded by $3\sigma$ confidence.}
\label{FigureFail}
\end{center}
\end{figure}

Fig.~\ref{FigureVV} shows a plot of $|V_{td}/V_{ts}|$ as a function of ${\cal D}_{11}^{'}$. It requires a value of ${\cal D}_{11}^{'}$ of approximately 16 which is sufficiently close to that in Fig.~\ref{FigureDprime}.

For the value of $|V_{ub}/V_{cb}|$ there is approximately a factor of $2$ between the prediction (higher) and the best value from Ref.~\citen{preMoriond} as seen in Fig.~\ref{FigureFail}.

Note that once the off-diagonal, third family elements in Eq.~(\ref{calD}) are taken as much smaller than the elements involved in the Cabibbo angle, and that the two CKM angles and the \gls{CP} phase are predicted by the present \gls{NMRTM}.

With regard to alternatives to \gls{NMRTM}({\bf D}), named earlier in Eqs.~(\ref{newscalarA}, \ref{newscalarB}, \ref{newscalarC}), the possibilities {\bf A} and {\bf C} modify the up-type mass matrix where we take flavor and mass eigenstates coincident. The final possibility {\bf B} does modify the down-type mass matrix (like {\bf D} does), but fails to permit the \gls{CP}-violation we prefer (seen in the Chen-Mahanthappa mechanism), as in the present {\bf D} model and in Ref.~\citen{Frampton:2009fw}.

\chapter{Quartification}
\noindent
{\it This Chapter is largely based on the work of Ref.~\citen{Eby:2011ph}}
\label{Quart}
\section{\texorpdfstring{A \glsentrytext{T'} Quiver Model}{Quiver}}

Now that we have managed to construct a functional \gls{NMRTM}, it may be of interest to examine the wider context of fundamental physics. While we have previously noted the group symmetries utilized by the \gls{MSM}, we have not sought to incorporate the (${\bf SU(3)_{\rm C}}\times {\bf SU(2)_{\rm L}}\times {\bf U(1)_{\rm Y}}$) groups into our model. In this chapter, we shall attempt to provide a suitable \gls{MSM}-like framework that is both compatible with the assignments of our \gls{T'} model and allows the particles to rotate under the appropriate groups.

For the purposes of this investigation we shall craft a model with ${\bf SU(3)}^N$. These types of group combination are sometimes termed quiver groups, due to the fact that the graphs used to diagram the various bifundamental representations resemble a series of arrows. 

We will begin by considering a quartification model using ${\bf SU(3)}^4$,\cite{Babu:2007fx} with bifundamental chiral fermions in the usual arrangement of bifundamentals, but find we are unable to make the necessary charge assignments to recover the requisite \gls{T'} family symmetry. This will lead us to add a sub-quiver of fermions to accommodate \gls{T'} quartification. We will give each \gls{irrep} under \gls{T'} a new set of assignments under the quartification groups comprised of singlets ($1$), triplets ($3$), and conjugate triplets ($\overline{3}$).

The quartification gauge group is
\beq
{\bf SU(3)_{\rm C}} \times {\bf SU(3)_{\rm L}} \times {\bf SU(3)_{\ell}} \times {\bf SU(3)_{\rm R}}~,
\eeq
which is assumed to break to the standard model at the ${\rm TeV}$ scale, and includes the common groups aligned with color, left-handed particles, leptons, and right-handed particles, respectively. We choose the family symmetry to be: (\gls{T'}$\times$\gls{Z2}) with the minimal anomaly-free bifundamental chiral fermions:
\beq
3[ (3,\bar{3},1,1) + (\bar{3},1,1,3) + (1,3,\bar{3},1) + (1,1,3,\bar{3})]~,
\eeq
where we assign the leptons to \glspl{irrep} as follows:
\beq
\setlength{\arraycolsep}{2pt}
\begin{array}{cccc}
\left. \begin{array}{c}
(13\bar3 1)_3\supset\left( \begin{array}{c} \nu_{\tau} \\ \tau^- \end{array} \right)_{{\rm L}} \\
(13\bar3 1)_2\supset\left( \begin{array}{c} \nu_{\mu} \\ \mu^- \end{array} \right)_{{\rm L}} \\
(13\bar3 1)_1\supset\left( \begin{array}{c} \nu_e \\ e^- \end{array} \right)_{{\rm L}} 
\end{array} \right\} 
L_{\rm L} (3, +1)~, &
\begin{array}{c}
~ (113\bar3 )_3\supset\tau^-_{{\rm R}}~ (1_1, -1)~ \\
~ (113\bar3 )_2\supset\mu^-_{{\rm R}} ~ (1_2, -1)~ \\
~ (113\bar3 )_1\supset e^-_{{\rm R}} ~ (1_3, -1)~,
\end{array}
&
{\rm and}
&
\begin{array}{c}
N^{(1)}_{\rm R} ~ (1_1, +1)~ \\
N^{(2)}_{\rm R} ~ (1_2, +1)~ \\
N^{(3)}_{\rm R} ~ (1_3, +1)~.\\
\end{array}
\end{array}
\label{leptons}
\eeq

For the left-handed quarks, we make the assignment,
\begin{equation}
\begin{array}{cc}
(3\bar3 11)_3\supset\left( \begin{array}{c} t \\ b \end{array} \right)_{{\rm L}}
~ {\cal Q}_{\rm L} ~~~~~~~~~~~ ({\bf 1_1}, +1)~ \\
\left. \begin{array}{c} (3\bar3 11)_2\supset\left( \begin{array}{c} c \\ s \end{array} \right)_{{\rm L}}
\\
(3\bar3 11)_1\supset\left( \begin{array}{c} u \\ d \end{array} \right)_{{\rm L}} \end{array} \right\}
Q_{\rm L} ~~~~~~~~ ({\bf 2_1}, +1)~.
\end{array}
\label{qL2}
\end{equation}

Finally we need assignments for the six right-handed quarks. They were previously assigned in Eq.~(\ref{QuarkAssign}) as,
\begin{equation}
\begin{array}{c}
t_{{\rm R}} ~~~~~~~~~~~~~~ ({\bf 1_1}, +1)~ \\
b_{{\rm R}} ~~~~~~~~~~~~~~ ({\bf 1_2}, -1)~ \\
\left. \begin{array}{c} c_{{\rm R}} \\ u_{{\rm R}} \end{array} \right\}
{\cal C}_{\rm R} ~~~~~~~~ ({\bf 2_3}, -1)~\\
\left. \begin{array}{c} s_{{\rm R}} \\ d_{{\rm R}} \end{array} \right\}
{\cal S}_{\rm R} ~~~~~~~~ ({\bf 2_2}, +1)~.
\end{array}
\label{qR2}
\end{equation}
However, this assignment in not available here since $t_{\rm R}$ and $b_{\rm R}$ are both in the same \gls{irrep}, $(\bar3113)_3$, and likewise for the first and second families. With no alteration of the model, we can only assign three of the six right-handed quarks. In our attempts to correct this problem, we attempted a number of possible alterations, but even adding a fifth ${\bf SU(3)}$ (this would have been Quintification) failed to alleviate the problem of insufficient \glspl{irrep} to close under known couplings.

We therefore need to add an anomaly-free sub-quiver representation,
\beq
3[(\bar{3}, 1, 3, 1)^{'} + (1, 1, \bar{3}, 3)^{'} + (3, 1, 1, \bar{3})^{'}]~,
\eeq
and proceed to reassign {\it all} fermions with \gls{Z2}$=-1$, including the corresponding subset in Eq.~(\ref{leptons}), and Eq.~(\ref{qR2}), to this sub-quiver:
\beq
\begin{array}{c}
b_{\rm R} ~~~~~~~~~~~~~~~~~~~ \subset (\bar{3}, 1, 3, 1)^{'}_3~ \\
{\cal C}_{\rm R} ~~~~~~~~~~~~~~~~~~~\subset (\bar{3}, 1, 3, 1)^{'}_{1,2}~ \\
\tau^-_{\rm R} ~~~~~~~~~~~~~~~~~ \subset (1, 1, \bar{3}, 3)^{'}_3~ \\
\mu^-_{\rm R} ~~~~~~~~~~~~~~~~~~ \subset (1, 1, \bar{3}, 3)^{'}_2~ \\
e^-_{\rm R} ~~~~~~~~~~~~~~~~~~\subset (1, 1, \bar{3}, 3)^{'}_1~.
\end{array}
\label{subquiver}
\eeq

\section{Yukawa Couplings}
We shall now introduce a notation for abbreviating the extended group designations of the \gls{T'} Quiver model. For each \gls{irrep} this notation utilizes a superscript to denote which ${\bf SU(3)}$ was assigned a $3$ and a subscript for each ${\bf SU(3)}$ assigned a $\bar{3}$. The benefit being that when combined into Yukawa terms, one can check that, for each term, every group in superscript should also be included in subscript on another (this notation will not apply to the Yukawa couplings, solely the objects rotating under our groups). We also list the assignments under \gls{T'} in parenthesis with a superscript $+$ or $-$ to distinguish between \gls{Z2}$=+1$ and \gls{Z2}$=-1$, respectively. In our first demonstration, the lepton Yukawas are
\begin{equation}
\Sigma_{i=1}^{i=3} Y_D^{(i)} L^{\rm L}_{\ell} (3^+) N^{\ell (i)}_{\rm R} (1_i^+) H_{\rm L}^{\rm R} (3^+)~,
\label{leptons1}
\end{equation}
for the neutrino terms and,
\begin{eqnarray}
&&\Sigma_{i=1}^{i=3} Y^{(i)}_{\ell} L_{\ell}^{\rm L} (3^+) \ell_{\rm R}^{\ell (i)}(1_i^+) H_{\rm L}^{\rm R} (3^-)~,
\label{leptons2}
\end{eqnarray}
for the charged terms. Where one can clearly see that in each term there is an $L$, $R$, and $\ell$ in both super- and subscript. Adopting the previous work from Ch.~\ref{EFM2}, the quark Yukawa couplings are
\begin{eqnarray}
&& Y_t {\cal Q}_{\rm L}^{\rm C} (1_1^+) t_{\rm C}^{\rm R} (1_1^+) H_{\rm R}^{\rm L} (1_1^+) + \nonumber \\
&& Y_b {\cal Q}_{\rm L}^{{\rm C}} (1_1^+) b_{\rm C}^{\ell} (1_2^-) H_{\ell}^{\rm L} (1_3^-) + \nonumber \\
&& Y_{{\cal Q} {\cal S}} {\cal Q}_{\rm L}^{\rm C} (1_1^+) {\cal S}_{\rm C}^{\rm R} (2^+) H_{\rm R}^{\rm L} (2_3^+) + \nonumber \\
&& Y_{{\cal C}} Q_{\rm L}^{\rm C} (2_1^+){\cal C}_{\rm C}^{\ell} (2_3^-) H_{\ell}^{\rm L} (3^-) + \nonumber \\
&& Y_{{\cal S}} Q_{\rm L}^{\rm C} (2_1^+) {\cal S}_{\rm C}^{\rm R} (2_2^+) H_{\rm R}^{\rm L} (3^+)~.
\end{eqnarray}

The Higgs scalar sector is sufficient to break to the \gls{MSM} and replicate the previously determined mixing matrices (Chs.~\ref{EFM1}, \ref{EFM2}). Note that, for example, the Cabibbo angle in Sec.~\ref{Original Model} follows because, after the breaking of (${\bf SU(3)_{\ell}} \times {\bf SU(3)_{\rm R}}$), the $H (3^-)$s have a common representation, and can thus act as the appropriate messengers between the charged leptons and the first two families of quarks. The Higgs \gls{T'} doublet, $2_3^+$ (Eq.~\ref{newYukawaD}), allows reproduction of the successful \gls{CKM} matrix derived in Ch.~\ref{EFM2}.

The Higgs \glspl{VEV} follow a form highly similar to that in Sec.~\ref{Original Model}:
\begin{eqnarray}
&& <H_{\rm L}^{\rm R} (3^{-})>=(\frac{m_{\tau}}{Y_{\tau}},\frac{m_{\mu}}{Y_{\mu}},\frac{m_{e}}{Y_{e}})~, \nonumber \\
&& <H_{\rm L}^{\rm R} (3^{+})>=V(-2,1 ,1)~, \nonumber \\
&& <H_{\rm R}^{\rm L} (2_3^+)>=V_{2_3} (1,1)~, \\
&& <H_{\rm R}^{\rm L} (1_{1}^{+})>=\frac{m_{t}}{Y_{t}}~, \nonumber \\
&& <H_{\rm L}^{\ell} (1_{3}^{-})>=\frac{m_{b}}{Y_{b}}~.  \nonumber 
\end{eqnarray}

We have now shown that it is possible to craft a model that successfully combines the predictiveness of the finite group \gls{T'}, with the familiar physics of the \gls{MSM}. While this framework contains no additional physics, or new predictions, it demonstrates that a combination of \gls{T'} and Lie groups is feasible. While at this point it is too early to claim that this is a sufficient replacement for the \gls{MSM}, it is sufficient to note that this framework demonstrates a unified symmetry and can act as a proof-of-concept for further attempts at unification.

\chapter{\texorpdfstring{\glsentrytext{T'} Model Dark Matter}{T' Dark Matter}}
\label{ch:T'DM}
\noindent
{\it This Chapter is largely based on the work of Ref.~\citen{Eby:2011qa}}
\section{Dark Matter Background}

As mentioned in Sec.~\ref{general}, dark matter remains one of the leading mysteries in modern physics. And, while the community has yet to reach a consensus on an explanation, there have been no shortages of suggested ideas. Fortunately, there have been a number of clues that have allowed us to better understand dark matter and, consequently, rule some possibilities out. Of course the true answer, need not be any single theory mentioned here, or elsewhere, and could be a combination of several, but most theories, and our calculations for this chapter, will assume (if only for simplicity) that our suggested candidate is the sole contributor.

The \glsreset{WIMP}\gls{WIMP} remains the best-known suggestion to this problem for several reasons. First is the so-called \gls{WIMP}-miracle, which notes that a particle with the appropriate relic abundance to explain dark matter would need a cross-section no larger than one typically seen on the weak scale. Additionally, this theory would indicate there are heavy, undiscovered particles (a common element in many \gls{BSM} models, including ours) who have had a significant impact on cosmological development. Many of these \gls{WIMP} candidates arise out of R-parity conserving \gls{SUSY} models, and typically come about as the lightest remaining \gls{SUSY} particle. A \gls{WIMP} is, as the name suggests, a rarely forming but massive stable particle, capable of interacting with know matter only by weak interactions and gravity. This would be an example of cold dark matter (non-relativistic), and would likely have gained its current stability via the mechanism called thermal freeze-out. As the universe cooled, higher energy particles or interactions would become less preferred until all that remained was a supply of dark matter. As the universe expanded the remaining annihilations would grow fewer as the particles diminished in number and were spread out.

Other ideas for dark matter include axions, a suggestion of Ref.~\citen{Peccei:1977hh} intended to solve the "Strong \gls{CP} Problem", and Massive Compact Halo Objects (sometimes abbreviated MACHOs). Searches for these objects continue, and there are numerous groups continuing to investigate these ideas and even more exotic theories.

Two additional topics of interest are some theories that have fallen out of favor. The first of these, Modified Newtonian Dynamics (MOND),\cite{Milgrom:1983ca} attempted to alter Newton's law of gravity to better accommodate the astronomical observations (an appealing idea since, to this point, there has been no short-range proof of dark matter) rather than resorting to the "missing mass" hypothesis that underlies this chapter. However, following the observation of the bullet cluster in Ref.~\citen{Clowe:2006eq} and some failures to explain galactic rotation curves, MOND has largely fallen out of favor. Another idea to explain observations has been hot dark matter (relativistic), primarily from neutrinos. While, at one time, there was considerable interest that neutrinos, always difficult to detect, were a significant cause of dark matter, this assumption led to significant changes to large-scale astronomical structure formation. As a consequence, they are now believed to play a relatively minor part of the universe's mass.

\section{The Valencia Mechanism and an Augmented Model}
\label{DMT' Model}
An ingenious new mechanism involving \gls{A4} model building has been discovered by a research group based in Valencia, Spain.\cite{Hirsch:2010ru,Boucenna:2011tj} Their research uses \gls{A4}, whose double cover is central to our present work, to add a small number of extra scalar fields, one of which, by virtue of a discrete \gls{Z2} analogous to R-symmetry in \gls{SUSY}, gives rise to stable dark matter. 

Their original model assigned all standard model leptons to different singlets of \gls{A4}, with the right-handed neutrinos and one of the newly added Higgs as the only triplets (their model's other Higgs was a singlet). These assignments were unconventional, as most \gls{A4} models, like the \gls{T'} model discussed in Sec.~\ref{Original Model}, utilize triplets in the lepton assignments.

In Refs.~\citen{Hirsch:2010ru,Boucenna:2011tj} a particular generator of \gls{A4} was used to give rise to a \gls{Z2} subgroup of \gls{A4} and stabilized the \gls{WIMP}. This \gls{Z2} established a particle sector that is discrete from the \gls{MSM} particles and inaccessible to it, except via the weak nuclear force and potentially gravity.

Since \gls{A4} alone has proved incapable of accommodating quarks in a like manner to leptons,\cite{Altarelli:2005yp,Altarelli:2005yx,Altarelli:2006kg} the Valencia group relegated the quark sector to "future work". An alternative approach, that we pursue, is to replace their \gls{A4} group with \gls{T'}, allowing the incorporation of quarks, a prediction of the Cabibbo angle, and controllable deviations from the \gls{TBM} angles.

\subsection{\texorpdfstring{Alterations to the \glsentrytext{MRTM}}{DMT'}}
To accommodate the quark sector, we adopt the (\gls{T'}$\times$\gls{Z2}) model formulated in Sec.~\ref{Original Model} and further analyzed in Ch.~\ref{EFM1}. This section will establish an extended model including elements of the Valencia Mechanism by incorporating a second \gls{Z2}, which we will label \gls{Z2}${\bf ^{'}}$ for clarity, while also adding scalar fields and heavy right-handed neutrinos that are odd under this new group; the lightest resultant odd scalar will be our dark matter \gls{WIMP}. This model contains a global symmetry of (\gls{T'}$\times$\gls{Z2}$\times$\gls{Z2}${\bf ^{'}}$) restricting the Yukawa couplings. One key difference from Ref.~\citen{Hirsch:2010ru,Boucenna:2011tj} is that our \gls{Z2}${\bf ^{'}}$ will not be subgroup of our added \gls{T'} and is instead an exterior addition.

The quark assignments below are unchanged from Eq.~(\ref{QuarkAssign}), and denote ${\cal{Q}}_{\rm L}=\dbinom{t}{b}_{\rm L}$, $Q_{\rm L}=\dbinom{c}{s}_{\rm L}~\&~\dbinom{u}{d}_{\rm L}$, ${\cal{C}}_{\rm R}= c_{\rm R}~\&~u_{\rm R}$, and ${\cal{S}}_{\rm R}= s_{\rm R}~\&~d_{\rm R}$. By setting all quarks to be even under \gls{Z2}${\bf ^{'}}$, past \gls{T'} predictions are preserved.

\begin{table}[ht]
\renewcommand{\arraystretch}{1.75}
\begin{center}
\begin{tabular}{c||c|c|c|c|c|c|}
Quarks & ${\cal{Q}}_{\rm L}$ & $Q_{\rm L}$ & $t_{\rm R}$ & $b_{\rm R}$ & ${\cal{C}}_{\rm R}$ & ${\cal{S}}_{\rm R}$ 
\\ \hline \hline
${\bf T^{'}}$ & $1_1$ & $2_1$ & $1_1$ & $1_2$ & $2_3$ & $2_2$ \\ \hline
${\bf Z_2}$ & $+$ & $+$ & $+$ & $-$ & $-$ & $+$ \\ \hline
${\bf Z_2^{'}}$ & $+$ & $+$ & $+$ & $+$ & $+$ & $+$ \\ \hline
\end{tabular}
\label{Tab:QuarkAssign}
\caption{Quark Group Assignments}
\end{center}
\end{table}

The leptons of Eq.~(\ref{LepAssign}) are retained unchanged, even under \gls{Z2}${\bf ^{'}}$, again keeping all the previous successes in Ch.~\ref{EFM1}. Inspired by Ref.~\citen{Hirsch:2010ru,Boucenna:2011tj}, we have incorporated an additional triplet of right-handed neutrinos, $N_T$. This triplet is odd under \gls{Z2}${\bf ^{'}}$ and is below summarized with the other lepton assignments.

\begin{table}[h]
\renewcommand{\arraystretch}{1.75}
\begin{center}
\begin{tabular}{c||c|c|c|c|c|c|c|c|}
Leptons & $L_{\rm L}$ & $\tau_{\rm R}$ & $\mu_{\rm R}$ & $e_{\rm R}$ & 
$N_{\rm R}^{(1)}$ & $N_{\rm R}^{(2)}$ & $N_{\rm R}^{(3)}$ & $N_T$ \\ \hline \hline
${\bf T^{'}}$ & $3$ & $1_1$ & $1_2$ & $1_3$ & $1_1$ & $1_2$ & $1_3$ & $3$ \\ \hline
${\bf Z_2}$ & $+$ & $-$ & $-$ & $-$ & $+$ & $+$ & $+$ & $+$ \\ \hline
${\bf Z_2^{'}}$ & $+$ & $+$ & $+$ & $+$ & $+$ & $+$ & $+$ & $-$ \\ \hline
\end{tabular}
\label{Tab:LeptonAssign}
\caption{Lepton Group Assignments}
\end{center}
\end{table}

The Higgs sector is also mostly the same as in Sec.~\ref{Original Model}, being \gls{Z2}${\bf ^{'}}$-even, with an added \gls{Z2}$^{'}$-odd, \gls{T'}-triplet, $H_3^{''}$. The five Higgs \glspl{irrep} of \gls{T'} are shown in the following table. Note that this makes for a total of 11 doublets under the gauge group ${\bf SU(2)_{\rm L}}$, one of which may serve as the \gls{MSM} Higgs.\cite{Frampton:2010uw} 

\begin{table}[h]
\renewcommand{\arraystretch}{1.75}
\begin{center}
\begin{tabular}{c||c|c|c|c|c|}
Higgs & $H_{1_1}$ & $H_{1_3}$ & $H_3$ & $H_3^{'}$ & $H_3^{''}$ \\ \hline \hline
${\bf T^{'}}$ & $1_1$ & $1_3$ & $3$ & $3$ & $3$ \\ \hline
${\bf Z_2}$ & $+$ & $-$ & $+$ & $-$ & $+$ \\ \hline
${\bf Z_2^{'}}$ & $+$ & $+$ & $+$ & $+$ & $-$ \\ \hline
\end{tabular}
\label{Tab:HiggsAssign}
\caption{Higgs Group Assignments}
\end{center}
\end{table}

The resultant Lagrangian and Yukawa couplings are:
\begin{align}
{\cal{L}}_{\rm Y} = \frac{1}{2} M_0 N_T N_T + \frac{1}{2} M_1 N_{\rm R}^{(1)} N_{\rm R}^{(1)} +
 M_{23} N_{\rm R}^{(2)} N_{\rm R}^{(3)} + \nonumber \\
Y_{e} L_{\rm L} e_{\rm R} H_3^{'} + Y_{\mu} L_{\rm L} \mu_{\rm R} H_3^{'} + 
Y_{\tau} L_{\rm L} \tau_{\rm R} H_3^{'} + \nonumber \\
Y_1 L_{\rm L} N_{\rm R}^{(1)} H_3 + Y_2 L_{\rm L} N_{\rm R}^{(2)} H_3 + 
Y_3 L_{\rm L} N_{\rm R}^{(3)} H_3 + \nonumber \\
Y_4 L_{\rm L} (N_T H_3^{''})_3 + Y_5 L_{\rm L} (N_T H_3^{''})_{3^{'}} + \nonumber \\
Y_t ({\cal{Q}}_{\rm L} t_{\rm R} H_{1_1}) + Y_b ({\cal{Q}}_{\rm L} b_{\rm R} H_{1_3}) + \nonumber \\
Y_{\cal C} (Q_{\rm L} {\cal{C}}_{\rm R} H^{'}_{3}) + Y_{\cal S} (Q_{\rm L} {\cal{S}}_{\rm R} H_{3}) + h.c. \ .
\end{align}

It is interesting to note that the terms containing the new right-handed neutrino triplet $N_T$, and new Higgs $H_3^{''}$, result in a multiplication of ($3 \times 3 \times 3$) under \gls{T'}. Surprisingly, this results in only two ($1_1$) singlets,\cite{TW} producing two additional Yukawa couplings, $Y_4$ and $Y_5$. This will prove important to our implementation of the Type-I seesaw mechanism. Intriguingly, should these new Yukawa couplings prove complex, they can naturally lead to leptogenesis.\footnote{It is notable that one decay mode of the triplet $N_T$ is into a light neutrino and dark matter.}

\subsection{Generalized Seesaw Mechanism}
At this point, we can summarize the \glspl{VEV} of our model's Higgs as follows,
\begin{equation}
\begin{array}{c}
<H_3>~=~(V_1,V_2,V_3)~,\\
<H_3^{'}>~=~(\frac{m_{\tau}}{Y_{\tau}},\frac{m_{\mu}}{Y_{\mu}},\frac{m_{e}}{Y_{e}})~,\\
<H_3^{''}>~=~(0,0,0)~,\\
<H_{1_{1}}>=(\frac{m_t}{Y_t}),~<H_{1_{3}}>=(\frac{m_b}{Y_b})~.
\label{VEVs}
\end{array}
\end{equation}

$<H_3^{'}>$ is tied to the charged lepton masses and remains disconnected from the neutrinos assuming the charged leptons are mass eigenstates. $<H_3^{''}>$ must have at least one component without a \gls{VEV} in order to create stable dark matter, but must also have 3 identical values in order for \gls{Z2}$^{'}$ to commute with (\gls{T'}$\times$\gls{Z2}), hence three zeroes. $<H_3>$ remains in general form to be specified using the seesaw mechanism. 

As seen in Sec.~\ref{Original Model}, we can use the \gls{TBM} form to generate a symmetry,
\begin{equation}
\begin{array}{c}
 M_{diag}=U_{\rm TBM} M_{\nu} U_{\rm TBM}^T~,\\ \\
 M_{\nu}=U_{\rm TBM}^T
\begin{pmatrix}
m_1 & 0 & 0 \\
0 & m_2 & 0 \\
0 & 0 & m_3
\end{pmatrix}
U_{\rm TBM}~,\\ \\
M_{\nu}=
\begin{pmatrix}
A ~~~ & B & ~~~ C \\
B & A & ~~~ C \\
C & C & ~~~ A+B-C
\end{pmatrix}.
\end{array}
\end{equation}

Next we will implement a generalized Type-I Seesaw Mechanism (the $(3,6)$ form, defined by 3 families and 6 ${\bf SU(2)}$ singlet fields),\cite{Schechter:1980gr} first noting the key equation in Ref.~\citen{Minkowski:1977sc}, given earlier in Eq.~(\ref{seesaw1}), which shows another way to determine $M_{\nu}$,
\begin{equation}
M_{\nu} = M_{\nu}^{\rm D} M_{\rm R}^{-1} (M_{\nu}^{\rm D})^{T}.
\label{eqn:neumass}
\end{equation}

The Dirac and Majorana mass matrices below are based on a generalized form of those used in Ref.~\citen{Frampton:2008ci}. Due to the 6 right-handed neutrino states, the Majorana matrix enlarges to $6\times6$, while the Dirac matrix becomes $3\times6$. The zero elements of the Dirac mass matrix are determined by the \glspl{VEV} of $H_3^{''}$.
\begin{equation}
\setlength{\arraycolsep}{4.5pt}
M_{\nu}^{\rm D}=
\begin{pmatrix}
0 & 0 & 0 & Y_1 V_1 & Y_2 V_3 & Y_3 V_2 \\
0 & 0 & 0 & Y_1 V_3 & Y_2 V_2 & Y_3 V_1 \\
0 & 0 & 0 & Y_1 V_2 & Y_2 V_1 & Y_3 V_3 
\end{pmatrix} 
,~M_{\rm R}=
\begin{pmatrix}
M_0 & 0 & 0 & 0 & 0 & 0 \\
0 & M_0 & 0 & 0 & 0 & 0 \\
0 & 0 & M_0 & 0 & 0 & 0 \\
0 & 0 & 0 & M_1 & 0 & 0 \\
0 & 0 & 0 & 0 & 0 & M_{23} \\
0 & 0 & 0 & 0 & M_{23} & 0 
\end{pmatrix}.
\end{equation}
These alterations to the seesaw mechanism will result in the following version of Eq.~(\ref{masses}),
\begin{equation}
\begin{array}{l}
m_1=A+B-2C=-9 x_{23}~, \\
m_2=A+B+C=0~, \\
m_3=A-B=6 x_1 + 3 x_{23}~.
\end{array}
\end{equation}
As these mass equation remain essentially unchanged, they show that the addition of a neutrino triplet to the \gls{MRTM} does not change the results of the seesaw mechanism and preserves the predictions of Chs.~\ref{intro}, \ref{EFM1}, \ref{EFM2}. Consequently, the \glspl{VEV} of $H_3$ will revert to the form, $<H_3>=V(1,-2,1)$.

\section{\texorpdfstring{\glsentrytext{T'} Dark Matter Predictions}{T' Dark Matter Predictions}}
The \gls{T'} \gls{WIMP} candidate is the lightest state with an assignment of \gls{Z2}${\bf ^{'}} = -1$. The \gls{Z2}${\bf ^{'}}$ odd states being $N_T$ and $H_3^{''}$. The neutrino triplet $N_T$, in particular, is expected to be very heavy from the seesaw mechanism discussed in the Sec.~\ref{DMT' Model}. It decays into an $H_3^{''}$ and a lepton, making it a good candidate for the leptogenesis mechanism.\cite{Fukugita:1986hr} 

The \gls{WIMP} candidate is therefore a superposition of the \gls{CP}-even neutral scalars contained in $H^{''}_3$, which has three ${\bf SU(2)_{\rm L}}$ doublets:
\begin{equation}
H_3^{''}(1)=
\begin{pmatrix}
h_1^+ \\
h_1^0 + i A_1 
\end{pmatrix} 
,~H_3^{''}(2)=
\begin{pmatrix}
h_2^+ \\
h_2^0 + i A_2 
\end{pmatrix}
,~H_3^{''}(3)=
\begin{pmatrix}
h_3^+ \\
h_3^0 + i A_3 
\end{pmatrix}.
\end{equation}
This set includes 6 charged scalars, 3 neutral \gls{CP}-even scalars, and 3 neutral \gls{CP}-odd scalars. Our dark matter candidate will be a superposition of the three real \gls{Z2}${\bf ^{'}}$-odd, \gls{CP}-even, neutral scalar states:
\begin{equation}
\Phi_{WIMP} =\alpha h_{1}^{0}+\beta h_{2}^{0}+\gamma h_{3}^{0}~.
\label{WIMPbits}
\end{equation}
An evaluation of the dark matter candidate coefficients, $\alpha$, $\beta$, and $\gamma$, requires knowledge of the coefficients in the Higgs scalar potential, shown in Appendix~\ref{sec:portent}, and is beyond the scope of this discussion.

\subsection{\texorpdfstring{Relic Density and \glsentrytext{WIMP} Mass}{Predictions}}
One of the most significant properties of a proposed particle is its mass, and by following the treatment laid out in Ref.~\citen{Kolb} we can use the measured relic abundance to determine this property, $M_{\Phi}$.

Beginning with the standard definition of dark matter abundance $\Omega_c=\rho/\rho_{{\rm cr}}$, and the standard assumptions that our particle is a cold relic (that freeze-out will occur when the particle is no longer relativistic), and that we will be focusing on the dominant s-wave coannihilation into \gls{MSM} \gls{QED} vector bosons, we can state
\begin{equation}
\begin{array}{ccccc}
\Omega_c
=
\dfrac{M_{\Phi} s_0 Y_{\infty}}{\rho_{{\rm cr},0}}
&
{\rm where}
&
\rho_{{\rm cr},0}=\dfrac{3 H_0^2}{8 \pi G}~,
&
s_0=\dfrac{2\pi^2}{45} g_{\star0} T_0^3~.
\end{array}
\end{equation}
$g_{\star}$ is a count of the number of relativistic degrees of freedom and is a common part of cosmological statistics. We have included a detailed discussion of the calculation to retrieve this factor in Appendix~\ref{sec:DoF}, but for our purposes $g_{\star}=119.375$ and $g_{\star0}=65/22\approx2.95$. After plugging in and dividing both sides by a scale factor of $\chi=100~{\rm km/s/Mpc}$ (typical in reporting of astronomical results) we find,
\begin{equation}
\Omega_c h^2=\dfrac{2 G g_{\star0}}{5 \chi^2} \biggl(\dfrac{2 \pi T_0}{3}\biggr)^3 M_{\Phi} Y_{\infty}~,
\end{equation}
which includes the cosmic microwave background temperature $T_0=2.726^{\circ}~{\rm K}$, and the gravitational constant $G=6.671\times10^{-39}~{\rm GeV}^{-2}$.\cite{Beringer:1900zz}

Next we need to determine $Y_{\infty}$ which can be approximated as 
\begin{equation}
\begin{array}{cccc}
Y_{\infty}\backsim \dfrac{H(M_{\Phi}) x_f}{s \langle\sigma_A |v|\rangle x}\bigg\rvert_{x=1}
&
{\rm where}
&
s=\dfrac{2\pi^2}{45} g_{\star} T^3~,
&
H(M_{\Phi})=H(x)\frac{M_{\Phi}^2}{T^2}~.
\end{array}
\end{equation}
In this notation $x\equiv M_{\Phi}/T$, with $x_f$ being defined at the freeze-out temperature ($x_f\gsim3$ for a cold relic). Next, $H(x)$ can be obtained from the 2nd Friedman Equation by assuming a flat universe:
\begin{equation}
\begin{array}{ccc}
H(x)=\sqrt{\dfrac{8 \pi G}{3} \rho}
&
{\rm where}
&
\rho=\dfrac{\pi^2}{30} g_{\star} T^4~.
\end{array}
\end{equation}
Combining these equations leads to
\begin{equation}
\begin{array}{ccc}
Y_{\infty}=\sqrt{\dfrac{45 G}{\pi g_{\star}}} \dfrac{x_f}{M_{\Phi} \langle \sigma_A |v|\rangle}
&
{\rm and}
&
\Omega_c h^2=\biggl(\dfrac{16 \pi^{\frac{5}{2}}}{9 \sqrt{5}}\biggr) \biggl(\dfrac{g_{\star0}}{\sqrt{g_{\star}}}\biggr) \biggl(\dfrac{T_0^3 G^{\frac{3}{2}} x_f}{\chi^2 \langle \sigma_A |v|\rangle}\biggr)~.
\end{array}
\label{tech1}
\end{equation}

In Ref.~\citen{Kolb} an approximation for $x_f$ is established (accurate within 5\% for any cold relic value) 
\begin{equation}
x_f=\ln{\bigg[\sqrt{\dfrac{45 g_{\star}}{32 G}} \dfrac{M_{\Phi} \langle \sigma_A |v|\rangle}{\pi^3}\bigg]}-\dfrac{3}{2}\ln{\bigg[\ln{\bigg\lbrace\sqrt{\dfrac{45 g_{\star}}{32 G}} \dfrac{M_{\Phi} \langle \sigma_A |v|\rangle}{\pi^3}\bigg\rbrace}\bigg]}~,
\label{tech2}
\end{equation}

For the annihilation cross-section, we will turn to Ref.~\citen{Cirelli:2005uq}, which lists a general form that we can customize. Recognizing that our dark matter candidate is a real scalar, inhabits an $SU(2)$ doublet, and, like the \gls{MSM} Higgs, has a hypercharge of $Y=1/2$ we see that,
\begin{equation}
\langle\sigma_A |v|\rangle\simeq\frac{3g^4 + (g^{'})^4 + 6g^2(g^{'})^2 + 
4\lambda^2}{128\pi M_{\Phi}^2}~,
\label{tech3}
\end{equation}
where $g$ and $g^{'}$ are the gauge coupling constants, defined as $g=\sqrt{4 \pi \alpha}/\sin{\theta_W}$ and $g^{'}=g\tan{\theta_W}$, respectively. Rather than solve the Higgs scalar potential (detailed in Appendix~\ref{sec:portent}), we make the assumption that the quartic coupling constant, $\lambda$, yields a very small contribution.

Now that all the pieces are in place, we can note that $\sin{\theta_W}^2=0.2231$ for the on-shell scheme,\cite{Beringer:1900zz} and recent data from Ref.~\citen{Ade:2013xsa} sets $\Omega_c h^2=0.11805$. This leads to a calculation of $M_{\Phi}\approx1.84~{\rm TeV}$

\subsection{Dark Matter Detection}

\begin{figure}[!ht]
\begin{center}
\includegraphics[height=90mm]{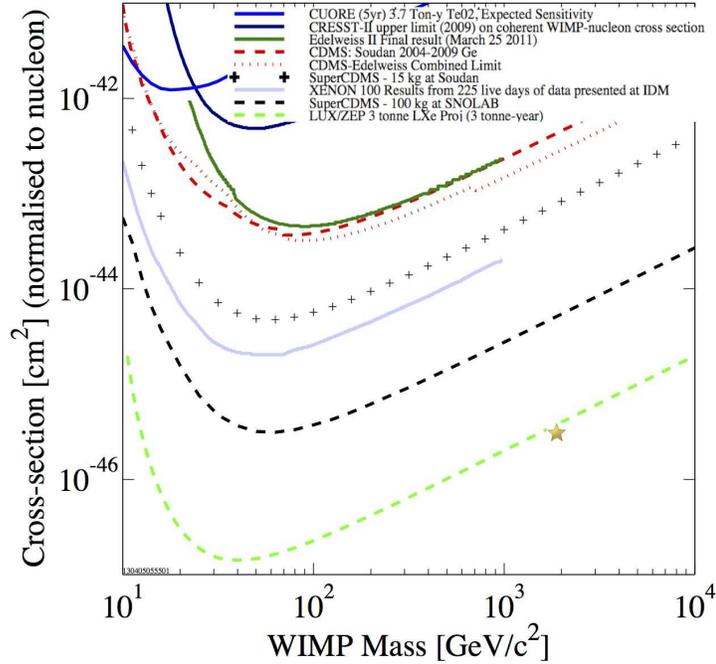}
\caption[WIMP cross-section vs. mass experimental limits]
{\normalsize
Generated by Ref.~\citen{DMsite}, this figure details the current and projected limits on \gls{WIMP} dark matter masses and cross-section. We have indicated our rough prediction for the \gls{T'} dark matter candidate with a star. While this analysis is not geared specifically towards our candidate, being designed for a \gls{WIMP} arising out of \gls{SUSY}, these limits should still be roughly applicable.}
\label{DMfig}
\end{center}
\end{figure}

A thorough discussion of the techniques and evidence for dark matter detection could fill volumes, so we will opt here for only a brief and superficial analysis. But it remains the case that currently there is little to no evidence for dark matter beyond the astrophysical evidence resulting in its discovery and confirmation.

Dark Matter detection usually falls into two categories: direct and indirect detection. Direct Detection would be any method of interacting with dark matter itself and includes searches from the LHC at CERN, as well as nuclear recoil experiments deep underground. Indirect Detectors search for signs of dark matter decay or annihilation. Most recently the PAMELA experiment and results from the Alpha Magnetic Spectrometer generated a great amount of excitement after announcing an excess of high-energy positrons,\cite{Cholis:2008qq,Aguilar:2013qda} leading some to suggest that they had seen dark matter decay products. While the cause of the anomaly remains uncertain, and could simply be a local astrophysical source, it remains a promising sign.

In addressing direct detection, we can combine Eqs.~(\ref{tech1}, \ref{tech2}, \ref{tech3}) to get a first order estimate for the coupling $\lambda$. Then using the derivation from Ref.~\citen{Hirsch:2010ru}, 
\begin{equation}
\sigma_{el}({\rm nucleon})\approx\lambda^2 \times \biggl( \dfrac{100 {\rm GeV}}{M_H} \biggr)^4 \times \biggl( \dfrac{50 {\rm GeV}}{M_{\Phi}} \biggr)^2 \times \biggl(5\times10^{-42} {\rm cm}^2\biggr)~,
\end{equation}
with the most recent measured value of the Higgs mass $M_H \approx 125~{\rm GeV}$, we can develop obtain an order-of-magnitude estimate for the dark matter-nucleon cross-section for our model of $\backsim3\times 10^{-46}~{\rm cm}^2$.

As Fig.~\ref{DMfig} shows, our prediction remains below the most stringent limits placed by current (or proposed) detectors. That said, each successive generation of detector has pushed dark matter cross-section limits further down, with the current best limits found at $20\mhyphen30~{\rm GeV}$. We hope that in the coming years detectors improve to the point where they will find our proposed \gls{WIMP}.

\chapter{Conclusions}
\section{Further Tests and Future Experiments}
\noindent
{\it This Section is largely based on data found in the work of Ref.~\citen{nuFuture}}

Part of the reason for the rapid advances in understanding of neutrinos over the past decade has been the continually growing number of neutrino experiments. In this section we will try and mention key current and future experiments and what aspects of our model they affect. 

Most of the recent excitement in neutrino physics has been over the rapid experimental precision in measurements of $\theta_{13}$. These measurements primarily come from the Daya Bay, Double Chooz, and RENO collaborations. While any of these experiments might receive updates in the future, they currently stand poised to have measured $\sigma(\sin^2(2\theta_{13}))<3\%$ by 2015. In addition, because $\theta_{13}$ is large, they may be able to provide $\Delta m_{31}^2$ measurements.

Next are $\theta_{23}$ and $\Delta m_{32}^2$. These parameters are currently being studied by the Super-Kamiokande and IceCube experiments, with Minos+ set to join in 2013. Looking to the future, the proposed INO and PINGU experiments may join this search near the end of the decade. As these are key to our predictions, we will be closely observing any new results.

The mass hierarchy question remains an unanswered for the time being. NO$\nu$A will begin taking data in late 2013 and may soon have some relevant data on this search. If, by the end of the decade, this factor remains unsettled, Daya Bay II, INO, or the LBNE may be able to make a final determination

The \gls{CP}-violating phase, $\delta_{{\rm CP}}$, has long been a mystery of the \gls{PMNS} matrix, as well as having one of the poorest constraints of the 28 \gls{MSM} parameters. Though simply assuming a value of $0$ simplifies calculations, it may yet have a non-zero value. Experiments have shown indications of what that value might be, but as their $1\sigma$ spreads always contain the entire region $0\mhyphen2\pi$, it remains a secondary concern. Currently T2K and NO$\nu$A are attempting to measure $\delta_{{\rm CP}}$, with Hyper-K (an update of Super-Kamiokande) and the LBNE to take the baton at the end of the decade.

\section{Outstanding Questions and Future Directions}
As we conclude, in the interest of candor, we note what work remains to be done and what limitations our model continues to face. 

Perhaps we should first mention the scope of our model. We have never claimed that the $T^{'}$ model in its current form can act as a grand unified theory. As such we seek to use it to better understand the mixing and masses of the leptons. We have made some movement toward generating a true unified theory in Ch.~\ref{Quart}, by demonstrating that such a model is possible in a Quiver Theory, but these are simply first steps. Ideally, the completed flavor symmetry should be reconcilable with the Lie Groups that make up the \gls{MSM} and more holistic theories.

Another outstanding issue remains a satisfying inclusion of the full \gls{CKM} matrix. Thus far we have shown that the ($2\times2$) Cabibbo matrix offers increased certainty but decreased utility, whereas our attempts to examine the full \gls{CKM} matrix are stymied by the added complexity. As fermion mixing is so integral to our model, this lack of coherence may be creating a number of problems. As mentioned earlier, our model produces an inaccurate value for $(m_d^2/m_s^2)$ and only the vaguest checks against the \gls{CKM} element values. To a degree this is due to by the lack of higher order corrections to the model, and by having multiple choices in forming the \gls{NMRTM} from Ch.~\ref{EFM2}.

Another issue remains the solar mass splitting. While experiments have clearly shown that the neutrino mass eigenstates, $m_1$ and $m_2$, have a separation of roughly $\Delta m_{21}^2=7.5\times 10^-5~{\rm eV}$, our models, thus far, have maintained these mass states are equal. As with the other listed issues, there are a number of potential perturbations that might be introduced to compensate for this initial assumption, but we are left with the dual problems of altering the neutrino properties while maintaining the symmetry and properly motivating this change without simply fitting to the data.

One last area of uncertainty is that of the assumptions we have made about neutrino properties, namely the \gls{NH}, Majorana behavior and others. Many models have a point of rigidity, a place where strong assumptions have been made that cannot be modified or altered without undoing the theory. In the past few years, a common tripping point for other theories has been $\theta_{13}\neq 0$. Even the original Valencia work was inflexible on this point (unlike our variation in Ch.~\ref{ch:T'DM}).\cite{Hirsch:2010ru} For our model, that point of rigidity may indeed be our assumption of the normal hierarchy. While an inverted hierarchy would in no way undermine the potential of our $T^{'}$ theory, our choice of \glspl{VEV} and our use of shared Higgs between leptons and quarks do not leave much room for alteration. At this point there remains no preferred hierarchy, only time, and experimental results, will tell if this assumption is borne out. 

On the same note, we should also mention our model's inability to predict the individual neutrino masses. Although only upper bounds for these masses exist, it would be preferable to suggest a value for experiments to reach for. In addition, we have always assumed that $\delta_{{\rm CP}}=0$. While this was done largely in the interest of simplifying the algebra to a solvable degree, it may not be the case. If this proves true we will need to accommodate this, and deal headlong with that more complicated reality.

\section{Summation}
Having now explained how our model works, both its capabilities and limitations, we should make some comments about our work in the context of the greater physics community. The work we have presented here is admittedly flawed, at times incomplete, and overly vague. And yet the number of completely verified and unanimously accepted theories in particle physics can likely be counted at less than a dozen for a generation. The real question to ask is: does this investigation advance our collective understanding of either the mathematical principles underlying our models, or the physical systems we are trying to describe. To both, we would answer: yes.

We have managed to use finite group symmetries to successfully explain a number of features of the quark and lepton mixing, we would hope that this model, and those like it, would prove the benefits to incorporating these ideas into future attempts at unification. In addition, the model as stated has proven remarkably resilient. We managed to predict that $\theta_{13}$ would prove to be nonzero, and that $\theta_{23}$ would be non-maximal and correlated together. While our model has limitations and may prove incorrect in the long run, the accuracy we have seen so far is immensely exciting. We would hope that it might prove the spark for new discoveries as we pass into an era of \gls{BSM} physics.

\appendix
\appendixpage
\noappendicestocpagenum
\chapter{The Higgs Scalar Potential}
\label{sec:portent}
\noindent
{\it This Appendix is adapted from Appendix A in Ref.~\citen{Eby:2011qa}}

As stated in Ch.~\ref{ch:T'DM}, below is the Higgs scalar potential up to quartic order, consisting of 218 terms and 77 hermitian conjugates. We will use $1_{1,2,3}$, to represent the three singlet representations of $T^{'}$; additionally $3_1$ and $3_2$ will be used to distinguish the two triplet products of two contracted $T^{'}$ triplets.

We have studied assiduously the set of equations $\partial V/ \partial v_i$, where the $v_i$ are the \glspl{VEV}, and the related requirements for a local minimum of positive Hessian eigenvalues. We find, after careful calculation, that the \glspl{VEV} in Eq.~(\ref{VEVs}) are allowed without fine-tuning.

Without further assumptions, one cannot determine the superposition coefficients $\alpha$, $\beta$, and $\gamma$ from Eq.~(\ref{WIMPbits}). It may be fruitful to seek an additional assumption to increase our model's predictivity. For the dedicated reader who wishes to pursue this interesting question, we provide the complete Higgs potential below.

\begin{center}
\begin{gather*}
V=\mu_{H_{1_{1}}}^2 H_{1_1}^{\dagger} H_{1_1}+\mu_{H_{1_{3}}}^2 H_{1_3}^{\dagger} H_{1_3}+
\mu_{H_{3}}^2 H_{3}^{\dagger} H_{3} \\+
\mu_{H_{3}^{'}}^2 H_{3}^{' \dagger} H_{3}^{'}+\mu_{H_{3}^{''}}^2 H_{3}^{'' \dagger} H_{3}^{''}  +
\lambda_{1} [H_{1_{1}}^{\dagger} H_{1_{1}}]_{1_{1}}^2 \\+
\lambda_{2} [H_{1_{3}}^{\dagger} H_{1_{3}}]_{1_{1}}^2+
\lambda_{3} [H_{1_{1}}^{\dagger} H_{1_{1}}]_{1_{1}} [H_{1_{3}}^{\dagger} H_{1_{3}}]_{1_{1}} \\+
\lambda_{4} [H_{1_{1}}^{\dagger} H_{1_{3}}^{\dagger}]_{1_{2}} [H_{1_{1}} H_{1_{3}}]_{1_{3}}+
\lambda_{5} [H_{1_{3}}^{\dagger} H_{1_{3}}^{\dagger}]_{1_{3}} [H_{1_{3}} H_{1_{3}}]_{1_{2}} \\+
\lambda_{6} [H_{1_{1}}^{\dagger} H_{1_{1}}]_{1_{1}} [H_{3}^{\dagger} H_{3}]_{1_{1}} +
\lambda_{7} [H_{1_{1}}^{\dagger} H_{1_{1}}]_{1_{1}} [H_{3}^{' \dagger} H_{3}^{'}]_{1_{1}} +
\lambda_{8} [H_{1_{1}}^{\dagger} H_{1_{1}}]_{1_{1}} [H_{3}^{'' \dagger} H_{3}^{''}]_{1_{1}} \\+
\lambda_{9} [H_{1_{3}}^{\dagger} H_{1_{3}}]_{1_{1}} [H_{3}^{\dagger} H_{3}]_{1_{1}} +
\lambda_{10} [H_{1_{3}}^{\dagger} H_{1_{3}}]_{1_{1}} [H_{3}^{' \dagger} H_{3}^{'}]_{1_{1}} +
\lambda_{11} [H_{1_{3}}^{\dagger} H_{1_{3}}]_{1_{1}} [H_{3}^{'' \dagger} H_{3}^{''}]_{1_{1}} \\+
\lambda_{12} ([H_{1_{1}}^{\dagger} H_{1_{1}}^{\dagger}]_{1_{1}} [H_{3} H_{3}]_{1_{1}} +h.c.)+
\lambda_{13} ([H_{1_{1}}^{\dagger} H_{1_{1}}^{\dagger}]_{1_{1}} [H_{3}^{'} H_{3}^{'}]_{1_{1}} +h.c.) \\+
\lambda_{14} ([H_{1_{1}}^{\dagger} H_{1_{1}}^{\dagger}]_{1_{1}} [H_{3}^{''} H_{3}^{''}]_{1_{1}} +h.c.) \\+
\lambda_{15} ([H_{1_{3}}^{\dagger} H_{1_{3}}^{\dagger}]_{1_{3}} [H_{3} H_{3}]_{1_{2}} +h.c.)+
\lambda_{16} ([H_{1_{3}}^{\dagger} H_{1_{3}}^{\dagger}]_{1_{3}} [H_{3}^{'} H_{3}^{'}]_{1_{2}} +h.c.) \\+
\lambda_{17} ([H_{1_{3}}^{\dagger} H_{1_{3}}^{\dagger}]_{1_{3}} [H_{3}^{''} H_{3}^{''}]_{1_{2}} +h.c.) \\+
\lambda_{18} [H_{1_{1}}^{\dagger} H_{3}]_{3} [H_{3}^{\dagger} H_{1_{1}}]_{3} +
\lambda_{19} [H_{1_{1}}^{\dagger} H_{3}^{'}]_{3} [H_{3}^{' \dagger} H_{1_{1}}]_{3} +
\lambda_{20} [H_{1_{1}}^{\dagger} H_{3}^{''}]_{3} [H_{3}^{'' \dagger} H_{1_{1}}]_{3} \\+
\lambda_{21} [H_{1_{3}}^{\dagger} H_{3}]_{3} [H_{3}^{\dagger} H_{1_{3}}]_{3} +
\lambda_{22} [H_{1_{3}}^{\dagger} H_{3}^{'}]_{3} [H_{3}^{' \dagger} H_{1_{3}}]_{3} +
\lambda_{23} [H_{1_{3}}^{\dagger} H_{3}^{''}]_{3} [H_{3}^{'' \dagger} H_{1_{3}}]_{3} \\+
\lambda_{24} ([H_{1_{3}}^{\dagger} H_{3}^{'}]_{3} [H_{3}^{\dagger} H_{1_{1}}]_{3} +h.c.) \\+
\lambda_{25} ([H_{1_{1}}^{\dagger} H_{3}]_{3} [H_{3}^{\dagger} H_{3}]_{3_{1}} +h.c.) +
\lambda_{26} ([H_{1_{1}}^{\dagger} H_{3}^{\dagger}]_{3} [H_{3} H_{3}]_{3_{1}} +h.c.) \\+
\lambda_{27} ([H_{1_{1}}^{\dagger} H_{3}]_{3} [H_{3}^{\dagger} H_{3}]_{3_{2}} +h.c.) +
\lambda_{28} ([H_{1_{1}}^{\dagger} H_{3}^{\dagger}]_{3} [H_{3} H_{3}]_{3_{2}} +h.c.) \\+
\lambda_{29} ([H_{1_{1}}^{\dagger} H_{3}]_{3} [H_{3}^{' \dagger} H_{3}^{'}]_{3_{1}} +h.c.)+
\lambda_{30} ([H_{1_{1}}^{\dagger} H_{3}^{'}]_{3} [H_{3}^{\dagger} H_{3}^{'}]_{3_{1}} +h.c.)\\+
\lambda_{31} ([H_{1_{1}}^{\dagger} H_{3}^{\dagger}]_{3} [H_{3}^{'} H_{3}^{'}]_{3_{1}} +h.c.)+
\lambda_{32} ([H_{1_{1}}^{\dagger} H_{3}^{' \dagger}]_{3} [H_{3} H_{3}^{'}]_{3_{1}} +h.c.)\\+
\lambda_{33} ([H_{1_{1}}^{\dagger} H_{3}]_{3} [H_{3}^{' \dagger} H_{3}^{'}]_{3_{2}} +h.c.)+
\lambda_{34} ([H_{1_{1}}^{\dagger} H_{3}^{'}]_{3} [H_{3}^{\dagger} H_{3}^{'}]_{3_{2}} +h.c.)\\+
\lambda_{35} ([H_{1_{1}}^{\dagger} H_{3}^{\dagger}]_{3} [H_{3}^{'} H_{3}^{'}]_{3_{2}} +h.c.)+
\lambda_{36} ([H_{1_{1}}^{\dagger} H_{3}^{' \dagger}]_{3} [H_{3} H_{3}^{'}]_{3_{2}} +h.c.)\\+ 
\lambda_{37} ([H_{1_{1}}^{\dagger} H_{3}]_{3} [H_{3}^{'' \dagger} H_{3}^{''}]_{3_{1}} +h.c.)+
\lambda_{38} ([H_{1_{1}}^{\dagger} H_{3}^{''}]_{3} [H_{3}^{\dagger} H_{3}^{''}]_{3_{1}} +h.c.)\\+
\lambda_{39} ([H_{1_{1}}^{\dagger} H_{3}^{\dagger}]_{3} [H_{3}^{''} H_{3}^{''}]_{3_{1}} +h.c.)+
\lambda_{40} ([H_{1_{1}}^{\dagger} H_{3}^{'' \dagger}]_{3} [H_{3} H_{3}^{''}]_{3_{1}} +h.c.)\\+
\lambda_{41} ([H_{1_{1}}^{\dagger} H_{3}]_{3} [H_{3}^{'' \dagger} H_{3}^{''}]_{3_{2}} +h.c.)+
\lambda_{42} ([H_{1_{1}}^{\dagger} H_{3}^{''}]_{3} [H_{3}^{\dagger} H_{3}^{''}]_{3_{2}} +h.c.)\\+
\lambda_{43} ([H_{1_{1}}^{\dagger} H_{3}^{\dagger}]_{3} [H_{3}^{''} H_{3}^{''}]_{3_{2}} +h.c.)+
\lambda_{44} ([H_{1_{1}}^{\dagger} H_{3}^{'' \dagger}]_{3} [H_{3} H_{3}^{''}]_{3_{2}} +h.c.) \\+
\lambda_{45} ([H_{1_{3}}^{\dagger} H_{3}^{'}]_{3} [H_{3}^{' \dagger} H_{3}^{'}]_{3_{1}} +h.c.) +
\lambda_{46} ([H_{1_{3}}^{\dagger} H_{3}^{' \dagger}]_{3} [H_{3}^{'} H_{3}^{'}]_{3_{1}} +h.c.) \\+
\lambda_{47} ([H_{1_{3}}^{\dagger} H_{3}^{'}]_{3} [H_{3}^{' \dagger} H_{3}^{'}]_{3_{2}} +h.c.) +
\lambda_{48} ([H_{1_{3}}^{\dagger} H_{3}^{' \dagger}]_{3} [H_{3}^{'} H_{3}^{'}]_{3_{2}} +h.c.) \\+
\lambda_{49} ([H_{1_{3}}^{\dagger} H_{3}^{'}]_{3} [H_{3}^{\dagger} H_{3}]_{3_{1}} +h.c.)+
\lambda_{50} ([H_{1_{3}}^{\dagger} H_{3}]_{3} [H_{3}^{' \dagger} H_{3}]_{3_{1}} +h.c.)\\+
\lambda_{51} ([H_{1_{3}}^{\dagger} H_{3}^{' \dagger}]_{3} [H_{3} H_{3}]_{3_{1}} +h.c.) +
\lambda_{52} ([H_{1_{3}}^{\dagger} H_{3}^{\dagger}]_{3} [H_{3}^{'} H_{3}]_{3_{1}} +h.c.) \\+
\lambda_{53} ([H_{1_{3}}^{\dagger} H_{3}^{'}]_{3} [H_{3}^{\dagger} H_{3}]_{3_{2}} +h.c.) +
\lambda_{54} ([H_{1_{3}}^{\dagger} H_{3}]_{3} [H_{3}^{' \dagger} H_{3}]_{3_{2}} +h.c.) \\+
\lambda_{55} ([H_{1_{3}}^{\dagger} H_{3}^{' \dagger}]_{3} [H_{3} H_{3}]_{3_{2}} +h.c.) +
\lambda_{56} ([H_{1_{3}}^{\dagger} H_{3}^{\dagger}]_{3} [H_{3}^{'} H_{3}]_{3_{2}} +h.c.) \\+
\lambda_{57} ([H_{1_{3}}^{\dagger} H_{3}^{'}]_{3} [H_{3}^{'' \dagger} H_{3}^{''}]_{3_{1}} +h.c.) +
\lambda_{58} ([H_{1_{3}}^{\dagger} H_{3}^{''}]_{3} [H_{3}^{' \dagger} H_{3}^{''}]_{3_{1}} +h.c.) \\+
\lambda_{59} ([H_{1_{3}}^{\dagger} H_{3}^{' \dagger}]_{3} [H_{3}^{''} H_{3}^{''}]_{3_{1}} +h.c.) +
\lambda_{60} ([H_{1_{3}}^{\dagger} H_{3}^{'' \dagger}]_{3} [H_{3}^{'} H_{3}^{''}]_{3_{1}} +h.c.) \\+
\lambda_{61} ([H_{1_{3}}^{\dagger} H_{3}^{'}]_{3} [H_{3}^{'' \dagger} H_{3}^{''}]_{3_{2}} +h.c.) +
\lambda_{62} ([H_{1_{3}}^{\dagger} H_{3}^{''}]_{3} [H_{3}^{' \dagger} H_{3}^{''}]_{3_{2}} +h.c.) \\+
\lambda_{63} ([H_{1_{3}}^{\dagger} H_{3}^{' \dagger}]_{3} [H_{3}^{''} H_{3}^{''}]_{3_{2}} +h.c.) +
\lambda_{64} ([H_{1_{3}}^{\dagger} H_{3}^{'' \dagger}]_{3} [H_{3}^{'} H_{3}^{''}]_{3_{2}} +h.c.) \\+
\lambda_{65} [H_{3}^{\dagger} H_{3}]_{1_{2}} [H_{3}^{\dagger} H_{3}]_{1_{3}} +
\lambda_{66} [H_{3}^{\dagger} H_{3}^{\dagger}]_{1_{2}} [H_{3} H_{3}]_{1_{3}} +
\lambda_{66}^{'} [H_{3}^{\dagger} H_{3}^{\dagger}]_{1_{3}} [H_{3} H_{3}]_{1_{2}} \\+
\lambda_{67} [H_{3}^{\dagger} H_{3}]_{1_1}^2 +
\lambda_{68} [H_{3}^{\dagger} H_{3}^{\dagger}]_{1_{1}} [H_{3} H_{3}]_{1_{1}} +
\lambda_{69} ([H_{3}^{\dagger} H_{3}]_{3_{1}} [H_{3}^{\dagger} H_{3}]_{3_{1}} +h.c.) \\+
\lambda_{70} [H_{3}^{\dagger} H_{3}]_{3_{1}} [H_{3}^{\dagger} H_{3}]_{3_{2}} +
\lambda_{71} [H_{3}^{\dagger} H_{3}^{\dagger}]_{3_{1}} [H_{3} H_{3}]_{3_{2}}  \\+
\lambda_{72} [H_{3}^{' \dagger} H_{3}^{'}]_{1_{2}} [H_{3}^{' \dagger} H_{3}^{'}]_{1_{3}} +
\lambda_{73} [H_{3}^{' \dagger} H_{3}^{' \dagger}]_{1_{2}} [H_{3}^{'} H_{3}^{'}]_{1_{3}} +
\lambda_{73}^{'} [H_{3}^{' \dagger} H_{3}^{' \dagger}]_{1_{3}} [H_{3}^{'} H_{3}^{'}]_{1_{2}} \\+
\lambda_{74} [H_{3}^{' \dagger} H_{3}^{'}]_{1_1}^2 +
\lambda_{75} [H_{3}^{' \dagger} H_{3}^{' \dagger}]_{1_{1}} [H_{3}^{'} H_{3}^{'}]_{1_{1}} +
\lambda_{76} ([H_{3}^{' \dagger} H_{3}^{'}]_{3_{1}} [H_{3}^{' \dagger} H_{3}^{'}]_{3_{1}} +h.c.) \\+
\lambda_{77} [H_{3}^{' \dagger} H_{3}^{'}]_{3_{1}} [H_{3}^{' \dagger} H_{3}^{'}]_{3_{2}} +
\lambda_{78} [H_{3}^{' \dagger} H_{3}^{' \dagger}]_{3_{1}} [H_{3}^{'} H_{3}^{'}]_{3_{2}}  \\+
\lambda_{79} [H_{3}^{'' \dagger} H_{3}^{''}]_{1_{2}} [H_{3}^{'' \dagger} H_{3}^{''}]_{1_{3}} +
\lambda_{80} [H_{3}^{'' \dagger} H_{3}^{'' \dagger}]_{1_{2}} [H_{3}^{''} H_{3}^{''}]_{1_{3}} +
\lambda_{80}^{'} [H_{3}^{'' \dagger} H_{3}^{'' \dagger}]_{1_{3}} [H_{3}^{''} H_{3}^{''}]_{1_{2}} \\+
\lambda_{81} [H_{3}^{'' \dagger} H_{3}^{''}]_{1_1}^2 +
\lambda_{82} [H_{3}^{'' \dagger} H_{3}^{'' \dagger}]_{1_{1}} [H_{3}^{''} H_{3}^{''}]_{1_{1}} +
\lambda_{83} ([H_{3}^{'' \dagger} H_{3}^{''}]_{3_{1}} [H_{3}^{'' \dagger} H_{3}^{''}]_{3_{1}} +h.c.) \\+
\lambda_{84} [H_{3}^{'' \dagger} H_{3}^{''}]_{3_{1}} [H_{3}^{'' \dagger} H_{3}^{''}]_{3_{2}} +
\lambda_{85} [H_{3}^{'' \dagger} H_{3}^{'' \dagger}]_{3_{1}} [H_{3}^{''} H_{3}^{''}]_{3_{2}}\\+
\lambda_{86} [H_{3}^{\dagger} H_{3}]_{1_{1}} [H_{3}^{' \dagger} H_{3}^{'}]_{1_{1}}+
\lambda_{87} ([H_{3}^{\dagger} H_{3}^{'}]_{1_{1}} [H_{3}^{\dagger} H_{3}^{'}]_{1_{1}} +h.c.)\\+
\lambda_{88} [H_{3}^{\dagger} H_{3}^{' \dagger}]_{1_{1}} [H_{3} H_{3}^{'}]_{1_{1}} +
\lambda_{89} ([H_{3}^{\dagger} H_{3}^{\dagger}]_{1_{1}} [H_{3}^{'} H_{3}^{'}]_{1_{1}} +h.c.)\\+
\lambda_{90} [H_{3}^{\dagger} H_{3}]_{1_{2}} [H_{3}^{' \dagger} H_{3}^{'}]_{1_{3}} +
\lambda_{91} ([H_{3}^{\dagger} H_{3}^{'}]_{1_{2}} [H_{3}^{\dagger} H_{3}^{'}]_{1_{3}} +h.c.)\\+
\lambda_{92} [H_{3}^{\dagger} H_{3}^{' \dagger}]_{1_{2}} [H_{3} H_{3}^{'}]_{1_{3}} +
\lambda_{93} ([H_{3}^{\dagger} H_{3}^{\dagger}]_{1_{2}} [H_{3}^{'} H_{3}^{'}]_{1_{3}} +h.c.)\\+
\lambda_{92}^{'} [H_{3}^{\dagger} H_{3}^{' \dagger}]_{1_{3}} [H_{3} H_{3}^{'}]_{1_{2}} +
\lambda_{93}^{'} ([H_{3}^{\dagger} H_{3}^{\dagger}]_{1_{3}} [H_{3}^{'} H_{3}^{'}]_{1_{2}} +h.c.)\\+
\lambda_{94} ([H_{3}^{\dagger} H_{3}]_{3_{1}} [H_{3}^{' \dagger} H_{3}^{'}]_{3_{1}} +h.c.)+
\lambda_{95} ([H_{3}^{\dagger} H_{3}^{'}]_{3_{1}} [H_{3}^{\dagger} H_{3}^{'}]_{3_{1}} +h.c.)\\+
\lambda_{96} [H_{3}^{\dagger} H_{3}]_{3_{1}} [H_{3}^{' \dagger} H_{3}^{'}]_{3_{2}} +
\lambda_{97} ([H_{3}^{\dagger} H_{3}^{'}]_{3_{1}} [H_{3}^{\dagger} H_{3}^{'}]_{3_{2}} +h.c.)\\+
\lambda_{98} [H_{3}^{\dagger} H_{3}^{' \dagger}]_{3_{1}} [H_{3} H_{3}^{'}]_{3_{2}} +
\lambda_{99} ([H_{3}^{\dagger} H_{3}^{\dagger}]_{3_{1}} [H_{3}^{'} H_{3}^{'}]_{3_{2}} +h.c.)\\+
\lambda_{100} [H_{3}^{\dagger} H_{3}]_{1_{1}} [H_{3}^{'' \dagger} H_{3}^{''}]_{1_{1}} +
\lambda_{101} ([H_{3}^{\dagger} H_{3}^{''}]_{1_{1}} [H_{3}^{\dagger} H_{3}^{''}]_{1_{1}} +h.c.)\\+
\lambda_{102} [H_{3}^{\dagger} H_{3}^{'' \dagger}]_{1_{1}} [H_{3} H_{3}^{''}]_{1_{1}} +
\lambda_{103} ([H_{3}^{\dagger} H_{3}^{\dagger}]_{1_{1}} [H_{3}^{''} H_{3}^{''}]_{1_{1}} +h.c.)\\+
\lambda_{104} [H_{3}^{\dagger} H_{3}]_{1_{2}} [H_{3}^{'' \dagger} H_{3}^{''}]_{1_{3}} +
\lambda_{105} ([H_{3}^{\dagger} H_{3}^{''}]_{1_{2}} [H_{3}^{\dagger} H_{3}^{''}]_{1_{3}} +h.c.)\\+
\lambda_{106} [H_{3}^{\dagger} H_{3}^{'' \dagger}]_{1_{2}} [H_{3} H_{3}^{''}]_{1_{3}}+
\lambda_{107} ([H_{3}^{\dagger} H_{3}^{\dagger}]_{1_{2}} [H_{3}^{''} H_{3}^{''}]_{1_{3}} +h.c.)\\+
\lambda_{106}^{'} [H_{3}^{\dagger} H_{3}^{'' \dagger}]_{1_{3}} [H_{3} H_{3}^{''}]_{1_{2}} +
\lambda_{107}^{'} ([H_{3}^{\dagger} H_{3}^{\dagger}]_{1_{3}} [H_{3}^{''} H_{3}^{''}]_{1_{2}} +h.c.)\\+
\lambda_{108} ([H_{3}^{\dagger} H_{3}]_{3_{1}} [H_{3}^{'' \dagger} H_{3}^{''}]_{3_{1}} +h.c.)+
\lambda_{109} ([H_{3}^{\dagger} H_{3}^{''}]_{3_{1}} [H_{3}^{\dagger} H_{3}^{''}]_{3_{1}} +h.c.)\\+
\lambda_{110} [H_{3}^{\dagger} H_{3}]_{3_{1}} [H_{3}^{'' \dagger} H_{3}^{''}]_{3_{2}} +
\lambda_{111} ([H_{3}^{\dagger} H_{3}^{''}]_{3_{1}} [H_{3}^{\dagger} H_{3}^{''}]_{3_{2}} +h.c.)\\+
\lambda_{112} [H_{3}^{\dagger} H_{3}^{'' \dagger}]_{3_{1}} [H_{3} H_{3}^{''}]_{3_{2}}+
\lambda_{113} ([H_{3}^{\dagger} H_{3}^{\dagger}]_{3_{1}} [H_{3}^{''} H_{3}^{''}]_{3_{2}} +h.c.)\\+
\lambda_{114} [H_{3}^{' \dagger} H_{3}^{'}]_{1_{1}} [H_{3}^{'' \dagger} H_{3}^{''}]_{1_{1}} +
\lambda_{115} ([H_{3}^{' \dagger} H_{3}^{''}]_{1_{1}} [H_{3}^{' \dagger} H_{3}^{''}]_{1_{1}} +h.c.)\\+
\lambda_{116} [H_{3}^{' \dagger} H_{3}^{'' \dagger}]_{1_{1}} [H_{3}^{'} H_{3}^{''}]_{1_{1}} +
\lambda_{117} ([H_{3}^{' \dagger} H_{3}^{' \dagger}]_{1_{1}} [H_{3}^{''} H_{3}^{''}]_{1_{1}} +h.c.)\\+
\lambda_{118} [H_{3}^{' \dagger} H_{3}^{'}]_{1_{2}} [H_{3}^{'' \dagger} H_{3}^{''}]_{1_{3}} +
\lambda_{119} ([H_{3}^{' \dagger} H_{3}^{''}]_{1_{2}} [H_{3}^{' \dagger} H_{3}^{''}]_{1_{3}} +h.c.)\\+
\lambda_{120} [H_{3}^{' \dagger} H_{3}^{'' \dagger}]_{1_{2}} [H_{3}^{'} H_{3}^{''}]_{1_{3}} +
\lambda_{121} ([H_{3}^{' \dagger} H_{3}^{' \dagger}]_{1_{2}} [H_{3}^{''} H_{3}^{''}]_{1_{3}} +h.c.)\\+
\lambda_{120}^{'} [H_{3}^{' \dagger} H_{3}^{'' \dagger}]_{1_{3}} [H_{3}^{'} H_{3}^{''}]_{1_{2}} +
\lambda_{121}^{'} ([H_{3}^{' \dagger} H_{3}^{' \dagger}]_{1_{3}} [H_{3}^{''} H_{3}^{''}]_{1_{2}} +h.c.)\\+
\lambda_{122} ([H_{3}^{' \dagger} H_{3}^{'}]_{3_{1}} [H_{3}^{'' \dagger} H_{3}^{''}]_{3_{1}} +h.c.)+
\lambda_{123} ([H_{3}^{' \dagger} H_{3}^{''}]_{3_{1}} [H_{3}^{' \dagger} H_{3}^{''}]_{3_{1}} +h.c.)\\+
\lambda_{124} [H_{3}^{' \dagger} H_{3}^{'}]_{3_{1}} [H_{3}^{'' \dagger} H_{3}^{''}]_{3_{2}} +
\lambda_{125} ([H_{3}^{' \dagger} H_{3}^{''}]_{3_{1}} [H_{3}^{' \dagger} H_{3}^{''}]_{3_{2}} +h.c.)\\+
\lambda_{126} [H_{3}^{' \dagger} H_{3}^{'' \dagger}]_{3_{1}} [H_{3}^{'} H_{3}^{''}]_{3_{2}}+
\lambda_{127} ([H_{3}^{' \dagger} H_{3}^{' \dagger}]_{3_{1}} [H_{3}^{''} H_{3}^{''}]_{3_{2}} +h.c.)
\end{gather*}
\end{center}

\chapter{Counting Relativistic Degrees of Freedom}
\label{sec:DoF}
A notable hurdle in a dark matter relic density calculation is determining the number of relativistic degrees of freedom. This factor, $g_{\star}$, can be calculated using the formula from Ref.~\citen{Kolb}:
\begin{equation}
g_{\star}=\sum_{i={\rm bosons}} g_i \biggl(\dfrac{T_i}{T_{\gamma}}\biggr)^4 + \dfrac{7}{8} \sum_{j={\rm fermions}} g_j \biggl(\dfrac{T_j}{T_{\gamma}}\biggr)^4~,
\end{equation}
where $T_{\gamma}$ is the photon temperature. In the modern universe, with a temperature of approximately $2.7^{\circ}$~K this number is quite small because so few objects fit the requirements, basically the photon and the neutrinos:
\beq
g_{\star 0}=(2_{\rm H}\times 1_{\rm b})_{\gamma_0}+(3_{e,\mu,\tau}\times 1_{\rm M} \times 1_{\rm L} \times \tfrac{7}{8}_{\rm f} \times \tfrac{4}{11}_{\rm E})_{\nu}\approx2.95~,
\eeq
where we have assumed Majorana neutrinos (Dirac neutrinos typically yield a value of $g_{\star 0}=3.91$. In this demonstration, we have signified with subscripts the causes for several contributing factors. The subscripts ${\rm b}$ and ${\rm f}$ indicate boson or fermion, respectively. The subscript ${\rm L}$ indicates the left-handed (or right-handed for subscript ${\rm R}$) helicity state. A bar indicates a factor from antiparticles, while a subscript ${\rm M}$ is for Majorana fermions (who are their own antiparticle). Here, the subscript ${\rm E}$ indicates a contribution from entropy, which only becomes a factor on very recent cosmological scales (long after dark matter freezes out). Two subscripts, unused here, but still important, are ${\rm C}$ for single colored variants and ${\rm Sp}$ from the massive bosons. Below, in Fig.~\ref{DMT':DoF}, we have included a tabulation for all of the potential relativistic degrees of freedom for the model detailed in Ch.~\ref{ch:T'DM} at some arbitrarily high temperature, yielding a value of $g_{\star}=119.375$ (The \gls{MSM} produces a value of $g_{{\rm SM}}=104.125$ for Majorana neutrinos and $g_{{\rm SM}}=106.75$ for Dirac neutrinos). Depending on when dark matter freeze-out occurs this leads to a range of possible values above the \gls{MSM}, up to $15.25$, but to find the extremes of our theory we assume the maximum allowed value.

\begin{table}[ht]
\renewcommand{\arraystretch}{1.75}
\begin{center}
\begin{tabular}{||c|c|c||}
\hline\hline
${\rm Particles}$ & ${\rm Multipliers}$ & ${\rm DoF}$ \\
\hline\hline
$\left( \begin{array}{ccc}
u & c & t \\
d & s & b 
\end{array}\right)$ 
& $\times \bar{2}\times 2_{{\rm L,R}}\times 3_{{\rm C}}\times \tfrac{7}{8}_{{\rm f}}$ & $63$ \\
$\left( \begin{array}{ccc}
e^{-} & \mu^{-} & \tau^{-} 
\end{array}\right)$ 
& $\times \bar{2}\times 2_{{\rm L,R}}\times  \tfrac{7}{8}_{{\rm f}}$ & $10.5$ \\
$\left( \begin{array}{ccc}
\nu_{e} & \nu_{\mu} & \nu_{\tau} 
\end{array}\right)_{{\rm L}}$ 
& $\times 1_{{\rm M}}\times 1_{{\rm L}}\times \tfrac{7}{8}_{{\rm f}}$ & $2.625$ \\
$\left( \begin{array}{cccc}
N^{1} & N^{2} & N^{3} & N_{T}
\end{array}\right)_{{\rm R}}$ 
& $\times 1_{{\rm M}}\times 1_{{\rm R}}\times \tfrac{7}{8}_{{\rm f}}$ & $5.25$ \\
$\left( \begin{array}{ccccc}
H_{1_1} & H_{1_3} & H_{3} & H_{3}^{'} & H_{3}^{''}
\end{array}\right)$ 
& $\times 1_{{\rm Sp}}\times 1_{{\rm b}}$ & $11$ \\
$\left( \begin{array}{ccc}
Z_{0} & W^{+} & W^{-} 
\end{array}\right)$ 
& $\times 3_{{\rm Sp}}\times 1_{{\rm b}}$ & $9$ \\
$\left( \begin{array}{c}
\gamma_{0}
\end{array}\right)$ 
& $\times 2_{{\rm L,R}}\times 1_{{\rm b}}$ & $2$ \\
$\left( \begin{array}{cccc}
g_{r\bar{b}} & g_{r\bar{g}} & g_{b\bar{g}} & g_{r\bar{r}-g\bar{g}}\\
g_{\bar{r}b} & g_{\bar{r}g} & g_{\bar{b}g} & g_{r\bar{r}+g\bar{g}-b\bar{b}}
\end{array}\right)$ 
& $\times 2_{{\rm L,R}}\times 1_{{\rm b}}$ & $16$ \\
\hline\hline
\multicolumn{2}{ |c| }{Total   $g_{\star}$} & $119.375$ \\
\hline\hline
\end{tabular}
\label{DMT':DoF}
\caption[Relativistic Degrees of Freedom]{Potential Relativistic Degrees of Freedom for our dark matter model}
\end{center}
\end{table}



\begin{thebibliography}{100}

\bibitem{Street:1937me} 
  J.~C.~Street and E.~C.~Stevenson,
  Phys.\ Rev.\  {\bf 52}, 1003 (1937).
  
\bibitem{Muons}
  R.~H.~March,
  "Muon, Discovery of", 
  p.321 in {\it Building Blocks of Matter: A Supplement to the Macmillan Encyclopedia of Physics},
  edited by J.~Bagger, J.~Rigden, and R.~Stuewer, Macmillan 2003.

\bibitem{Glashow:1961tr} 
  S.~L.~Glashow,
  Nucl.\ Phys.\  {\bf 22}, 579 (1961).
    
\bibitem{Weinberg:1967tq} 
  S.~Weinberg,
  Phys.\ Rev.\ Lett.\  {\bf 19}, 1264 (1967).
  
\bibitem{Salam:1968rm} 
  A.~Salam,
  Conf.\ Proc.\ C {\bf 680519}, 367 (1968).
  
\bibitem{Arnison:1983rp} 
  G.~Arnison {\it et al.}  [UA1 Collaboration],
  Phys.\ Lett.\ B {\bf 122}, 103 (1983).
    
\bibitem{Banner:1983jy} 
  M.~Banner {\it et al.}  [UA2 Collaboration],
  Phys.\ Lett.\ B {\bf 122}, 476 (1983).
  
\bibitem{Arnison:1983mk} 
  G.~Arnison {\it et al.}  [UA1 Collaboration],
  Phys.\ Lett.\ B {\bf 126}, 398 (1983).
    
\bibitem{Bagnaia:1983zx} 
  P.~Bagnaia {\it et al.}  [UA2 Collaboration],
  Phys.\ Lett.\ B {\bf 129}, 130 (1983).
  
\bibitem{Augustin:1974xw} 
  J.~E.~Augustin {\it et al.}  [SLAC-SP-017 Collaboration],
  Phys.\ Rev.\ Lett.\  {\bf 33}, 1406 (1974).
    
\bibitem{Aubert:1974js} 
  J.~J.~Aubert {\it et al.}  [E598 Collaboration],
  Phys.\ Rev.\ Lett.\  {\bf 33}, 1404 (1974).
  
\bibitem{Herb:1977ek} 
  S.~W.~Herb, D.~C.~Hom, L.~M.~Lederman, J.~C.~Sens, H.~D.~Snyder, J.~K.~Yoh, J.~A.~Appel and B.~C.~Brown {\it et al.},
  Phys.\ Rev.\ Lett.\  {\bf 39}, 252 (1977).
  
\bibitem{Abe:1995hr} 
  F.~Abe {\it et al.}  [CDF Collaboration],
  Phys.\ Rev.\ Lett.\  {\bf 74}, 2626 (1995)
  [hep-ex/9503002].
  
\bibitem{Abachi:1995iq} 
  S.~Abachi {\it et al.}  [D0 Collaboration],
  Phys.\ Rev.\ Lett.\  {\bf 74}, 2632 (1995)
  [hep-ex/9503003].
  
\bibitem{Berger:1978rr} 
  C.~Berger {\it et al.}  [PLUTO Collaboration],
  Phys.\ Lett.\ B {\bf 82}, 449 (1979).
  
\bibitem{Berger:1979cj} 
  C.~Berger {\it et al.}  [PLUTO Collaboration],
  Phys.\ Lett.\ B {\bf 86}, 418 (1979).
    
\bibitem{Barber:1979yr} 
  D.~P.~Barber, U.~Becker, H.~Benda, A.~Boehm, J.~G.~Branson, J.~Bron, D.~Buikman and J.~Burger {\it et al.},
  Phys.\ Rev.\ Lett.\  {\bf 43}, 830 (1979).
  
\bibitem{Brandelik:1979bd} 
  R.~Brandelik {\it et al.}  [TASSO Collaboration],
  Phys.\ Lett.\ B {\bf 86}, 243 (1979).
  
\bibitem{Aad:2012tfa} 
  G.~Aad {\it et al.}  [ATLAS Collaboration],
  Phys.\ Lett.\ B {\bf 716}, 1 (2012)
  [arXiv:1207.7214 [hep-ex]].
    
\bibitem{Chatrchyan:2012ufa} 
  S.~Chatrchyan {\it et al.}  [CMS Collaboration],
  Phys.\ Lett.\ B {\bf 716}, 30 (2012)
  [arXiv:1207.7235 [hep-ex]].
  
\bibitem{Beringer:1900zz} 
  J.~Beringer {\it et al.}  [Particle Data Group Collaboration],
  Phys.\ Rev.\ D {\bf 86}, 010001 (2012).
  
\bibitem{Fukuda:1998fd} 
  Y.~Fukuda {\it et al.}  [Super-Kamiokande Collaboration],
  Phys.\ Rev.\ Lett.\  {\bf 81}, 1158 (1998)
  [Erratum-ibid.\  {\bf 81}, 4279 (1998)]
  [hep-ex/9805021].
  
\bibitem{Rubin:1970zza} 
  V.~C.~Rubin, W.~K.~Ford, Jr. and ,
  Astrophys.\ J.\  {\bf 159}, 379 (1970).
  
\bibitem{Begeman:1991iy} 
  K.~G.~Begeman, A.~H.~Broeils, R.~H.~Sanders and ,
  Mon.\ Not.\ Roy.\ Astron.\ Soc.\  {\bf 249}, 523 (1991).
    
\bibitem{Zwicky:1933gu} 
  F.~Zwicky,
  Helv.\ Phys.\ Acta {\bf 6}, 110 (1933).
  
\bibitem{Clowe:2006eq} 
  D.~Clowe, M.~Bradac, A.~H.~Gonzalez, M.~Markevitch, S.~W.~Randall, C.~Jones, D.~Zaritsky and ,
  Astrophys.\ J.\  {\bf 648}, L109 (2006)
  [astro-ph/0608407].
  
\bibitem{Goldberg:1983nd} 
  H.~Goldberg,
  Phys.\ Rev.\ Lett.\  {\bf 50}, 1419 (1983)
  [Erratum-ibid.\  {\bf 103}, 099905 (2009)].
  
\bibitem{Hubble} 
  E.~Hubble, 
  in {\it Proceedings of the National Academy of Sciences}, 
  {\bf 15}, No. 3 (1929).

\bibitem{Perlmutter:1998np} 
  S.~Perlmutter {\it et al.}  [Supernova Cosmology Project Collaboration],
  Astrophys.\ J.\  {\bf 517}, 565 (1999)
  [astro-ph/9812133].
  
\bibitem{Riess:1998cb} 
  A.~G.~Riess {\it et al.}  [Supernova Search Team Collaboration],
  Astron.\ J.\  {\bf 116}, 1009 (1998)
  [astro-ph/9805201].
  
\bibitem{Caldwell:1997ii} 
  R.~R.~Caldwell, R.~Dave, P.~J.~Steinhardt and ,
  Phys.\ Rev.\ Lett.\  {\bf 80}, 1582 (1998)
  [astro-ph/9708069].
  
\bibitem{Frampton:2011sp} 
  P.~H.~Frampton, K.~J.~Ludwick, R.~J.~Scherrer and ,
  Phys.\ Rev.\ D {\bf 84}, 063003 (2011)
  [arXiv:1106.4996 [astro-ph.CO]].
  
\bibitem{Ade:2013xsa} 
  P.~A.~R.~Ade {\it et al.}  [Planck Collaboration],
  arXiv:1303.5062 [astro-ph.CO].
  
\bibitem{Siegel:1988yz} 
  W.~Siegel,
  hep-th/0107094.
  
\bibitem{Dimopoulos:1981yj} 
  S.~Dimopoulos, S.~Raby, F.~Wilczek and ,
  Phys.\ Rev.\ D {\bf 24}, 1681 (1981).
    
\bibitem{Ibanez:1981yh} 
  L.~E.~Ibanez, G.~G.~Ross and ,
  Phys.\ Lett.\ B {\bf 105}, 439 (1981).
      
\bibitem{Marciano:1981un} 
  W.~J.~Marciano, G.~Senjanovic and ,
  Phys.\ Rev.\ D {\bf 25}, 3092 (1982).
  
\bibitem{Peccei:1977hh} 
  R.~D.~Peccei, H.~R.~Quinn and ,
  Phys.\ Rev.\ Lett.\  {\bf 38}, 1440 (1977).
  
\bibitem{Ramond2}
  P.~Ramond, 
  "Group Theory", 
  Cambridge University Press (2010).

\bibitem{Carr:2007qw} 
  P.~D.~Carr, P.~H.~Frampton and ,
  hep-ph/0701034.
  
\bibitem{Frampton:2009pr} 
  P.~H.~Frampton, T.~W.~Kephart, R.~M.~Rohm and ,
  Phys.\ Lett.\ B {\bf 679}, 478 (2009)
  [arXiv:0904.0420 [hep-ph]].
  
\bibitem{Pontecorvo:1957qd} 
  B.~Pontecorvo,
  Sov.\ Phys.\ JETP {\bf 7}, 172 (1958)
  [Zh.\ Eksp.\ Teor.\ Fiz.\  {\bf 34}, 247 (1957)].

\bibitem{Maki:1962mu} 
  Z.~Maki, M.~Nakagawa, S.~Sakata and ,
  Prog.\ Theor.\ Phys.\  {\bf 28}, 870 (1962).

\bibitem{Aguilar:2001ty} 
  A.~Aguilar-Arevalo {\it et al.}  [LSND Collaboration],
  Phys.\ Rev.\ D {\bf 64}, 112007 (2001)
  [hep-ex/0104049].
  
\bibitem{AguilarArevalo:2012va} 
  A.~A.~Aguilar-Arevalo {\it et al.}  [MiniBooNE Collaboration],
  arXiv:1207.4809 [hep-ex].
  
\bibitem{Abazajian:2012ys} 
  K.~N.~Abazajian, M.~A.~Acero, S.~K.~Agarwalla, A.~A.~Aguilar-Arevalo, C.~H.~Albright, S.~Antusch, C.~A.~Arguelles and A.~B.~Balantekin {\it et al.},
  arXiv:1204.5379 [hep-ph].
  
\bibitem{Dirac:1928hu} 
  P.~A.~M.~Dirac,
  Proc.\ Roy.\ Soc.\ Lond.\ A {\bf 117}, 610 (1928).
  
\bibitem{Majorana:1937vz} 
  E.~Majorana,
  Nuovo Cim.\  {\bf 14}, 171 (1937).
  
\bibitem{Furry:1939qr} 
  W.~H.~Furry,
  Phys.\ Rev.\  {\bf 56}, 1184 (1939).
  
\bibitem{Frampton:1999yn} 
  P.~H.~Frampton, S.~L.~Glashow and ,
  Phys.\ Lett.\ B {\bf 461}, 95 (1999)
  [hep-ph/9906375].
  
\bibitem{Frampton:2002yf} 
  P.~H.~Frampton, S.~L.~Glashow, D.~Marfatia and ,
  Phys.\ Lett.\ B {\bf 536}, 79 (2002)
  [hep-ph/0201008].
    
\bibitem{Frampton:2002qc} 
  P.~H.~Frampton, S.~L.~Glashow, T.~Yanagida and ,
  Phys.\ Lett.\ B {\bf 548}, 119 (2002)
  [hep-ph/0208157].
    
\bibitem{Barger:1998ta} 
  V.~D.~Barger, S.~Pakvasa, T.~J.~Weiler, K.~Whisnant and ,
  Phys.\ Lett.\ B {\bf 437}, 107 (1998)
  [hep-ph/9806387].
    
\bibitem{Harrison:1994iv} 
  P.~F.~Harrison, D.~H.~Perkins, W.~G.~Scott and ,
  Phys.\ Lett.\ B {\bf 349}, 137 (1995).
    
\bibitem{Harrison:2002er} 
  P.~F.~Harrison, D.~H.~Perkins, W.~G.~Scott and ,
  Phys.\ Lett.\ B {\bf 530}, 167 (2002)
  [hep-ph/0202074].
  
\bibitem{Fuki:2006ag} 
  K.~Fuki, M.~Yasue and ,
  Phys.\ Rev.\ D {\bf 73}, 055014 (2006)
  [hep-ph/0601118].
  
\bibitem{Frampton:2008ci} 
  P.~H.~Frampton, S.~Matsuzaki and ,
  arXiv:0806.4592 [hep-ph].
  
\bibitem{Frampton:2008bz} 
  P.~H.~Frampton, T.~W.~Kephart, S.~Matsuzaki and ,
  Phys.\ Rev.\ D {\bf 78}, 073004 (2008)
  [arXiv:0807.4713 [hep-ph]].

\bibitem{Ma:2001dn} 
  E.~Ma, G.~Rajasekaran and ,
  Phys.\ Rev.\ D {\bf 64}, 113012 (2001)
  [hep-ph/0106291].

\bibitem{Babu:2002dz} 
  K.~S.~Babu, E.~Ma, J.~W.~F.~Valle and ,
  Phys.\ Lett.\ B {\bf 552}, 207 (2003)
  [hep-ph/0206292].
  
\bibitem{Ma:2005mw} 
  E.~Ma,
  Mod.\ Phys.\ Lett.\ A {\bf 20}, 2601 (2005)
  [hep-ph/0508099].
  
\bibitem{Ma:2005pd} 
  E.~Ma,
  Phys.\ Lett.\ B {\bf 632}, 352 (2006)
  [hep-ph/0508231].

\bibitem{Ma:2005qf} 
  E.~Ma,
  Phys.\ Rev.\ D {\bf 73}, 057304 (2006)
  [hep-ph/0511133].

\bibitem{Altarelli:2005yp} 
  G.~Altarelli, F.~Feruglio and ,
  Nucl.\ Phys.\ B {\bf 720}, 64 (2005)
  [hep-ph/0504165].

\bibitem{Altarelli:2005yx} 
  G.~Altarelli, F.~Feruglio and ,
  Nucl.\ Phys.\ B {\bf 741}, 215 (2006)
  [hep-ph/0512103].

\bibitem{Altarelli:2006kg} 
  G.~Altarelli, F.~Feruglio, Y.~Lin and ,
  Nucl.\ Phys.\ B {\bf 775}, 31 (2007)
  [hep-ph/0610165].

\bibitem{Ma:2006sk} 
  E.~Ma, H.~Sawanaka, M.~Tanimoto and ,
  Phys.\ Lett.\ B {\bf 641}, 301 (2006)
  [hep-ph/0606103].
  
\bibitem{Ma:2006ip} 
  E.~Ma,
  Mod.\ Phys.\ Lett.\ A {\bf 21}, 1917 (2006)
  [hep-ph/0607056].
  
\bibitem{Adhikary:2006wi} 
  B.~Adhikary, B.~Brahmachari, A.~Ghosal, E.~Ma, M.~K.~Parida and ,
  Phys.\ Lett.\ B {\bf 638}, 345 (2006)
  [hep-ph/0603059].
    
\bibitem{Cabibbo:1963yz} 
  N.~Cabibbo,
  Phys.\ Rev.\ Lett.\  {\bf 10}, 531 (1963).
  
\bibitem{Kobayashi:1973fv} 
  M.~Kobayashi, T.~Maskawa and ,
  Prog.\ Theor.\ Phys.\  {\bf 49}, 652 (1973).
  
\bibitem{Lee:2006pr} 
  T.~D.~Lee,
  hep-ph/0605017.
  
\bibitem{Frampton:1994rk} 
  P.~H.~Frampton, T.~W.~Kephart and ,
  Int.\ J.\ Mod.\ Phys.\ A {\bf 10}, 4689 (1995)
  [hep-ph/9409330].
  
\bibitem{Frampton:2000mq} 
  P.~H.~Frampton, T.~W.~Kephart and ,
  Phys.\ Rev.\ D {\bf 64}, 086007 (2001)
  [hep-th/0011186].

\bibitem{Chen:2007afa} 
  M.~-C.~Chen, K.~T.~Mahanthappa and ,
  Phys.\ Lett.\ B {\bf 652}, 34 (2007)
  [arXiv:0705.0714 [hep-ph]].
      
\bibitem{Feruglio:2007uu} 
  F.~Feruglio, C.~Hagedorn, Y.~Lin, L.~Merlo and ,
  Nucl.\ Phys.\ B {\bf 775}, 120 (2007)
  [Erratum-ibid.\  {\bf 836}, 127 (2010)]
  [hep-ph/0702194].
  
\bibitem{Frampton:2007et} 
  P.~H.~Frampton, T.~W.~Kephart and ,
  JHEP {\bf 0709}, 110 (2007)
  [arXiv:0706.1186 [hep-ph]].

\bibitem{Fairbairn:1963}
  W.M. Fairbairn, T. Fulton and W.H. Klink, 
  J.\ Math.\ Phys.\ {\bf 5,} 1038 (1964).
  
\bibitem{Eby:2008uc} 
  D.~A.~Eby, P.~H.~Frampton, S.~Matsuzaki and ,
  Phys.\ Lett.\ B {\bf 671}, 386 (2009)
  [arXiv:0810.4899 [hep-ph]].
  
\bibitem{preMoriond}
{\tt http://utfit.org/UTfit/ResultsWinter2013PreMoriond}.

\bibitem{Eby:2011aa} 
  D.~A.~Eby, P.~H.~Frampton and ,
  Phys.\ Rev.\ D {\bf 86}, 117304 (2012)
  [arXiv:1112.2675 [hep-ph]].
  
\bibitem{Nakamura:2010zzi} 
  K.~Nakamura {\it et al.}  [Particle Data Group Collaboration],
  J.\ Phys.\ G {\bf 37}, 075021 (2010).

\bibitem{Abe:2011sj} 
  K.~Abe {\it et al.}  [T2K Collaboration],
  Phys.\ Rev.\ Lett.\  {\bf 107}, 041801 (2011)
  [arXiv:1106.2822 [hep-ex]].
  
\bibitem{Dufour:2011zz} 
  F.~Dufour [T2K Collaboration],
  J.\ Phys.\ Conf.\ Ser.\  {\bf 335}, 012053 (2011).
  
\bibitem{Hartz:2012np} 
  M.~Hartz [T2K Collaboration],
  arXiv:1201.1846 [hep-ex].
  
\bibitem{Frank:2011zz} 
  E.~Frank [T2K Collaboration],
  PoS IDM {\bf 2010}, 103 (2011).
  
\bibitem{Izmaylov:2011np} 
  A.~Izmaylov [T2K Collaboration],
  arXiv:1112.0273 [hep-ex].
    
\bibitem{T2Ktalk}
  T.~Nakaya, 
  in Proceedings of Neutrino 2012 Conference, Kyoto, 2012 
  [Nucl.\ Phys.\ B\ Proc.\ Suppl. (to be published)].

\bibitem{MINOStalk}
  R.~Nichol,
  in Proceedings of Neutrino 2012 Conference, Kyoto, 2012 
  [Nucl.\ Phys.\ B\ Proc.\ Suppl. (to be published)].
  
\bibitem{Adamson:2012rm} 
  P.~Adamson {\it et al.}  [MINOS Collaboration],
  Phys.\ Rev.\ Lett.\  {\bf 108}, 191801 (2012)
  [arXiv:1202.2772 [hep-ex]].
  
\bibitem{Adamson:2011qu} 
  P.~Adamson {\it et al.}  [MINOS Collaboration],
  Phys.\ Rev.\ Lett.\  {\bf 107}, 181802 (2011)
  [arXiv:1108.0015 [hep-ex]].
  
\bibitem{Holin:2012np} 
  A.~Holin [MINOS Collaboration],
  PoS EPS {\bf -HEP2011}, 088 (2011)
  [arXiv:1201.3645 [hep-ex]].
  
\bibitem{Habig:2011zz} 
  A.~Habig [MINOS Collaboration],
  Nucl.\ Phys.\ Proc.\ Suppl.\  {\bf 218}, 320 (2011).
  
\bibitem{Orchanian:2011qq} 
  M.~Orchanian [MINOS Collaboration],
  arXiv:1109.6795 [hep-ex].
  
\bibitem{Evans:2010zza} 
  J.~Evans [MINOS Collaboration],
  PoS ICHEP {\bf 2010}, 298 (2010).
  
\bibitem{DCtalk}
  M.~Ishitsuka,
  in Proceedings of Neutrino 2012 Conference, Kyoto, 2012 
  [Nucl.\ Phys.\ B\ Proc.\ Suppl.\ (to be published)].

\bibitem{DC}
  H.~De Kerret,
  in {\it Proceedings of LowNu} 
  (Seoul National University, Seoul, 2011), 
  {\tt http://workshop.kias.re.kr/lownu11/}.

\bibitem{Palomares:2009wz} 
  C.~Palomares [Double-Chooz Collaboration],
  PoS EPS {\bf -HEP2009}, 275 (2009)
  [arXiv:0911.3227 [hep-ex]].
  
\bibitem{Abe:2011fz} 
  Y.~Abe {\it et al.}  [Double-Chooz Collaboration],
  Phys.\ Rev.\ Lett.\  {\bf 108}, 131801 (2012)
  [arXiv:1112.6353 [hep-ex]].
  
\bibitem{Palomares:2011zz} 
  C.~Palomares [Double-Chooz Collaboration],
  J.\ Phys.\ Conf.\ Ser.\  {\bf 335}, 012055 (2011).
  
\bibitem{An:2012eh} 
  F.~P.~An {\it et al.}  [DAYA-BAY Collaboration],
  Phys.\ Rev.\ Lett.\  {\bf 108}, 171803 (2012)
  [arXiv:1203.1669 [hep-ex]].
  
\bibitem{DBtalk}
  D.~Dwyer,
  in Proceedings of Neutrino 2012 Conference, Kyoto, 2012 
  [Nucl.\ Phys.\ B\ Proc.\ Suppl. (to be published)].
  
\bibitem{Ahn:2012nd} 
  J.~K.~Ahn {\it et al.}  [RENO Collaboration],
  Phys.\ Rev.\ Lett.\  {\bf 108}, 191802 (2012)
  [arXiv:1204.0626 [hep-ex]].
  
\bibitem{RENOtalk}
  S.-B.~Kim,
  in Proceedings of Neutrino 2012 Conference, Kyoto, 2012 
  [Nucl.\ Phys.\ B\ Proc.\ Suppl. (to be published)].

\bibitem{Fogli:2012ua} 
  G.~L.~Fogli, E.~Lisi, A.~Marrone, D.~Montanino, A.~Palazzo, A.~M.~Rotunno and ,
  Phys.\ Rev.\ D {\bf 86}, 013012 (2012)
  [arXiv:1205.5254 [hep-ph]].
  
\bibitem{Tortola:2012te} 
  D.~V.~Forero, M.~Tortola, J.~W.~F.~Valle and ,
  Phys.\ Rev.\ D {\bf 86}, 073012 (2012)
  [arXiv:1205.4018 [hep-ph]].
    
\bibitem{GonzalezGarcia:2012sz} 
  M.~C.~Gonzalez-Garcia, M.~Maltoni, J.~Salvado, T.~Schwetz and ,
  JHEP {\bf 1212}, 123 (2012)
  [arXiv:1209.3023 [hep-ph]].
  
\bibitem{Harrison:2005dj} 
  P.~F.~Harrison, W.~G.~Scott and ,
  Phys.\ Lett.\ B {\bf 628}, 93 (2005)
  [hep-ph/0508012].
  
\bibitem{Fuki:2006xw} 
  K.~Fuki, M.~Yasue and ,
  Nucl.\ Phys.\ B {\bf 783}, 31 (2007)
  [hep-ph/0608042].
  
\bibitem{Ge:2011qn} 
  S.~-F.~Ge, D.~A.~Dicus, W.~W.~Repko and ,
  Phys.\ Rev.\ Lett.\  {\bf 108}, 041801 (2012)
  [arXiv:1108.0964 [hep-ph]].
  
\bibitem{Ashie:2005ik} 
  Y.~Ashie {\it et al.}  [Super-Kamiokande Collaboration],
  Phys.\ Rev.\ D {\bf 71}, 112005 (2005)
  [hep-ex/0501064].
  
\bibitem{SKtalk}
  Y.~Itow,
  in Proceedings of Neutrino 2012 Conference, Kyoto, 2012 
  [Nucl.\ Phys.\ B\ Proc.\ Suppl. (to be published)].

\bibitem{Case:1956zz} 
  K.~M.~Case, R.~Karplus, C.~N.~Yang and ,
  Phys.\ Rev.\  {\bf 101}, 874 (1956).

\bibitem{Eby:2009ii} 
  D.~A.~Eby, P.~H.~Frampton, S.~Matsuzaki and ,
  Phys.\ Rev.\ D {\bf 80}, 053007 (2009)
  [arXiv:0907.3425 [hep-ph]].

\bibitem{Chen:2009gf} 
  M.~-C.~Chen, K.~T.~Mahanthappa and ,
  Phys.\ Lett.\ B {\bf 681}, 444 (2009)
  [arXiv:0904.1721 [hep-ph]].

\bibitem{Frampton:2009fw} 
  P.~H.~Frampton, S.~Matsuzaki and ,
  Phys.\ Lett.\ B {\bf 679}, 347 (2009)
  [arXiv:0902.1140 [hep-ph]].
  
\bibitem{Eby:2011ph} 
  D.~A.~Eby, P.~H.~Frampton, X.~-G.~He, T.~W.~Kephart and ,
  Phys.\ Rev.\ D {\bf 84}, 037302 (2011)
  [arXiv:1103.5737 [hep-ph]].
  
\bibitem{Babu:2007fx} 
  K.~S.~Babu, T.~W.~Kephart, H.~Pas and ,
  Phys.\ Rev.\ D {\bf 77}, 116006 (2008)
  [arXiv:0709.0765 [hep-ph]].
  
\bibitem{Eby:2011qa} 
  D.~A.~Eby, P.~H.~Frampton and ,
  Phys.\ Lett.\ B {\bf 713}, 249 (2012)
  [arXiv:1111.4938 [hep-ph]].
  
\bibitem{Milgrom:1983ca} 
  M.~Milgrom,
  Astrophys.\ J.\  {\bf 270}, 365 (1983).
  
\bibitem{Hirsch:2010ru} 
  M.~Hirsch, S.~Morisi, E.~Peinado, J.~W.~F.~Valle and ,
  Phys.\ Rev.\ D {\bf 82}, 116003 (2010)
  [arXiv:1007.0871 [hep-ph]].

\bibitem{Boucenna:2011tj} 
  M.~S.~Boucenna, M.~Hirsch, S.~Morisi, E.~Peinado, M.~Taoso, J.~W.~F.~Valle and ,
  JHEP {\bf 1105}, 037 (2011)
  [arXiv:1101.2874 [hep-ph]].

\bibitem{Frampton:2010uw} 
  P.~H.~Frampton, C.~M.~Ho, T.~W.~Kephart, S.~Matsuzaki and ,
  Phys.\ Rev.\ D {\bf 82}, 113007 (2010)
  [arXiv:1009.0307 [hep-ph]].
  
\bibitem{TW}
  A.D.~Thomas and G.V.~Wood,
  "Group Tables",
  Shiva Publishing Ltd. 1980.

\bibitem{Schechter:1980gr} 
  J.~Schechter, J.~W.~F.~Valle and ,
  Phys.\ Rev.\ D {\bf 22}, 2227 (1980).
  
\bibitem{Minkowski:1977sc} 
  P.~Minkowski,
  Phys.\ Lett.\ B {\bf 67}, 421 (1977).
  
\bibitem{Fukugita:1986hr} 
  M.~Fukugita, T.~Yanagida and ,
  Phys.\ Lett.\ B {\bf 174}, 45 (1986).
  
\bibitem{Kolb}
  E.W.~Kolb and M.S.~Turner, 
  “The Early Universe”, 
  Addison-Wesley (1993).

\bibitem{Cirelli:2005uq} 
  M.~Cirelli, N.~Fornengo, A.~Strumia and ,
  Nucl.\ Phys.\ B {\bf 753}, 178 (2006)
  [hep-ph/0512090].
  
\bibitem{DMsite}
  R. Gaitskell, J. Filippini, and D. Speller (2012),
  {\tt http://dendera.berkeley.edu/plotter/entryform.html}

\bibitem{Cholis:2008qq} 
  I.~Cholis, D.~P.~Finkbeiner, L.~Goodenough, N.~Weiner and ,
  JCAP {\bf 0912}, 007 (2009)
  [arXiv:0810.5344 [astro-ph]].
  
\bibitem{Aguilar:2013qda} 
  M.~Aguilar {\it et al.}  [AMS Collaboration],
  Phys.\ Rev.\ Lett.\  {\bf 110}, no. 14 (2013).
  
\bibitem{nuFuture}
  K.~Heeger,
  Fundamental Symmetries and Neutrinos Working Group 
  (Oak Ridge National Laboratory, Oak Ridge, 2012), 
  {\tt http://www.phy.ornl.gov/funsym/}.
  
\end{thebibliography}

\begingroup
\raggedright
\sloppy

\endgroup
\end{document}